\documentclass[12pt]{article}
\usepackage{amsmath}
\usepackage{graphicx}
\usepackage{caption}
\usepackage{enumerate}
\usepackage{natbib}
\usepackage{array}
\usepackage{algorithmic}
\usepackage{booktabs} 
\usepackage{subcaption}
\usepackage{pdflscape}  
\usepackage{threeparttable} 
\usepackage[plain]{algorithm}

\usepackage{setspace}
\newcommand*{\doublerule}{\hrule width \hsize height 0.5pt \kern 0.5mm \hrule width \hsize height 0.5pt}

\usepackage[colorlinks = true,
linkcolor = red,
urlcolor  = blue,
filecolor={red},
citecolor = blue,
anchorcolor = blue]{hyperref}
\usepackage{xurl}
\usepackage{siunitx}
\usepackage{mathtools}
\usepackage{etoolbox}
\usepackage{pdflscape}
\usepackage{adjustbox}
\usepackage{afterpage}
\usepackage{capt-of}

\newcommand{\proofend}{$\hfill\Box{~}$}
\newenvironment{Proof}{\noindent {\em{\bf Proof.}}}{\proofend\\}
\newcommand{\set}[1]{\left\{#1\right\}}

\usepackage{xcolor}

\DeclareMathOperator*{\argmin}{arg\,min}
\usepackage{caption}\newcommand\independent{\protect\mathpalette{\protect\independenT}{\perp}}
\def\independenT#1#2{\mathrel{\rlap{$#1#2$}\mkern2mu{#1#2}}}\usepackage{graphicx,latexsym,amssymb,amsmath}
\usepackage{subcaption}
\usepackage{bbm, bm}

\def\R{{\mathbb{R}}}

\def\P{{\Pr}}
\def\E{{\mathbb{E}}}
\usepackage{newfloat,comment}
\newtheorem{Lemma}{Lemma} 
\newtheorem{Assumption}{Assumption}
\newtheorem{Theorem}{Theorem}

\newtheorem{Corollary}{Corollary}

\begin{document}
\renewcommand{\baselinestretch}{1.2}
\title{Inference after discretizing time-varying unobserved heterogeneity\footnote{The authors thank Hugo Freeman for sharing his code from \cite{freeman2023linear} and Wei Miao for research assistance. Jad Beyhum thanks Juan Carlos Escanciano for encouraging him to pursue research on double machine learning applied to panel data. The authors are also grateful to Christophe Bruneel, Clément de Chaisemartin, Ferre De Graeve, Jan De Loecker, Geert Dhaene, Raffaella Giacomini, Hugo Freeman, Artūras Juodis, Dennis Kristensen, Ovidijus Stauskas, Andrew Shephard, Aleksey Tetenov, Frank Verboven, Jeffrey Wooldridge, Andrei Zeleneev, as well as seminar participants at Maastricht University, BI Oslo, the University of Glasgow, the University of Toronto, and University College London, for their helpful comments. They thank Martin Weidner for supporting a research visit to Oxford.
Jad Beyhum gratefully acknowledges financial support from the Research Fund KU Leuven (grant STG/23/014) and from the Research Foundation – Flanders (FWO) under project G031125N. Martin Mugnier gratefully acknowledges financial support from the French National Research Agency (ANR) under the “Investissements d'Avenir” program (grant ANR-17-EURE-0001) and the European Research Council (grant ERC-2018-CoG-819086-PANEDA).}}
  \author{Jad Beyhum\hspace{.2cm}\\
     Department of Economics, KU Leuven, Belgium\\
    and \\
    Martin Mugnier \\
    Paris School of Economics, France}

  \maketitle
\vspace{-1cm}
\begin{abstract}
Approximating time-varying unobserved heterogeneity by discrete types has become increasingly popular in economics. Yet, provably valid post-clustering inference for target parameters in models that do not impose an exact group structure is still lacking. This paper fills this gap in the leading case of a linear panel data model with nonseparable two-way unobserved heterogeneity. Building on insights from the double machine learning literature, we propose a simple inference procedure based on a bias-reducing moment. Asymptotic theory and simulations suggest excellent performance. In the application on fiscal policy we revisit, the novel approach yields conclusions in line with economic theory.
\end{abstract}


    \newpage

\setstretch{1.5}

\section{Introduction}
Accounting for unobserved heterogeneity is often critical for credible identification in both reduced-form and structural economic analyses. Among the various available dimension reduction devices, economists have increasingly relied on clustering 
techniques. This strategy has proven particularly effective in panel data models with time-varying unobserved heterogeneity. Early contributions developed valid inference procedures under the assumption of well-separated groups \citep[e.g.,][]{bonhomme2015grouped}. Recent work considers more realistic settings with continuous heterogeneity, using clustering only to approximate the unobserved structure through discretization \citep{bonhomme2022discretizing, freeman2023linear}. This approach has recently gained traction in empirical studies \citep[e.g.,][]{bonhomme2019distributional,jolivet2024structural,mahler2024lifestyle}. In these settings, however, while consistent two-step estimators and their rates of convergence are available, valid inference procedures are still lacking, so that practitioners are unable to assess the statistical significance of their results.\footnote{\cite{bonhomme2022discretizing} does have inference results when the unobserved heterogeneity is time-invariant, but lacks them in the case where it is time-varying, which we study here.}

This paper takes a first step toward filling this gap by establishing the asymptotic normality and unbiasedness, at the parametric rate, of a novel estimator for the slope coefficient in a linear semiparametric version of \cite{bonhomme2022discretizing}'s model. Specifically,  for units $i=1,\dots,N$ and dates $t=1,\dots,T$, we assume that there exist unobserved fixed effects $\alpha_i \in \mathbb{R}^{K_\alpha}$ and $\gamma_t \in \mathbb{R}^{K_\gamma}$ such that
\begin{equation}\label{eq:model}
\left\{
\begin{array}{l}
y_{it}  =x_{it}^\top \beta + f(\alpha_i, \gamma_t) + v_{it}, \\
 x_{itk} = h_k(\alpha_i, \gamma_t) + u_{itk}, \quad k=1,\dots,K,
\end{array}
\right. \quad 
\end{equation}
where $y_{it} \in \mathbb{R}$ is an observed outcome variable, $x_{it} = (x_{it1}, \ldots, x_{itK})^\top \in \mathbb{R}^K$ is a vector of observed covariates, $\E[x_{it}v_{it}]=0$, and $f(\alpha_i,\gamma_t)=\E[y_{it}-x_{it}^\top \beta|\alpha_i,\gamma_t]$ and $h_k(\alpha_i,\gamma_t)= \E[x_{itk}|\alpha_i,\gamma_t],\ k \in \{1, \dots, K\}$, are unknown deterministic mappings from $\mathbb{R}^{K_\alpha} \times \mathbb{R}^{K_\gamma}$ to $\mathbb{R}$. The main features of the model are that the mappings $f$ and $(h_k)_{k\in \{1,\dots,K\}}$ are smooth, the fixed effects are low-dimensional, and the error terms $v_{it}\in\R$ and $u_{it} = (u_{it1}, \ldots, u_{itK})^\top\in\R^K$ are uncorrelated and sufficiently weakly dependent across $i$ and $t$.\footnote{The existence of such fixed effects can be motivated by the literature on exchangeable arrays. If $(y_{it}, x_{it}^\top)^\top,\ i = 1, \ldots, N,\ t = 1, \ldots, T$, are exchangeable arrays, then the Aldous--Hoover--Kallenberg representation theorems guarantee that $\alpha_i$ and $\gamma_t$ exist and the residuals $v_{it}$ and $u_{it}$ are i.i.d. However, the low-dimensionality of the fixed effects, the smoothness of the functions $f$ and $(h_k)_{k \in \{1, \ldots, K\}}$, and the absence of correlation between $u_{it}$ and $v_{it}$ are not implied by the Aldous--Hoover--Kallenbeng representation.} Our interest lies in the unknown regression parameter $\beta\in\R^K$. Importantly, some fixed effects may appear in one equation only.\footnote{For example, suppose that $K=1$,
$y_{it} = x_{it1} \beta_1 + f^y(\alpha_i^y, \gamma_t^y) + v_{it} $, and $x_{it1} = h^x(\alpha_i^x, \gamma_t^x) + u_{it1}.$
This can be written as \eqref{eq:model}
with $\alpha_i=((\alpha_i^y)^\top,(\alpha_i^x)^\top)^\top$, $\gamma_t=((\gamma_t^y)^\top, (\gamma_t^x)^\top)^\top$, $f(\alpha_i,\gamma_t)= f^y(\alpha_i^y, \gamma_t^y)$, and $h_1(\alpha_i,\gamma_t)= h^x(\alpha_i^x, \gamma_t^x).$
    } 
  Moreoever, the fixed effects' contribution to the outcome, $f(\alpha_i, \gamma_t)$, can be flexibly correlated with the covariates through $(h_k(\alpha_i, \gamma_t))_{k \in \{1, \dots, K\}}$.
    In applications, $\alpha_i$ typically represents consumers' preferences, workers' abilities, or states' political structures, while $\gamma_t$ captures macroeconomic shocks and business cycles. We consider asymptotic regimes such that $N$ and $T$ grow to infinity while $K$ is fixed.

We propose a novel two-step estimation procedure that combines well-established methods from the existing literature. The first step constructs a discrete approximation of the unobserved heterogeneity using k-means clustering of unit-specific and time-specific informative moments (e.g., cross-section or time-series average of the data). The second step is a linear regression with additively separable two-way grouped fixed effects specific to each cluster, estimated by ordinary least squares (OLS). The main contribution of this paper is to formally establish that this simple combination can enjoy parametric-rate asymptotic normality and unbiasedness because it leverages bias-reducing Neyman-orthogonal moments. Such moments are standard in the double machine learning literature \citep[see][]{chernozhukov2018double} and, in the present context, mitigate the influence of estimation errors originating from the clustering step.\footnote{Robust moments, though used in the interactive fixed effects literature \citep{pesaran2006estimation,westerlund2015cross,beyhum2023factor,freeman2022linear}, are novel in the present non-interactive fixed effects setting.} As a by-product, standard OLS inference routines can be applied to the second-step regression. 

We establish the asymptotic normality of a cross-fitted version of the proposed estimator under the condition $\max(N, T) = o(\min(N, T)^3)$, up to logarithmic factors. Cross-fitting is a resampling technique from the double machine learning literature that simplifies the theoretical analysis. We evaluate the finite-sample performance of the proposed estimator and its cross-fitted version by means of Monte Carlo simulations. The baseline estimator exhibits excellent finite sample properties and slightly outperforms its cross-fitted variant, which itself significantly improves upon benchmark estimators. Notably, the confidence intervals from both estimators achieve nearly nominal coverage even when $T$ is much smaller than $N$, a common feature in many microeconomic datasets. We apply the methodology to assess the fiscal response of U.S.~states to resource revenues and find results closely aligned with predictions from economic theory. The proposed estimators are implemented in the R package \href{https://github.com/Wei-M-Wei/pcluster}{\texttt{pcluster}}.

Inference procedures have been proposed for models with a fixed effects structure corresponding to special cases of ours. A first related strand of literature is that on panel data models with interactive fixed effects. \cite{pesaran2006estimation}, \cite{greenaway2012asymptotic}, and \cite{westerlund2015cross} assume that both $f$ and $(h_k)_{k \in \{1, \ldots, K\}}$ are interactive and therefore known. \cite{bai2009panel} imposes that $f$ is known and interactive but does not model covariates. A second related strand of literature is that of the special case of grouped fixed effects models \citep[see, among others,][]{bonhomme2015grouped,chetverikov2022spectralpostspectralestimatorsgrouped,mugnier2024simple}, in which an exact group structure is assumed. Following \cite{bonhomme2022discretizing}, we do not assume that the fixed effects follow a group pattern but instead use clustering as an approximation device. 

Several recent papers consider models with nonseparable fixed effects (this structure is also called a ``nonlinear factor model" in the literature). Most closely related are \cite{freeman2023linear} and \cite{bonhomme2022discretizing}. \cite{freeman2023linear} study the same outcome model as us, leaving the relationship between the covariates and the fixed effects unrestricted. \cite{bonhomme2022discretizing} consider nonlinear versions of both the outcome and covariate models with a parametric likelihood specification of the distribution of $y_{it}$ given $x_{it},\alpha_i,\gamma_t$. Both \cite{freeman2023linear} and \cite{bonhomme2022discretizing} derive convergence rates but do not establish asymptotic normality, thus falling short of providing the inference tools we develop here. Our novel estimation procedure shares similarities with these two papers. The first-step clustering is the same as that of \cite{bonhomme2022discretizing}, but unlike the latter paper, we rely on additively separable two-way grouped fixed effects in the second step. \cite{freeman2023linear} employ the same second step as we do, but the first steps differ between the two papers.

Next, \cite{feng2020causal}, \cite{deaner2025inferring}, and \cite{athey2025identification} consider estimation of the average treatment effect on the treated of a binary treatment when the potential outcome in the absence of treatment follows a nonlinear factor model. In contrast to our proposal, these approaches allow for heterogeneous treatment effects but cannot be applied when the covariate of interest is continuous and do not rely on clustering.\footnote{Relatedly, it has been shown that synthetic control methods are valid under a linear factor model with a growing number of factors, which can be seen as the approximation of a nonlinear factor model, see \cite{arkhangelsky2021synthetic,arkhangelsky2023large}.} Finally, \cite{zeleneev2020identification} proposes estimators for linear and nonlinear network models with nonseparable fixed effects and obtains rates of convergence. Unlike in our work, regressors with an exact two-way structure are allowed in the latter paper.

The rest of the paper is organized as follows. Section~\ref{sec:est} introduces the two-step estimation procedure and discusses its link with double machine learning. We then describe the cross-fitted variant of the estimator and provide our main large sample result in Section~\ref{sec:th}. Section~\ref{sec:sim} displays the results of the Monte Carlo simulations. The application to fiscal policy is developed in Section~\ref{sec:app}. Section~\ref{sec:conc} concludes. Several key lemmas, all proofs, and results from additional Monte Carlo simulations are presented in the Appendix.

\section{Two-step estimation and double machine learning}\label{sec:est}

\subsection{Two-step estimation}\label{subsec.est}
We begin by providing some intuition for the estimation strategy. By plugging the covariate model into the outcome model, we have
\begin{equation*}\label{to_cluster} 
\begin{aligned}
    y_{it} &= h_{K+1}(\alpha_i,\gamma_t)+e_{it},
\end{aligned}\end{equation*}
where $h_{K+1}(\alpha_i,\gamma_t)= f(\alpha_i,\gamma_t) + \sum_{k=1}^K \beta_k h_k(\alpha_i,\gamma_t)$ and $e_{it}=u_{it}^\top \beta+ v_{it}$. Next, note that $\E[v_{it}|\alpha_i,\gamma_t]= \E[y_{it}-x_{it}^\top \beta-f(\alpha_i,\gamma_t)|\alpha_i,\gamma_t]=0,$ because $f(\alpha_i,\gamma_t)=\E[y_{it}-x_{it}^\top \beta|\alpha_i,\gamma_t]$. This, the fact that $x_{it}=h(\alpha_i,\gamma_t)+u_{it}$, and the contemporaneous exogeneity assumption $\E[x_{it}v_{it}]=0$, then imply
$$
\E[u_{itk}v_{it}]=0, \quad k=1,\ldots,K,
$$
so that $\beta$ is the slope coefficient in the linear regression of $e_{it}$ on $u_{it}$. Since $e_{it}$ and $u_{it}$ are both unobserved, this linear regression is infeasible but it suggests the following two-step estimation procedure for $\beta$: (i) estimate $e_{it}$ and $u_{it}$, and (ii) linearly regress the estimates of $e_{it}$ on those of $u_{it}$.

To estimate $e_{it}$ and $u_{it}$, we start by constructing a discrete approximation of unobserved heterogeneity across units and dates.\footnote{\label{foot:2}An alternative approach would be to discretize solely across one dimension (either units or dates). However, as noted in \cite{bonhomme2022discretizing} and \cite{freeman2023linear}, this leads to slower rates of convergence. See also \cite{beyhum2023factor} for a similar argument in panel data models with interactive fixed effects. Simulations in Section~\ref{sec:sim} confirm that discretizing along a single dimension yields much worse performance.} 
Following \cite{bonhomme2022discretizing}, we focus on the popular k-means clustering algorithm applied to cross-sectional and time-series averages of the data. This approach can be expected to perform well if such averages are informative about the underlying unobserved heterogeneity in a way that can be exploited by the discretization method (see Section~\ref{sec:asymp_assumptions} below). Other algorithms are discussed below. Estimates of $(e_{it},u_{it}^\top)^\top$ are  then obtained from the residuals of the linear projection of $y_{it}$ and $x_{it}$ on cluster-specific additively separable two-way fixed effects. The two main steps of the proposed estimation procedure are formally described below.
Let $z_{it}:=(x_{it}^\top,y_{it})^\top$.
\paragraph{Step 1 (Two-way clustering).} 
 Let $G$ and $C$ denote the number of unit and time clusters, respectively (a rule to select them is outlined below). Let $\|\cdot\|$ denote the Euclidean norm. 


\medskip

\noindent \textit{Clustering algorithm for units.} Let $a_i:=\frac{1}{T} \sum_{t=1}^T z_{it}, \ i\in\{1,\dots,N\}$. Compute
$$\left(\widehat{a}(1),\dots,\widehat{a}(G),g_1,\dots,g_N\right)\in\argmin\limits_{\begin{array}{c}a(1),\dots,a(G)\in\R^{K+1}\\\tilde g_1,\dots,\tilde g_N\in\{1,\dots,G\}\end{array}}\sum_{i=1}^N \left\| a_i-a(\tilde g_i)\right\|^2.$$

\medskip

\noindent \textit{Clustering algorithm for dates.} Let $b_t:=\frac{1}{N} \sum_{i=1}^N z_{it},\ t\in\{1,\dots,T\}$. Compute
$$\left(\widehat{b}(1),\dots,\widehat{b}(C),c_1,\dots,c_T\right)\in\argmin\limits_{\begin{array}{c}b(1),\dots,b(C)\in\R^{K+1}\\\tilde c_1\dots,\tilde c_T\in\{1,\dots,C\}\end{array}}\sum_{t=1}^T \left\| b_t-b(\tilde c_t)\right\|^2. $$
The procedures deliver unit and time cluster labels $g_1,\dots,g_N$ and $c_1,\dots,c_T$, respectively. Since some fixed effects can enter the outcome model but not the covariate model, or vice versa, it is crucial to include both $y_{it}$ and $x_{it}$ as inputs of each clustering algorithm. Else, such fixed effects would not be accounted for. 
Fast computational routines exist to find exact solutions to both k-means clustering problems for data sets of moderate sizes \citep[e.g.,][]{Merle1997AnIP,Aloise2009AnIC}, and local minima for others (e.g., Hartigan--Wong’s algorithm).\footnote{If the quality of the local minima raises suspicion, we recommend using hierarchical clustering approaches as outlined in Appendix~\ref{subsec.hiech_app} as a sensitivity analysis, though we leave the verification of their approximation properties for further research.}
\paragraph{Step 2 (Two-way grouped fixed effect estimator).}  The estimators of $e_{it}$ and $u_{it}$ are 
\begin{align*}
    \widehat{e}_{it}&:=y_{it}-\bar{y}_{g_it}- \bar{y}_{ic_t}+\bar{y}_{g_ic_t}, \\
    \widehat{u}_{it}&:=x_{it}-\bar{x}_{g_it}- \bar{x}_{ic_t}+\bar{x}_{g_ic_t},
\end{align*}
where, for any variable $w_{it}$, we define
\begin{align*}\bar{w}_{g_it}&:=\frac{1}{N_{g_i}}\sum_{j=1}^N \mathbf{1}\{g_j=g_i\} w_{jt},\\
\bar{w}_{ic_t}&:=\frac{1}{T_{c_t}}\sum_{s=1}^T \mathbf{1}\{c_s=c_t\} w_{is},\\
\bar{w}_{g_ic_t}&:=\frac{1}{N_{g_i}T_{c_t}}\sum_{j=1}^N\sum_{s=1}^T \mathbf{1}\{g_j=g_i\}\mathbf{1}\{c_s=c_t\} w_{js},
\end{align*}
with $N_{g_i}:= \sum_{j=1}^N \mathbf{1}\{g_j=g_i\}$ and $T_{c_t} :=\sum_{s=1}^T \mathbf{1}\{c_s=c_t\}$.
These estimators correspond to within-group transformations applied to $y_{it}$ and $x_{it}$ in a similar fashion to the standard within transformations in standard linear panel data models with two-way fixed effects. The final estimator of $\beta$ is the ordinary least squares estimator of $\widehat e_{it}$ on $\widehat u_{it}$,
\begin{equation*}\label{eq:OLS}
\widehat{\beta}=\left(\sum_{i=1}^N\sum_{t=1}^T\widehat{u}_{it}\widehat{u}_{it}^\top \right)^{-1}\sum_{i=1}^N\sum_{t=1}^T\widehat{u}_{it}\widehat{e}_{it},
\end{equation*}
which is numerically equivalent to the two-way grouped fixed effects regression coefficient
$$\argmin_{\beta\in\R^K}\min_{\delta\in\R^{N\times C}} \min_{\nu\in\R^{G\times T}} \sum_{i=1}^N\sum_{t=1}^T \left(y_{it}-x_{it}^\top\beta -\delta_{i,c_t}-\nu_{g_i,t}\right)^2.$$

In contrast, \cite{bonhomme2022discretizing} considers estimators with either only unit cluster fixed effects of the form $\delta_{i,c_t}$, or interacted unit and time clusters of the form $ \xi_{g_i,c_t}$. The additively separable grouped fixed effects structure that we use here delivers better rates of convergence, see  \citet[Sect. 2.2.1]{freeman2023linear} for a heuristic discussion. The grouped fixed effects estimator in \cite{freeman2023linear} uses a similar second step but a different first-step clustering procedure; essentially, their proposal uses clusters that approximate $f(\alpha_i,\gamma_t)$ but not $h(\alpha_i,\gamma_t)$.\footnote{The comparison between our estimator and that of \cite{freeman2023linear} is similar to the relation between, respectively, the Double Lasso and the Post-Lasso estimators, see \citet[Chap. 4]{chernozhukov2024applied} for a discussion. The post-Lasso estimator is not asymptotically normal since it only approximates the best linear predictor in the outcome equation. Approximating the best linear predictor in both the outcome and covariate equations, as done by the double Lasso, is key for inference.} In Section \ref{sec:sim}, we compare our approach with these alternative estimators in simulations. 

Since the two-way grouped fixed effects estimator relies on linear regression, usual standard errors (with a degree of freedom correction) can be used.  We note that extending the approach to accommodate a model with unit- or time-heterogeneous slopes ($\beta_i$ or $\beta_t$) is relatively straightforward.

\paragraph{Choice of the number of clusters.}\label{sec:choice} 
To choose the number of clusters $G$ and $C$, we use the data-driven selection procedure developed by \cite{bonhomme2022discretizing}.  Let 
$Q_g(G):=\frac{1}{N}\sum_{i=1}^N \left\| a_i-a(g_i)\right\|^2$
and 
$Q_c(C):=\frac{1}{T}\sum_{t=1}^T \left\| b_t-b( c_t)\right\|^2$ denote the k-means objective functions evaluated at their maxima.
The quantities $Q_g(G)$ and $Q_c(C)$ measure the approximation errors made through the clustering. Let
$ \widehat{V}_g:=\frac{1}{NT^2}\sum_{i=1}^N \sum_{t=1}^T\left\|z_{it}- a_i\right\|^2$ and $
\widehat{V}_c:=\frac{1}{N^2T}\sum_{t=1}^T\sum_{i=1}^N \left\|z_{it}- b_t\right\|^2$
denote empirical dispersions, which measure the fundamental noise level in the inputs of the clustering procedures. The data-driven choice of the number of clusters is $\widehat{G}:=\min_{G\ge 1}\{G:\ Q_g(G)\le \widehat{V}_g\}$ and  $\widehat{C}:=\min_{C\ge 1}\{C:\ Q_c(C)\le \widehat{V}_c\}$. It aims at
balancing the approximation error and the input noise. We provide some theoretical guarantees in Section~\ref{subsec.dd}.

\paragraph{On the clustering algorithm.}
As in \cite{bonhomme2022discretizing}, the baseline approach clusters on cross-section and time-series averages using a k-means algorithm. Intuitively, this procedure requires that these averages be informative about the fixed effects. This leads to an ``injectivity'' condition, formalized in Assumption~\ref{as.clus} below, which imposes that the limit of the averages is injective in the fixed effects. Such an assumption can be relaxed or avoided. One solution is to use moments beyond averages, leading to weaker restrictions. Another approach, studied in Appendix~\ref{subsec.hiech_app}, uses hierarchical clustering on the pseudo-distance of \cite{zhang2017estimating}, avoiding averaging the data before clustering. We focus on k-means clustering of averages in the main text because of its simplicity and excellent performance in simulations.

\subsection{Link with double machine learning}\label{subsec:dml}  

Two-step estimation procedures whose second steps are based on Neyman-orthogonal moments lie at the heart of the double machine learning literature \citep[e.g.,][]{chernozhukov2018double}. Such moments are bias-reducing because they limit the influence of the errors in estimating the nuisance parameters in the first step and, therefore, make inference possible. This robustness property arises because the difference between the empirical counterpart of the Neyman-orthogonal moment and the infeasible empirical moment based on the true values of the nuisance parameters decomposes into sums of either products of estimation errors or products of an estimation error and an error term; see in particular the discussion in Section 1 of \cite{chernozhukov2018double}. It turns out that the moment on which our second-step estimator is based exhibits the same type of robustness properties. 
To see this, note that the second-step estimator solves the empirical moment equation
\begin{equation}\label{Neyman_moment}\frac{1}{NT}\sum_{i=1}^N\sum_{t=1}^T \widehat{u}_{it}(\widehat{e}_{it} -\widehat{u}_{it}^\top \beta)=0.\end{equation}
Moment \eqref{Neyman_moment} approximates the empirical moment equation 
\begin{equation}\label{Oracle_moment} \frac{1}{NT} \sum_{i=1}^N \sum_{t=1}^T u_{it} (e_{it}-u_{it}^\top \beta)=0,\end{equation}
 solved by an infeasible ``oracle'' OLS estimator knowing $u_{it}$ and $e_{it}$. Notice that 
\begin{align*}
   \frac{1}{NT} \sum_{i=1}^N\sum_{t=1}^T \widehat{u}_{it}(\widehat{e}_{it} -\widehat{u}_{it}^\top \beta) -  \frac{1}{NT} \sum_{i=1}^N \sum_{t=1}^T u_{it} (e_{it}-u_{it}^\top \beta)=a^*+b^*+c^*,
\end{align*}
where 
\begin{align*}
    a^*&:=  \frac{1}{NT} \sum_{i=1}^N\sum_{t=1}^T (\widehat{u}_{it}-u_{it})(\widehat{e}_{it}-e_{it} -(\widehat{u}_{it}-u_{it})^\top \beta),\\
    b^*&:= \frac{1}{NT} \sum_{i=1}^N\sum_{t=1}^T (\widehat{u}_{it}-u_{it})v_{it},\\
    c^*&:= \frac{1}{NT} \sum_{i=1}^N\sum_{t=1}^T u_{it}(\widehat{e}_{it}-e_{it} -(\widehat{u}_{it}-u_{it})^\top \beta).
\end{align*}
Hence, the difference between the moments \eqref{Neyman_moment} and \eqref{Oracle_moment} is the sum of a term $a^*$, corresponding to the sum of the products of two estimation errors, and two terms $b^*$ and $c^*$ which are sums of products of an estimation error and an error term. All of these terms are, therefore, sums of products of ``small terms" and will thus be asymptotically negligible. This explains why the proposed estimator can be asymptotically normal. 


\section{Asymptotic theory}\label{sec:th}

In this section, we provide theoretical 
guarantees for a cross-fitted variant of the estimator. In Section \ref{sec:cf}, we motivate and discuss the use of cross-fitting. Section~\ref{sec:OLS_cs} introduces the cross-fitted version of the two-step estimator. Section~\ref{sec:asymp_assumptions} provides sufficient conditions for its asymptotic normality. Section~\ref{sec:asymp_results} formally presents the large sample result. Section \ref{subsec.dd} contains some results regarding the data-driven choice of the number of clusters.

\subsection{On the use of cross-fitting}\label{sec:cf}
Deriving the limiting distribution of the least-squares estimator $\widehat{\beta}$ is challenging, as it requires controlling the dependence between the clusters estimated in the first step and the error terms of the data used in the second step. This difficulty is a common feature of many two-step estimators based on highly nonlinear black-box first-step estimators.\footnote{In particular, without a control of the dependence between the two steps, one cannot use concentration arguments on $u_{it}$ and $v_{it}$ to bound the terms $b^*$ and $c^*$ introduced in Section \ref{subsec:dml}.}

This type of issue has also been encountered in the literature on double machine learning \citep{chernozhukov2018double}. The solution taken in this research area is to use cross-fitting. The data is split into different folds, and the first-step and second-step estimations are performed on different folds. The role of the folds is then reversed, and the second-step estimators over the different folds are averaged to improve efficiency. Under independent observations, this mechanically eliminates the dependence between the first-step estimator and the data used in the second step, therefore solving the aforementioned problem. 

In this section, we follow this strategy to establish the asymptotic normality at the parametric $\sqrt{NT}$-rate of a cross-fitted version of the estimator that learns  clusters and estimates the slope coefficient from separate batches of the data. 

We emphasize that cross-fitting is merely a proof device, and we recommend using $\widehat{\beta}$ in practice. Indeed, Monte Carlo simulations in Section~\ref{sec:sim} demonstrate that the original estimator $\widehat{\beta}$ outperforms its cross-fitted version, which already performs very well.
Cross-fitting has been shown not to improve estimator performance in simulations across various settings \citep{dukes2021inference, chen2022debiased, vansteelandt2024assumption, wang2024doubly, shi2024off}. Moreover, it has been demonstrated that cross-fitting is not always essential for achieving asymptotic results in double machine learning when the learners adhere to a natural leave-one-out stability property \citep{chen2022debiased} or the lasso is used \citep{chernozhukov2015post}. These findings suggest that in certain contexts, cross-fitting is not only unnecessary but may even be counterproductive. Our simulation results indicate that k-means clustering is one such learner where cross-fitting can be omitted without compromising performance.\footnote{An intuition of why cross-fitting is not needed in practice is as follows. In our case, the clusters are fully determined by cross-section and time series averages of the variables. A particular observation should be only very weakly dependent on these averages (unless the variables have heavy tails) so that the clustering step does not overfit. In contrast, if one replaces k-means on averages by hierarchical clustering on a pseudo-distance as studied in Appendix \ref{subsec.hiech_app}, the clusters do not depend only on the averages, and we see in simulations that the estimator without cross-fitting does not perform well. }

\subsection{Alternative estimator with cross-fitting}\label{sec:OLS_cs}
To describe the alternative estimator based on cross-fitting, let us
consider a simple cross-fitting scheme with only four folds:
\begin{align*}
    \mathcal{O}_1&:=\{1,\dots,\lfloor N/2\rfloor\}\times \{1,\dots,\lfloor T/2\rfloor\}=:\mathcal{N}_1\times \mathcal{T}_1,\\
    \mathcal{O}_2&:=\{1,\dots,\lfloor N/2\rfloor\}\times \{\lfloor T/2\rfloor+1,\dots,T\}=: \mathcal{N}_2\times \mathcal{T}_2,\\
        \mathcal{O}_3&:=\{\lfloor N/2\rfloor+1,\dots,N\}\times \{1,\dots,\lfloor T/2\rfloor\}=: \mathcal{N}_3\times \mathcal{T}_3,\\
            \mathcal{O}_4&:=\{\lfloor N/2\rfloor+1,\dots,N\}\times \{\lfloor T/2\rfloor+1,\dots,T\}=: \mathcal{N}_4\times \mathcal{T}_4.
\end{align*}
We also use the notation $N_d:=|\mathcal{N}_d|$ and $T_d:=|\mathcal{T}_d|$ and note that $\mathcal N_1=\mathcal N_2$, $\mathcal N_3=\mathcal N_4$, $\mathcal T_1=\mathcal T_2$, and $\mathcal T_3=\mathcal T_4$. This type of division in four folds is appropriate for panel data and also appears in \cite{freeman2023linear}.\footnote{In unreported simulations, we have not found any improvement resulting from increasing the number of folds.}

We briefly outline the construction of the cross-fitted estimator, denoted $\widehat{\beta}^{\rm CF}$. For a detailed presentation, we refer to Appendix \ref{sub.app_CF}. For an observation $(i,t) \in \mathcal{O}_d$, we estimate $u_{it}$ and $e_{it}$ as follows:

\begin{enumerate}
    \item Estimate $G_d$ unit clusters $g_i^d$ for $i \in \mathcal{N}_d$ using time series averages $a_i^d$, computed from the fold with the same units as fold $d$ but different dates.
    \item Estimate $C_d$ unit clusters $c_t^d$ for $t \in \mathcal{T}_d$ using cross-section averages $b_t^d$, computed from the fold with the same dates as fold $d$ but different units.
    \item Use these time and unit clusters on the data of fold $\mathcal{O}_d$ to obtain the estimators $\widehat{u}_{it}^d$ and $\widehat{e}_{it}^d$ through within-group transformation.
\end{enumerate}
The final estimator $\widehat{\beta}^{\rm CF}$ is the pooled OLS estimator from regressing $\widehat{e}_{it}^d$ on $\widehat{u}_{it}^d$. This procedure determines cluster memberships using data distinct from that used for within-transformations, thereby simplifying the theoretical analysis while maintaining efficiency across the entire dataset.

\subsection{Assumptions}\label{sec:asymp_assumptions} 
Consider the following assumptions.
\begin{Assumption}[Heterogeneity]\label{as.fac}
 The functions $(h_k)_{k\in\{1,\dots,K+1\}}$ are bounded and twice differentiable with second-order derivatives bounded uniformly in the support of $(\alpha_i,\gamma_t)$.
\end{Assumption}

\begin{Assumption}[Injectivity]\label{as.clus} 
For all $d\in\{1,\dots,4\}$:
\begin{enumerate}[\textup{(}i\textup{)}]  
\item\label{clusi} There exist a Lipschitz-continuous function $\varphi_d^\alpha$ and a sequence $\{r_\alpha\}$ such that $$\max_{i\in\mathcal N_d}\left\|a_i^d-\varphi_d^\alpha(\alpha_i)\right\|^2=O_P\left(\frac{r_\alpha}{T}\right)$$ as $N,T$ tend to infinity. Moreover, there exists a Lipschitz-continuous function $\psi_d^\alpha$ such that, for all $i\in\mathcal N_d$, $\alpha_{i}=\psi_d^\alpha(\varphi_d^\alpha(\alpha_i))$.
\item\label{clusii} There exist a Lipschitz-continuous function $\varphi_d^\gamma$ and a sequence $\{r_\gamma\}$ such that $$\max_{t\in\mathcal T_d}\left\|b_t^d-\varphi_d^\gamma(\gamma_t)\right\|^2=O_P\left(\frac{r_\gamma}{N}\right)$$ as $N,T$ tend to infinity. Moreover, there exists a Lipschitz-continuous function $\psi_d^\gamma$ such that, for all $t\in\mathcal T_d$, $\gamma_{t}=\psi_d^\gamma(\varphi_d^\gamma(\gamma_t))$.
\end{enumerate}
\end{Assumption}
Assumption \ref{as.fac} is a mild regularity condition on $(h_k)_{k\in\{1,\dots,K+1\}}$. Assumption~\ref{as.clus} is similar to Assumption 2 in \cite{bonhomme2022discretizing}. It is best understood in the case of pointwise limits, where $\text{plim}_{T\to \infty}a_i^d=\varphi_d^\alpha(\alpha_i)$ and $\text{plim}_{N\to \infty}b_t^d=\varphi_d^\gamma(\gamma_t)$, which can be justified by laws of large numbers. Assumption~\ref{as.clus} then requires that the probability limits are injective and imposes some rate of convergence of the sample averages to these limits.

Let us first discuss the injectivity property. It requires that units (resp.~time periods) with similar values of time-series (resp.~cross-sectional) averages of $z_{it}$ have similar values of unit-specific (resp.~time-specific) fixed effects and vice versa, with equality in the limit. Intuitively, such an injectivity property suggests that matching on observed panel data averages is sufficient to control for unobserved heterogeneity (i.e., matching on the fixed effects). It is also useful to analyze the injectivity assumption in an example. Consider the case where $K=d_\alpha=d_\gamma=1$, $\beta=0$ and $f(\alpha_i,\gamma_t)=h_1(\alpha_i,\gamma_t)=\alpha_i\gamma_t$. Then, under weak regularity conditions, $\varphi_d^\alpha(\alpha_i) =(\alpha_i,\alpha_i)^\top \E[\gamma_t]$ and injectivity fails to hold only if $\E[\gamma_t]=0$, showing that failure of injectivity is the exception rather than the norm in this setting.\footnote{Also note that this assumption is related to the full rank condition in the common correlated effects literature \citep{pesaran2006estimation}, which guarantees that cross-section averages allow for the recovery of the factors.} As noted earlier, the injectivity property can be avoided by using different clustering approaches such as hierarchical clustering applied on a pseudo-distance matrix, as we study in Appendix \ref{subsec.hiech_app}.

Next, in Assumption~\ref{as.clus}, the rate of convergence of $a_i^d$ and $b_i^d$ to their probability limits in sup-norm is controlled by the sequences $r_\alpha$ and $r_\gamma$. Concentration inequalities \citep{boucheron2013concentration} can be used to show that the bounds hold for particular values of $r_\alpha$ and $r_\gamma$ under different dependence settings and conditions on the tails of the distribution of $z_{it}$. For instance, we show in Lemma \ref{lmm.suff_c} in Appendix \ref{sec.suff_app} that if, conditional on $\alpha_i$, $(z_{it})_{t\in\mathcal{T}_d}$ are independent sub-Gaussian random variables with with common mean $\E[z_{it}|\alpha_i]$ and sub-Gaussian norm bounded uniformly in $t$ and the value of $\alpha_i$, then the bound on $\max_{i\in\mathcal N_d}\left\|a_i^d-\varphi_d^\alpha(\alpha_i)\right\|^2$ in Assumption \ref{as.clus}\eqref{clusi} holds with $r_\alpha=\log(N)$. Under analogous conditions, the bound in Assumption \ref{as.clus}\eqref{clusii} holds with $r_\gamma=\log(T)$. As a result, $r_\alpha$ and $r_\gamma$ will typically be negligible with respect to $N$ and $T$, respectively.


The following assumption collects standard dependence, moment, and non-collinearity conditions that prove helpful in establishing the limiting distribution of the estimator. Let $\mathcal F_{NT}$ denote the sigma-algebra generated by $\{\alpha_i,\gamma_t:(i,t)\in\set{1\ldots,N}\times\set{1,\ldots,T}\}$. 

\begin{Assumption}[Dependence, moments, and non-collinearity]\label{as.errors}
~
\begin{enumerate}[\textup{(}i\textup{)}]  
\item\label{errorsi} 
Conditional on $\mathcal F_{NT}$, $(v_{it})_{(i,t)\in\set{1,\ldots,N}\times\set{1,\ldots,T}}$ and $(u_{it})_{(i,t)\in\set{1,\ldots,N}\times\set{1,\ldots,T}}$ are independent sequences of independently distributed mean-zero random vectors. 
\item\label{errorsii} There exist positive constants $\delta>0$ and $M>0$ such that, almost-surely,  
$$\E\left[\vert v_{it}\vert^{2+\delta}+\vert u_{itk}\vert ^{2+\delta}|\mathcal F_{NT}\right]\leq M$$ 
for all $i,t,k$.
\item\label{errorsiii} There exist positive definite matrices $\Sigma_U$ and $\Omega$ such that, as $N$ and $T$ tend to infinity, 
\begin{align*}
&\frac{1}{NT}\sum_{i=1}^N\sum_{t=1}^Tu_{it}u_{it}^\top\overset{p}{\to} \Sigma_U, \\
&\frac{1}{NT}\sum_{i=1}^N\sum_{t=1}^T\E\left[v_{it}^2u_{it}u_{it}^\top|\mathcal F_{NT}\right]\overset{a.s.}{\to} \Omega.
\end{align*}
\end{enumerate}
\end{Assumption}
Assumption~\ref{as.errors}\eqref{errorsi} rules out conditional cross-section or time-series dependence in the error terms, and requires errors to have zero conditional mean, i.e., that they are are mean-independent of the fixed effects, a standard assumption in the panel data literature. It implies that the data from the different folds are independent conditional on the fixed effects. Though it may be arguably strong, relaxing it would require obtaining a precise control of the dependence between the clustering algorithm's outcome and the error terms, which, as noted earlier, is particularly challenging with black-box methods such as k-means. In the simulations reported in Section \ref{sec:sim}, we find that the estimator still performs very well under time series correlation. Note that the assumption of i.i.d.~errors is commonly made in papers studying sophisticated panel data models; see, for instance, \cite{moon2015linear}, \cite{chen2021quantile}, 
\cite{bonhomme2022discretizing}, and \cite{freeman2023linear}. Similar to our work, these papers derive their main theoretical results under this assumption but provide simulation evidence suggesting that the restriction may not be necessary. 

Assumption~\ref{as.errors}\eqref{errorsii} requires the idiosyncratic component of each equation to admit slightly more than an uniformly bounded conditional second moment across units, time periods, and regressors. This is useful to verify a Lindeberg--Feller condition and apply a central limit theorem to the dominant term in the estimator. 

The first part of Assumption \ref{as.errors}\eqref{errorsiii} is a standard asymptotic non-collinearity condition on the covariates in the second-step regression. Together with the second part of Assumption \ref{as.errors}\eqref{errorsiii}, it ensures that the estimator possesses a non-degenerate limiting distribution.



The following assumption specifies the relative rates at which $N$, $T$, and the numbers of clusters $G_d$ and $C_d$ can grow.

\begin{Assumption}[Asymptotics]
    \label{as.rates}
For all $d\in\{1,\dots,4\}$, as $N,T,G_d,C_d$ tend to infinity,
\begin{enumerate}[\textup{(}i\textup{)}]  
\item\label{ratesi} $r_\alpha^4 N=o(T^3)$, $r_\gamma^4 T=o(N^3)$.
\item\label{ratesii}
$G_d=o(N)$, $C_d= o(T)$.
\end{enumerate}
\end{Assumption}

As noted above, under the standard conditions of Lemma \ref{lmm.suff_c}, $r_\alpha=\log(N)$ and $r_\gamma =\log(T)$ and Assumption~\ref{as.rates}\eqref{ratesi} becomes $\max(N,T)=o(\min(N,T)^3)$ up to logarithmic terms. The latter is weaker than the rate conditions on $N$ and $T$ typically found in the literature on panel data models with interactive fixed effects. For instance, \cite{bai2009panel} imposes $\max(N,T)=o(\min(N,T)^2)$ to derive asymptotic normality, while the estimators in \cite{westerlund2015cross} are biased as $T/N$ goes to a constant. This improvement is substantial, as it is obtained while relaxing the modeling assumption that $g$ and $h_k$ are interactive. We relax the condition in \cite{bai2009panel} thanks to the use of the orthogonal moment stemming from the covariate equations, while we improve on \cite{westerlund2015cross} by estimating both unit and time-specific fixed effects in the first step, while \cite{westerlund2015cross} only estimate the factors (corresponding to the time-specific fixed effects in an interactive fixed effects model); see also Footnote \ref{foot:2} for a related discussion.
In contrast, the rate condition \eqref{ratesi} is stronger than that for grouped fixed effects models such as in \cite{bonhomme2015grouped}, where $T$ can grow at an arbitrary polynomial rate with respect to $N$. This is because we do not assume that the data has an exact group structure, and instead use clustering as an approximation device.

Assumption~\ref{as.rates}\eqref{ratesii} stipulates that both the number of unit clusters and time clusters must be negligible with respect to $N$ and $T$, respectively. Intuitively, this is necessary because, otherwise, the within transformations applied to the data to estimate $e_{it}$ and $u_{it}$ would create non-negligible time series and cross-section dependence in the generated regressors of the second step, precluding the estimator from being $\sqrt{NT}$-consistent.

The last assumption concerns the approximation error of an infeasible ``oracle'' approximation procedure that would directly cluster the unobserved unit and time fixed effects. We follow \cite{bonhomme2022discretizing} and define such approximation errors as, for all $d\in\{1,\ldots,4\}$,
\begin{equation*}
    B_\alpha^d(G_d):= \min\limits_{\begin{array}{c}\alpha(1),\dots,\alpha(G_d)\in\R^{K_\alpha}\\\tilde g_i\in\{1,\dots,G_d\},\ i\in\mathcal{N}_{d}\end{array}}\frac{1}{N_d}\sum_{i\in\mathcal{N}_{d}} \left\|\alpha_i-\alpha(\tilde g_i)\right\|^2
\end{equation*}
and
\begin{equation*}
    B_\gamma^d(C_d):= \min\limits_{\begin{array}{c}\gamma(1),\dots,\gamma(C_d)\in\R^{K_\gamma}\\\tilde c_t\in\{1,\dots,C_d\},\ t\in\mathcal{T}_{d}\end{array}}\frac{1}{T_d}\sum_{t\in\mathcal{T}_{d}} \left\|\gamma_t-\gamma(\tilde c_t)\right\|^2.
\end{equation*}
Lemma \ref{lm.clus} in Section~\ref{sec:asymp_results} below suggests that, due to the injectivity condition (Assumption \ref{as.clus}), the k-means clustering algorithm used in the first step achieves an approximation error close to the infeasible oracle k-means algorithm (that is $ B_\alpha^d(G_d)$, $B_\gamma^d(C_d)$). Next, we require this approximation error of the clustering algorithm to be small enough for the estimator to be asymptotically normal. This is subsumed in the next assumption below.
\begin{Assumption}[Approximation error]
    \label{as.approx}
For all $d\in\{1,\dots,4\}$, as $N,T,G_d,C_d$ tend to infinity,
\[
B_\alpha^d(G_d)=o_P\left((NT)^{-1/4}\right) \text{ and } B_\gamma^d(C_d)=o_P\left((NT)^{-1/4}\right).
\]
\end{Assumption}
Assumption~\ref{as.approx} requires the oracle approximation error resulting from discretizing the unobserved heterogeneity to decrease sufficiently fast as the sample size increases. 
Intuitively, this condition requires the number of clusters to increase at a rate governed by the difficulty of the approximation problem, which itself depends on the dimensions of the fixed effects $K_\alpha$ and $K_\gamma$. As discussed in \cite{freeman2023linear} and \cite{bonhomme2022discretizing}, a precise dependence of the approximation error on $K_\alpha$ and $K_\gamma$ can be obtained under further regularity conditions on the distribution of $\alpha_i$ and $\gamma_t$.
\begin{Lemma}[\cite{graf20002quantization}]\label{lm.graf}
Let $\alpha_i$ and $\gamma_t$ be i.i.d.~random vectors with compact supports. Then, for all $d\in\{1,\ldots,4\}$, as $N,T,G_d,C_d$ tend to infinity we have
\[
    B_\alpha^d(G_d)=O_P\left((G_d)^{-\frac{2}{K_\alpha}}\right) \text{ and }  B_\gamma^d(C_d)=O_P\left((C_d)^{-\frac{2}{K_\gamma}}\right).
\]
\end{Lemma}
Lemma~\ref{lm.graf} shows that the approximation error decreases at a rate inversely proportional to the dimension of the underlying fixed effects. The assumption that $\alpha_i$ and $\gamma_t$ are i.i.d with compact support is only a sufficient condition that may not be necessary. While it may be restrictive for some applications 
and the result might hold under departures from this assumption, proving the validity of such an extension is beyond the scope of this paper. In the Monte Carlo study, the estimator continues to perform well when the time-specific fixed effects exhibit autocorrelation and have an unbounded support. We note that the assumption of i.i.d.~fixed effects with compact support is invoked in Assumption S2(i) in \cite{bonhomme2022discretizing}. Using Lemma \ref{lm.graf}, we obtain the following corollary, which gives sufficient conditions for Assumption \ref{as.approx}.
\begin{Corollary}\label{cor}
    Let $\alpha_i$ and $\gamma_t$ be i.i.d.~random vectors with compact supports. Then, Assumption \ref{as.approx} holds if for all $d\in\{1,\ldots,4\}$, as $N,T,G_d,C_d$ tend to infinity, we have
    \[
    (NT)^{K_\alpha/8}=o(G_d) \text{ and } (NT)^{K_\gamma/8}=o(C_d).
    \]
\end{Corollary}
Note that, when $N$ and $T$ grow at the same rate, the rate conditions of Corollary \ref{cor} and Assumption \ref{as.rates}\eqref{ratesii} can only hold together if $K_\alpha\le 3$ and $K_\gamma\le 3,$ so that we are imposing a restriction on the dimensions of the fixed effect spaces.

\subsection{Asymptotic results}\label{sec:asymp_results}
Our first asymptotic result is Lemma~\ref{lm.clus} below. It states that the clustering algorithm groups together units (resp.~time periods) with similar unit (resp.~time) fixed effects, up to the oracle approximation error. 
A similar type of result is Lemma 1 in \cite{bonhomme2022discretizing}.
\begin{Lemma}\label{lm.clus}
Let Assumption~\ref{as.clus} hold. Then, for every $d\in\{1,\ldots,4\}$, as $N,T,G_d,C_d$ tend to infinity we have
\begin{enumerate}[\textup{(}i\textup{)}]  
\item\label{approxi}
    $\frac{1}{N_d}\sum_{i\in\mathcal{N}_d} \left\|\alpha_i-\frac{1}{N^d_{g_i^d}}\sum_{j\in\mathcal N_d}\mathbf{1}\{g_j^d= g_i^d\}\alpha_j\right\|^2=O_P\left(\frac{r_\alpha}{T} +
B_\alpha^d(G_d)\right)$,
\item\label{approxii}
    $\frac{1}{T_d}\sum_{t\in\mathcal{T}_d} \left\|\gamma_t-\frac{1}{T^d_{c_t^d}}\sum_{s\in\mathcal T_d} \mathbf{1}\{c_s^d= c_t^d\}\gamma_s\right\|^2=O_P\left(\frac{r_\gamma}{N} +
B_\gamma^d(C_d)\right)$.
\end{enumerate}
\end{Lemma}
Lemma~\ref{lm.clus} suggests that injectivity ensures that if the approximation errors resulting from discretizing the unobserved heterogeneity based on the unobserved heterogeneity itself, $B_\alpha(G_d)$ and $B_\gamma(C_d)$, are small, then the approximation errors resulting from discretizing the unobserved heterogeneity based on discretizing time-series or cross-sectional averages of the data are small as well,  as $N,T$ tend to infinity.   

Next, we state the main result of the paper, that is, the asymptotic normality of the cross-fitted version of the two-step estimator. 
\begin{Theorem}\label{th.ols}
Let Assumptions~\ref{as.fac}--
\ref{as.approx} hold. Then, as $N,T,G_d,C_d$ tend to infinity, we have 
$$\sqrt{NT}(\widehat{\beta}^{\rm CF}-\beta)\xrightarrow[]{d}\mathcal{N}(0,\Sigma_U^{-1}\Omega\Sigma_U^{-1}),$$
where $\Sigma_U$ and $\Omega$ are defined in Assumption \ref{as.errors}\eqref{errorsiii}. 
\end{Theorem}
Theorem~\ref{th.ols} justifies inference on $\beta$ based on Gaussian approximations of the asymptotic distribution. This contrasts with the properties of grouped fixed effects estimators in nonlinear likelihood models \citep{bonhomme2022discretizing}. Indeed,  classification noise affects the properties of second-step estimators in general through an incidental parameter bias. Theorem~\ref{th.ols} shows that under a linear structure and using a Neyman-orthogonal moment, one can construct an estimator that is free of such bias and thus allows the researcher to avoid using potentially computationally difficult and not proven valid bias reduction or bootstrap techniques for inference.

\subsection{Theory for the data-driven choice of the number of clusters}\label{subsec.dd}
We now turn to discussing some theory for the data-driven selection rules for the number of clusters. Similarly to Section \ref{subsec.est}, define $Q_g^d(G):=\frac{1}{N}\sum_{i\in\mathcal{N}_d} \left\| a_i^d-\widehat{a}^d(g_i^d)\right\|^2$
and 
$Q_c^d(C):=\frac{1}{T_d}\sum_{t\in\mathcal{T}_d} \left\| b_t^d-b^d( c_t^d)\right\|^2$, where the cluster centers $a^d(\cdot)$ and $b^d(\cdot)$ are formally defined in Appendix \ref{sub.app_CF}. Let 
$\widehat{V}_g^d:= \frac{1}{N_dT_d^2}\sum_{i\in\mathcal{N}_d} \sum_{t\in\mathcal{T}_d}\left\|z_{it}- a_i^d\right\|^2$ and $\widehat{V}_c^d:=\frac{1}{N_d^2T_d}\sum_{t\in\mathcal{T }_d}\sum_{i\in\mathcal{N}_d} \left\|z_{it}- b_t^d\right\|^2$ denote estimators of the variance of $a_i^d$ and $b_t^d$, respectively.
The data-driven selection rules are
$\widehat{G}_d:=\min_{G\ge 1}\{G:\ Q_g^d(G)\le \widehat{V}_g^d\}$ and $\widehat{C}_d:=\min_{C\ge 1}\{C:\ Q_c^d(C)\le \widehat{V}_c^d\}$.
The following lemma gives conditions under which $\widehat{G}_d$ and $\widehat{C}_d$ yield an approximation error decaying at a rate satisfying Assumption \ref{as.approx}.

\begin{Lemma}\label{lmm.data.driven}
Let Assumptions \ref{as.clus} and \ref{as.rates}\eqref{ratesi} hold.
  Suppose that, for all $d\in\{1,\dots,4\}$, $\widehat{V}_g^d=O_P(1/T)$ and $\widehat{V}_c^d=O_P(1/N).$
  Then, as $N$ and $T$ tend to infinity, $B_\alpha^d(\widehat{G}_d)=o_P((NT)^{-1/4})$ and $B_\gamma^d(\widehat{C}_d)=o_P((NT)^{-1/4})$.
\end{Lemma}
The condition $\widehat{V}_g^d=O_P(1/T)$ and $\widehat{V}_c^d=O_P(1/N)$ is natural since $\widehat{V}_g^d$ and $\widehat{V}_c^d$ are variance estimators for time series and cross-section averages, respectively.

\section{Simulations}\label{sec:sim} 
We consider Monte Carlo simulations to evaluate the finite sample performance of the estimator $\widehat{\beta}$ and its cross-fitted version $\widehat{\beta}^{\rm CF}$. All results in this section are averages over 10,000 replications. In all simulations, we use 30 random starting values and the Hartigan-Wong algorithm to optimize the k-means objective functions.\footnote{The results are not sensitive to the implementation of k-means.}

\paragraph{DGP.} First, we describe the data-generating processes (DGPs). We consider the sample sizes $N=50$ and $T\in\{10,20,30,40,50\}$. There is a single regressor, that is, $K=1$, and we set $\beta=1$. The unit fixed effects $\alpha_i$ are i.i.d.~$\text{Gamma}(1,1)$ random variables (so that $K_\alpha=1$). The time fixed effects $\gamma_t$ are one-dimensional, that is $K_\gamma=1$, and follow an AR$(1)$ process with parameter $\rho\in \{0,0.7\}$ and disturbances drawn from a Gamma distribution with shape parameter $(1-\rho)^2/(1-\rho^2)$ and scale parameter $(1-\rho)/(1-\rho^2)$.\footnote{The process is initialized with a $\text{Gamma}(1,1)$ distribution, and we discard the first $10,000$ observations as a burn-in period.} Here, $\rho$ controls the degree of serial correlation in $\gamma_t$. When $\rho=0$, $\gamma_t$ simply follows an i.i.d. $\text{Gamma}(1,1)$ distribution. 

The error terms $u_{it1}$ and $v_{it}$ also follow AR$(1)$ processes. Specifically, we set $u_{i11}\sim\mathcal{N}(0,1)$ and $v_{i1}\sim \mathcal{N}(0,1)$, and, for all $i\in\{1,\dots,N\}$ and $t\in\{2,\dots,T\}$, $$ u_{it1} =\kappa u_{i(t-1)1} + \mathcal{N}(0,(1-\kappa^2))\ \text{  and  }\ v_{it} = \kappa v_{i(t-1)} + \mathcal{N}(0,(1-\kappa^2)), $$ where $\kappa$ is set to either $0$ or $0.7$ and controls the level of time-series dependence in the error terms.

 For the functions $f$ and $h_1$, we consider two DGPs:
 \begin{itemize} 
 \item[] \textbf{DGP 1.} $\begin{array}{cc}
    f(\alpha_i,\gamma_t)&= \left(0.5\times\alpha_i^{10}+0.5\times\gamma_t^{10} \right)^{1/10},\\
    h_1(\alpha_i,\gamma_t)&= \left(0.5\times\alpha_i^{10}+0.5\times\gamma_t^{10}  \right)^{1/5}.
\end{array}$
 \item[] \textbf{DGP 2.} $\begin{array}{cc}
    f(\alpha_i,\gamma_t)&= \alpha_i^2 + \alpha_i\gamma_t+\sin(\alpha_i\gamma_t),\\
    h_1(\alpha_i,\gamma_t)&=\gamma_t^2+ \alpha_i\gamma_t+\sin(\alpha_i\gamma_t).
\end{array}$
 \end{itemize}
DGP 1 is inspired by the constant elasticity of substitution (CES) specification for time-varying unobserved heterogeneity proposed in \citet[page 631]{bonhomme2022discretizing}.

\paragraph{Estimators.}
We start by evaluating the baseline estimator $\widehat{\beta}$, where the number of clusters $G$ and $C$ are chosen according to the rule outlined in Section \ref{sec:est}.\footnote{When $T\in\{10,20\},$ for a small fraction of replications, this rule yields values of $\widehat{G}$ and $\widehat{C}$ such that the number of degrees of freedom of the estimator is 0. To circumvent this problem, when the data-driven rule implies a number of unit clusters (resp. time clusters) larger than $4N/5$ (resp. $4T/5)$, we replace it by $4N/5$ (resp. $4T/5$). This ensures that the number of degrees of freedom remains strictly positive across all replications.}   For inference, we use heteroskedasticity autocorrelation consistent standard errors clustered at the level of each unit à la \cite{Arellano1987}, that is, the standard error for $\widehat{\beta}$ is
$$\text{se}(\widehat{\beta}):=\sqrt{\frac{NT}{NT-NC-TG}}\left(\frac{1}{NT}\sum_{i=1}^N\sum_{t=1}^T\widehat{u}_{it1}^2 \right)^{-1}\left(\frac{1}{NT}\sum_{i=1}^N\left(\sum_{t=1}^T(\widehat{e}_{it}-\widehat{\beta}\widehat{u}_{it1})\widehat{u}_{it1}\right)^2\right)^{1/2},$$
where the factor $ \sqrt{\frac{NT}{NT-NC-TG}}$ is a degrees-of-freedom correction.

Then, in the same designs, we study the cross-fitted estimator $\widehat{\beta}^{\rm CF}$. For all $d\in\{1,\dots,4\}$, we set $G_d$ and $C_d$ in each fold according to the data-driven rule described in Section \ref{sec:est}. The standard errors are heteroskedasticity and autocorrelation consistent standard errors and computed as 
$$\text{se}(\widehat{\beta}^{\rm CF}):=\sqrt{\frac{NT}{df}}\left(\frac{1}{NT}\sum_{i=1}^N\sum_{t=1}^T\left(\widehat{u}_{it1}^{d^*_{it}}\right)^2 \right)^{-1}\left(\frac{1}{NT}\sum_{i=1}^N\left(\sum_{t=1}^T(\widehat{e}_{it}^{d^*_{it}}-\widehat{\beta}^{\rm CF}\widehat{u}^{d^*_{it}}_{it1})\widehat{u}^{d^*_{it}}_{it1}\right)^2\right)^{1/2},$$
where $d^{*}_{it}= \sum_{d=1}^4  d\mathbf{1}\{(i,t)\in\mathcal{O}_d\}$ is the fold corresponding to observation $(i,t)$ and $df:=\sum_{d=1}^4 \left(N_dT_d- N_dC_d-T_dG_d\right)$ is the number of degrees of freedom. 

We compare our estimators against seven alternative approaches. The first benchmark, denoted $\widehat{\beta}^{\text{Bai}}$, corresponds to the estimator proposed by \cite{bai2009panel}, using $\lfloor T^{1/2} \rfloor$ factors. As shown by \cite{freeman2023linear}, this estimator is consistent under our model. 

We next assess the two-step grouped fixed effects estimator introduced by \cite{freeman2023linear}, denoted $\widehat{\beta}^{\text{GFE}}$. Our implementation follows their methodology, clustering only the first five loadings and factors using a hierarchical clustering procedure with a minimum single linkage algorithm. 

We also include the classical two-way fixed effects estimator, denoted $\widehat{\beta}^{\rm TWFE}$, as well as the factor-augmented regression estimator, $\widehat{\beta}^{\rm FA}$, proposed by \cite{greenaway2012asymptotic} and \cite{westerlund2015cross}, where the number of factors is selected using the eigenvalue ratio estimator of \cite{ahn2013eigenvalue}. Additionally, we consider the pooled CCE estimator, $\widehat{\beta}^{\rm CCE}$, introduced by \cite{pesaran2006estimation}.

Finally, we evaluate the two estimators proposed by \cite{bonhomme2022discretizing} in general parametric likelihood models with nonseparable two-way fixed effects, for which no inference results are available.\footnote{Specifically, we consider here the estimators of \cite{bonhomme2022discretizing} corresponding to the special case where the likelihood is that of the linear model with Gaussian homoscedastic errors.} Let the number of clusters and the actual clusters be computed as in Section~\ref{sec:est}. The first estimator, denoted $\widehat{\beta}^{\rm 1}$, is defined as
\begin{equation*}
\widehat{\beta}^{\rm 1} := \left(\sum_{i=1}^N\sum_{t=1}^T\widetilde{u}_{it}^1\left(\widetilde{u}_{it}^1\right)^\top \right)^{-1}\sum_{i=1}^N\sum_{t=1}^T\widetilde{u}_{it}^1\widetilde{e}_{it}^1=\argmin_{\beta\in\R^K}\min_{\delta\in\R^{N\times C}}\sum_{i=1}^N\sum_{t=1}^T \left(y_{it}-x_{it}^\top\beta -\delta_{i,c_t}\right)^2,
\end{equation*}
where
$
\widetilde{e}_{it}^1 := y_{it} - \bar{y}_{g_it}$ and
$\widetilde{u}_{it}^1 := x_{it} - \bar{x}_{g_it}$.
 This estimator has only unit cluster fixed effects, i.e., it applies a within-transformation with respect to unit clusters only. The second estimator, denoted $\widehat{\beta}^{\rm 2}$, is given by
\begin{equation*}
\widehat{\beta}^{\rm 2} := \left(\sum_{i=1}^N\sum_{t=1}^T\widetilde{u}_{it}^2\left(\widetilde{u}_{it}^2\right)^\top \right)^{-1}\sum_{i=1}^N\sum_{t=1}^T\widetilde{u}_{it}^2\widetilde{e}_{it}^2=\argmin_{\beta\in\R^K}\min_{\xi\in\R^{G\times C}}  \sum_{i=1}^N\sum_{t=1}^T \left(y_{it}-x_{it}^\top\beta-\xi_{g_i,c_t}\right)^2,
\end{equation*}
where
$
\widetilde{e}_{it}^2 := y_{it} - \bar{y}_{g_ic_t}, $
$
\widetilde{u}_{it}^2 := x_{it} - \bar{x}_{g_ic_t},
$
and, for any variable $w_{it}$,
$$
\bar{w}_{g_ic_t} := \frac{1}{N_{g_i}T_{c_t}}\sum_{j=1}^N\sum_{s=1}^T \mathbf{1}\{g_j=g_i\} \mathbf{1}\{c_s=c_t\}w_{js}.$$
This estimator has interacted unit and time cluster fixed effects, that is it performs the within-transformation with respect to the interaction of unit and time clusters.

For all alternative estimators, we use unit-clustered heteroskedasticity-robust standard errors with a degrees-of-freedom correction. 

\paragraph{Results.} The results for all estimators are reported in Tables \ref{tab.res} and \ref{tab.res.alt}. The columns ``Bias" and ``Var" report the estimators' bias and variance. The columns ``Cov" and ``Wid" present the coverage and width of the 95\% confidence intervals based on a Gaussian approximation based on the aforementioned standard errors. For the estimator $\widehat{\beta}$, we also report the average values of $\widehat{G}$ and $\widehat{C}$ in the columns with the same names. 

We find that the estimators, $\widehat{\beta}$ and $\widehat{\beta}^{\rm CF}$, exhibit small bias across all designs. The baseline estimator, $\widehat{\beta}$, achieves coverage levels close to the nominal 95\% across nearly all sample sizes. In comparison, the cross-fitted estimator has slightly lower performance, though it remains reasonably close to $\widehat{\beta}$. In contrast, all benchmark estimators have much larger bias and variance and clearly lower coverage. Concerning 
$\widehat{\beta}^{\rm Bai}$ and $\widehat{\beta}^{\rm GFE}$, we conjecture that this is because the latter two estimators are designed to approximate the term $f(\alpha_i,\gamma_t)$ but not $h(\alpha_i,\gamma_t)$, thereby losing the robustness property discussed in Section~\ref{subsec:dml}.\footnote{It can also be noted that the performance of $\widehat{\beta}^{\rm Bai}$ and $\widehat{\beta}^{\rm GFE}$ deteriorates when $\kappa=0.7$ instead of $0$. We conjecture that this is due to the fact that $\widehat{\beta}^{\rm Bai}$ is asymptotically unbiased and principal components analysis is fixed-T consistent under i.i.d. errors but not otherwise \citep[see][]{bai2003inferential,bai2009panel}.}  The estimators $\widehat{\beta}^{\rm TWFE}$, $ \widehat{\beta}^{\rm FA}$ and $ \widehat{\beta}^{\rm CCE}$ are biased because they estimate either an additive two-way fixed effects or a linear factor structure, which are not enough to capture the nonlinearities of our DGPs.
Interestingly, inference based on $\widehat{\beta}^{\rm 1}$ and $\widehat{\beta}^{\rm 2}$ does not seem to be correct, demonstrating the advantage of our proposal using additive two-way group fixed effects in the second step. Based on these results, we recommend that practitioners primarily use the baseline estimator, $\widehat{\beta}$. 

These findings affirm that cross-fitting primarily serves as a theoretical construct to facilitate asymptotic proofs, offering limited practical benefits in finite samples. The slightly weaker performance of $\widehat{\beta}^{\rm CF}$ can be intuitively attributed to its use of only half the observations for clustering. This robustness underscores the practical value of our approach in such contexts. Moreover, our results reveal that the estimators maintain strong performance under time-series dependence, suggesting that the i.i.d. assumption on the errors is also primarily a theoretical convenience.

Appendix \ref{sec:sim_add} contains additional simulation results. First, in Appendix \ref{subsec:sim_add1}, we study, in the same simulation designs, the finite-sample performance of the estimators when $N=500$ and $T\in\{10,20,30,40,50\}$. We also find that $\widehat{\beta}$ and $\widehat{\beta}^{\rm CF}$ have very good performance in such a large $N$ small $T$ setting common in real-world datasets and outperform the alternatives. Second, in Appendix \ref{subsec:sim_add2}, we study the sensitivity of the baseline estimator to the number of clusters and find that it is remarkably robust.

\begin{table}
    \caption{Simulation results for $\widehat{\beta}$, $\widehat{\beta}^{\rm CF}$, $\widehat{\beta}^{\rm Bai}$, $\widehat{\beta}^{\rm GFE}$ with $N=50$\textsuperscript{a}}
    \label{tab.res}
    \adjustbox{max width=1.\textwidth}{
  \begin{threeparttable}
    \centering   
    \renewcommand{\arraystretch}{1.3} 
    \setlength{\tabcolsep}{4pt} 
    \begin{tabular}{c|cccccc|cccc|cccc|cccc}
    \hline 
    \hline
    \renewcommand{\arraystretch}{1.3}
     & \multicolumn{6}{c|}{$\widehat{\beta}$} & \multicolumn{4}{c|}{$\widehat{\beta}^{\rm CF}$}&\multicolumn{4}{c|}{$\widehat{\beta}^{\rm Bai}$} &\multicolumn{4}{c}{$\widehat{\beta}^{\rm GFE}$}\\
    \hline
    \renewcommand{\arraystretch}{1.}
         $T$ & Bias & Var & Cov & Wid  &$\widehat{G}$ & $\widehat{C}$& Bias & Var & Cov & Wid& Bias & Var & Cov & Wid& Bias & Var & Cov & Wid\\
        \hline 
        \multicolumn{19}{c}{DGP 1,\ $\rho=0$,\ $\kappa=0$}\\
        \hline
10 & 0.010 & 0.003 & 0.972 & 0.260 & 4.259 & 2.732 & 0.017 & 0.004 & 0.985 & 0.336 & 0.089 & 0.010 & 0.482 & 0.152 & 0.080 & 0.011 & 0.560 & 0.223\\
20 & 0.008 & 0.001 & 0.958 & 0.161 & 4.551 & 3.625 & 0.015 & 0.002 & 0.962 & 0.184 & 0.065 & 0.005 & 0.514 & 0.107 & 0.046 & 0.005 & 0.715 & 0.184\\
30 & 0.006 & 0.001 & 0.957 & 0.126 & 4.889 & 4.308 & 0.013 & 0.001 & 0.951 & 0.139 & 0.047 & 0.003 & 0.555 & 0.091 & 0.030 & 0.003 & 0.803 & 0.159\\
40 & 0.005 & 0.001 & 0.951 & 0.106 & 5.213 & 4.834 & 0.012 & 0.001 & 0.941 & 0.116 & 0.040 & 0.002 & 0.556 & 0.080 & 0.021 & 0.002 & 0.856 & 0.142\\
50 & 0.004 & 0.001 & 0.953 & 0.094 & 5.430 & 5.405 & 0.011 & 0.001 & 0.942 & 0.102 & 0.034 & 0.001 & 0.568 & 0.073 & 0.016 & 0.001 & 0.880 & 0.129\\
\hline
        \multicolumn{19}{c}{DGP 1,\ $\rho=0.7$,\ $\kappa=0$}\\
        \hline
10 & 0.007 & 0.003 & 0.968 & 0.268 & 5.552 & 2.482 & 0.009 & 0.004 & 0.984 & 0.331 & 0.068 & 0.008 & 0.572 & 0.165 & 0.056 & 0.010 & 0.659 & 0.230\\
20 & 0.006 & 0.001 & 0.960 & 0.164 & 5.930 & 3.391 & 0.009 & 0.002 & 0.968 & 0.187 & 0.056 & 0.005 & 0.564 & 0.112 & 0.038 & 0.005 & 0.749 & 0.185\\
30 & 0.005 & 0.001 & 0.953 & 0.128 & 6.132 & 4.109 & 0.010 & 0.001 & 0.956 & 0.142 & 0.044 & 0.003 & 0.581 & 0.093 & 0.027 & 0.003 & 0.807 & 0.158\\
40 & 0.004 & 0.001 & 0.956 & 0.109 & 6.349 & 4.785 & 0.009 & 0.001 & 0.955 & 0.118 & 0.036 & 0.002 & 0.602 & 0.081 & 0.020 & 0.002 & 0.855 & 0.141\\
50 & 0.004 & 0.001 & 0.957 & 0.096 & 6.487 & 5.393 & 0.009 & 0.001 & 0.950 & 0.103 & 0.033 & 0.001 & 0.591 & 0.074 & 0.017 & 0.002 & 0.875 & 0.128\\
\hline
        \multicolumn{19}{c}{DGP 1,\ $\rho=0.7$,\ $\kappa=0.7$}\\
        \hline
10 & 0.014 & 0.006 & 0.966 & 0.365 & 7.829 & 2.495 & 0.016 & 0.006 & 0.984 & 0.425 & 0.116 & 0.009 & 0.359 & 0.148 & 0.069 & 0.010 & 0.632 & 0.234\\
20 & 0.009 & 0.003 & 0.962 & 0.250 & 7.548 & 3.351 & 0.015 & 0.004 & 0.968 & 0.274 & 0.120 & 0.006 & 0.214 & 0.093 & 0.066 & 0.005 & 0.643 & 0.193\\
30 & 0.008 & 0.002 & 0.958 & 0.204 & 7.385 & 4.096 & 0.015 & 0.002 & 0.961 & 0.220 & 0.113 & 0.004 & 0.171 & 0.074 & 0.062 & 0.004 & 0.625 & 0.165\\
40 & 0.007 & 0.002 & 0.955 & 0.177 & 7.460 & 4.706 & 0.014 & 0.002 & 0.956 & 0.189 & 0.102 & 0.004 & 0.170 & 0.065 & 0.056 & 0.003 & 0.620 & 0.148\\
50 & 0.006 & 0.001 & 0.956 & 0.157 & 7.403 & 5.331 & 0.012 & 0.002 & 0.954 & 0.169 & 0.091 & 0.003 & 0.184 & 0.060 & 0.052 & 0.003 & 0.618 & 0.136\\

        \hline 
        \multicolumn{19}{c}{DGP 2,\ $\rho=0$,\ $\kappa=0$}\\
        \hline
10 & 0.011 & 0.004 & 0.980 & 0.308 & 3.824 & 3.831 & 0.020 & 0.005 & 0.991 & 0.436 & 0.072 & 0.006 & 0.431 & 0.132 & 0.181 & 0.022 & 0.416 & 0.296\\
20 & 0.006 & 0.002 & 0.966 & 0.176 & 4.325 & 5.026 & 0.016 & 0.002 & 0.970 & 0.211 & 0.055 & 0.002 & 0.380 & 0.083 & 0.142 & 0.014 & 0.441 & 0.242\\
30 & 0.004 & 0.001 & 0.962 & 0.134 & 4.722 & 5.892 & 0.012 & 0.001 & 0.960 & 0.153 & 0.046 & 0.001 & 0.353 & 0.066 & 0.120 & 0.009 & 0.440 & 0.206\\
40 & 0.003 & 0.001 & 0.957 & 0.112 & 5.074 & 6.588 & 0.010 & 0.001 & 0.952 & 0.125 & 0.041 & 0.001 & 0.336 & 0.056 & 0.105 & 0.007 & 0.441 & 0.183\\
50 & 0.003 & 0.001 & 0.960 & 0.098 & 5.395 & 7.203 & 0.009 & 0.001 & 0.953 & 0.108 & 0.039 & 0.001 & 0.306 & 0.051 & 0.095 & 0.006 & 0.446 & 0.165\\
        \hline 
        \multicolumn{19}{c}{DGP 2,\ $\rho=0.7$,\ $\kappa=0$}\\
        \hline
10 & 0.007 & 0.004 & 0.978 & 0.296 & 5.337 & 3.264 & 0.009 & 0.004 & 0.988 & 0.379 & 0.062 & 0.005 & 0.536 & 0.156 & 0.125 & 0.024 & 0.579 & 0.277\\
20 & 0.006 & 0.002 & 0.969 & 0.175 & 5.619 & 4.554 & 0.009 & 0.002 & 0.975 & 0.204 & 0.056 & 0.003 & 0.432 & 0.096 & 0.118 & 0.016 & 0.552 & 0.227\\
30 & 0.004 & 0.001 & 0.962 & 0.134 & 5.876 & 5.413 & 0.007 & 0.001 & 0.968 & 0.151 & 0.047 & 0.002 & 0.413 & 0.075 & 0.105 & 0.011 & 0.526 & 0.196\\
40 & 0.003 & 0.001 & 0.957 & 0.112 & 6.131 & 6.156 & 0.007 & 0.001 & 0.964 & 0.125 & 0.041 & 0.001 & 0.388 & 0.063 & 0.094 & 0.008 & 0.509 & 0.175\\
50 & 0.003 & 0.001 & 0.958 & 0.098 & 6.342 & 6.764 & 0.007 & 0.001 & 0.959 & 0.109 & 0.038 & 0.001 & 0.362 & 0.056 & 0.088 & 0.007 & 0.494 & 0.159\\
        \hline 
        \multicolumn{19}{c}{DGP 2,\ $\rho=0.7$,\ $\kappa=0.7$}\\
        \hline
10 & 0.014 & 0.007 & 0.972 & 0.388 & 6.850 & 3.274 & 0.023 & 0.008 & 0.987 & 0.475 & 0.111 & 0.008 & 0.311 & 0.150 & 0.177 & 0.034 & 0.499 & 0.302\\
20 & 0.008 & 0.003 & 0.967 & 0.261 & 6.500 & 4.534 & 0.017 & 0.004 & 0.972 & 0.287 & 0.098 & 0.005 & 0.182 & 0.089 & 0.177 & 0.022 & 0.412 & 0.255\\
30 & 0.006 & 0.002 & 0.960 & 0.209 & 6.555 & 5.394 & 0.013 & 0.003 & 0.966 & 0.228 & 0.084 & 0.003 & 0.153 & 0.068 & 0.174 & 0.018 & 0.329 & 0.221\\
40 & 0.004 & 0.002 & 0.960 & 0.179 & 6.631 & 6.157 & 0.011 & 0.002 & 0.964 & 0.194 & 0.073 & 0.002 & 0.141 & 0.056 & 0.167 & 0.014 & 0.275 & 0.199\\
50 & 0.004 & 0.001 & 0.961 & 0.159 & 6.767 & 6.775 & 0.010 & 0.002 & 0.961 & 0.171 & 0.066 & 0.002 & 0.133 & 0.049 & 0.159 & 0.012 & 0.248 & 0.181\\
\hline
    \end{tabular}
    \begin{tablenotes}
      \small
      \item[a]Results are based on $10,000$ simulations. DGP 1 and 2 and all considered estimators are described in Section~\ref{sec:sim}.
    \end{tablenotes}
  \end{threeparttable}
  }
\end{table}

\begin{table}[h!]
\centering
    \caption{Simulation results for $\widehat{\beta}^{\rm TWFE}$, $\widehat{\beta}^{\rm FA}$, $\widehat{\beta}^{\rm CCE}$, $\widehat{\beta}^{\rm 1}$, $\widehat{\beta}^{\rm 2}$ with $N=50$\textsuperscript{a}}
    \label{tab.res.alt}
    \adjustbox{max width=1.\textwidth}{
        \begin{threeparttable}
    \renewcommand{\arraystretch}{1.3} 
    \setlength{\tabcolsep}{4pt} 
    \begin{tabular}{c|cccc|cccc|cccc|cccc|cccc}
    \hline 
    \hline
    \renewcommand{\arraystretch}{1.3}
     & \multicolumn{4}{c|}{$\widehat{\beta}^{\rm TWFE}$} & \multicolumn{4}{c|}{$\widehat{\beta}^{\rm FA}$}&\multicolumn{4}{c|}{$\widehat{\beta}^{\rm CCE}$} &\multicolumn{4}{c|}{$\widehat{\beta}^{\rm 1}$}&\multicolumn{4}{c}{$\widehat{\beta}^{\rm 2}$}\\
    \hline
    \renewcommand{\arraystretch}{1.}
         $T$ & Bias & Var & Cov & Wid  & Bias & Var & Cov & Wid& Bias & Var & Cov & Wid& Bias & Var & Cov & Wid& Bias & Var & Cov & Wid\\
        \hline 
        \multicolumn{21}{c}{DGP 1,\ $\rho=0$,\ $\kappa=0$}\\
       \hline
10 & 0.102 & 0.003 & 0.377 & 0.164 & 0.224 & 0.012 & 0.125 & 0.136 & 0.014 & 0.003 & 0.930 & 0.205 & 0.072 & 0.007 & 0.574 & 0.163 & 0.098 & 0.005 & 0.420 & 0.150\\
20 & 0.107 & 0.002 & 0.150 & 0.120 & 0.272 & 0.009 & 0.042 & 0.105 & 0.017 & 0.001 & 0.901 & 0.131 & 0.067 & 0.004 & 0.520 & 0.124 & 0.096 & 0.003 & 0.262 & 0.111\\
30 & 0.108 & 0.001 & 0.066 & 0.103 & 0.289 & 0.007 & 0.033 & 0.098 & 0.018 & 0.001 & 0.880 & 0.104 & 0.058 & 0.003 & 0.517 & 0.106 & 0.087 & 0.002 & 0.206 & 0.094\\
40 & 0.109 & 0.001 & 0.030 & 0.094 & 0.293 & 0.006 & 0.030 & 0.094 & 0.019 & 0.001 & 0.850 & 0.089 & 0.051 & 0.002 & 0.518 & 0.095 & 0.082 & 0.002 & 0.171 & 0.083\\
50 & 0.109 & 0.001 & 0.017 & 0.088 & 0.296 & 0.006 & 0.026 & 0.091 & 0.019 & 0.000 & 0.826 & 0.079 & 0.046 & 0.002 & 0.526 & 0.087 & 0.076 & 0.002 & 0.158 & 0.075\\
\hline
        \multicolumn{21}{c}{DGP 1,\ $\rho=0.7$,\ $\kappa=0$}\\
        \hline
10 & 0.062 & 0.004 & 0.623 & 0.170 & 0.119 & 0.015 & 0.462 & 0.168 & 0.010 & 0.003 & 0.930 & 0.205 & 0.034 & 0.006 & 0.737 & 0.173 & 0.060 & 0.005 & 0.618 & 0.158\\
20 & 0.078 & 0.003 & 0.391 & 0.120 & 0.183 & 0.018 & 0.239 & 0.122 & 0.013 & 0.001 & 0.911 & 0.131 & 0.044 & 0.004 & 0.676 & 0.127 & 0.075 & 0.004 & 0.430 & 0.113\\
30 & 0.085 & 0.002 & 0.253 & 0.102 & 0.219 & 0.016 & 0.136 & 0.107 & 0.015 & 0.001 & 0.886 & 0.104 & 0.045 & 0.004 & 0.637 & 0.108 & 0.078 & 0.003 & 0.319 & 0.095\\
40 & 0.090 & 0.002 & 0.164 & 0.092 & 0.237 & 0.014 & 0.093 & 0.100 & 0.015 & 0.001 & 0.872 & 0.089 & 0.044 & 0.003 & 0.616 & 0.096 & 0.077 & 0.002 & 0.254 & 0.084\\
50 & 0.094 & 0.002 & 0.104 & 0.085 & 0.254 & 0.012 & 0.066 & 0.097 & 0.017 & 0.000 & 0.840 & 0.079 & 0.042 & 0.003 & 0.609 & 0.088 & 0.075 & 0.002 & 0.206 & 0.076\\
\hline
        \multicolumn{21}{c}{DGP 1,\ $\rho=0.7$,\ $\kappa=0.7$}\\
     \hline
10 & 0.085 & 0.007 & 0.592 & 0.226 & 0.137 & 0.016 & 0.436 & 0.210 & 0.014 & 0.005 & 0.925 & 0.261 & -0.064 & 0.015 & 0.601 & 0.271 & -0.009 & 0.011 & 0.698 & 0.231\\
20 & 0.091 & 0.005 & 0.466 & 0.175 & 0.176 & 0.015 & 0.272 & 0.158 & 0.015 & 0.003 & 0.920 & 0.195 & -0.010 & 0.009 & 0.734 & 0.208 & 0.037 & 0.007 & 0.703 & 0.174\\
30 & 0.095 & 0.003 & 0.364 & 0.148 & 0.202 & 0.014 & 0.180 & 0.134 & 0.017 & 0.002 & 0.912 & 0.163 & 0.008 & 0.006 & 0.763 & 0.176 & 0.052 & 0.005 & 0.636 & 0.146\\
40 & 0.097 & 0.003 & 0.287 & 0.133 & 0.222 & 0.013 & 0.125 & 0.120 & 0.018 & 0.002 & 0.900 & 0.142 & 0.015 & 0.005 & 0.771 & 0.157 & 0.057 & 0.004 & 0.575 & 0.129\\
50 & 0.098 & 0.002 & 0.227 & 0.121 & 0.236 & 0.012 & 0.099 & 0.112 & 0.018 & 0.001 & 0.892 & 0.128 & 0.018 & 0.004 & 0.781 & 0.142 & 0.058 & 0.003 & 0.537 & 0.116\\
\hline
        \multicolumn{21}{c}{DGP 2,\ $\rho=0$,\ $\kappa=0$}\\
        \hline
10 & 0.376 & 0.038 & 0.065 & 0.307 & -0.266 & 0.195 & 0.168 & 0.568 & 0.043 & 0.004 & 0.844 & 0.211 & 0.401 & 0.130 & 0.239 & 0.351 & 0.408 & 0.105 & 0.118 & 0.312\\
20 & 0.401 & 0.028 & 0.007 & 0.299 & -0.089 & 0.369 & 0.055 & 0.580 & 0.050 & 0.002 & 0.692 & 0.140 & 0.313 & 0.067 & 0.146 & 0.264 & 0.324 & 0.049 & 0.034 & 0.227\\
30 & 0.410 & 0.022 & 0.001 & 0.296 & 0.038 & 0.446 & 0.030 & 0.584 & 0.054 & 0.001 & 0.549 & 0.116 & 0.255 & 0.038 & 0.116 & 0.217 & 0.269 & 0.026 & 0.017 & 0.185\\
40 & 0.420 & 0.020 & 0.001 & 0.298 & 0.162 & 0.468 & 0.015 & 0.588 & 0.057 & 0.001 & 0.428 & 0.103 & 0.220 & 0.027 & 0.099 & 0.189 & 0.236 & 0.018 & 0.008 & 0.160\\
50 & 0.423 & 0.018 & 0.000 & 0.298 & 0.231 & 0.458 & 0.007 & 0.579 & 0.059 & 0.001 & 0.331 & 0.095 & 0.189 & 0.018 & 0.086 & 0.166 & 0.208 & 0.011 & 0.006 & 0.142\\
        \hline 
        \multicolumn{21}{c}{DGP 2,\ $\rho=0.7$,\ $\kappa=0$}\\
        \hline
10 & 0.236 & 0.052 & 0.386 & 0.262 & -0.221 & 0.072 & 0.433 & 0.398 & 0.036 & 0.004 & 0.840 & 0.211 & 0.247 & 0.151 & 0.532 & 0.293 & 0.266 & 0.115 & 0.386 & 0.261\\
20 & 0.299 & 0.048 & 0.128 & 0.249 & -0.169 & 0.190 & 0.236 & 0.449 & 0.046 & 0.002 & 0.713 & 0.140 & 0.245 & 0.106 & 0.414 & 0.232 & 0.264 & 0.076 & 0.175 & 0.203\\
30 & 0.326 & 0.042 & 0.043 & 0.253 & -0.072 & 0.301 & 0.149 & 0.489 & 0.049 & 0.002 & 0.612 & 0.115 & 0.218 & 0.076 & 0.358 & 0.201 & 0.237 & 0.051 & 0.098 & 0.174\\
40 & 0.346 & 0.038 & 0.016 & 0.260 & 0.024 & 0.378 & 0.097 & 0.520 & 0.052 & 0.001 & 0.508 & 0.102 & 0.197 & 0.056 & 0.314 & 0.180 & 0.216 & 0.037 & 0.065 & 0.155\\
50 & 0.358 & 0.035 & 0.006 & 0.264 & 0.103 & 0.414 & 0.070 & 0.531 & 0.055 & 0.001 & 0.428 & 0.093 & 0.180 & 0.046 & 0.290 & 0.164 & 0.200 & 0.029 & 0.044 & 0.141\\
        \hline 
        \multicolumn{21}{c}{DGP 2,\ $\rho=0.7$,\ $\kappa=0.7$}\\
        \hline
10 & 0.299 & 0.063 & 0.353 & 0.317 & -0.244 & 0.086 & 0.451 & 0.472 & 0.063 & 0.009 & 0.781 & 0.269 & 0.086 & 0.196 & 0.416 & 0.405 & 0.154 & 0.139 & 0.561 & 0.337\\
20 & 0.337 & 0.054 & 0.186 & 0.293 & -0.247 & 0.141 & 0.268 & 0.492 & 0.061 & 0.005 & 0.722 & 0.203 & 0.150 & 0.129 & 0.530 & 0.316 & 0.197 & 0.087 & 0.496 & 0.258\\
30 & 0.354 & 0.046 & 0.094 & 0.286 & -0.200 & 0.213 & 0.173 & 0.508 & 0.061 & 0.003 & 0.674 & 0.172 & 0.151 & 0.087 & 0.562 & 0.267 & 0.191 & 0.057 & 0.404 & 0.217\\
40 & 0.371 & 0.041 & 0.048 & 0.284 & -0.135 & 0.280 & 0.118 & 0.516 & 0.061 & 0.003 & 0.624 & 0.152 & 0.150 & 0.067 & 0.549 & 0.236 & 0.185 & 0.042 & 0.327 & 0.192\\
50 & 0.376 & 0.037 & 0.024 & 0.283 & -0.069 & 0.336 & 0.088 & 0.528 & 0.062 & 0.002 & 0.574 & 0.138 & 0.136 & 0.050 & 0.550 & 0.211 & 0.171 & 0.031 & 0.273 & 0.171\\
\hline
    \end{tabular}
        \begin{tablenotes}
      \small
      \item[a] Results are based on $10,000$ simulations. DGP 1 and 2 and all considered estimators are described in Section~\ref{sec:sim}.
    \end{tablenotes}
    \end{threeparttable}}
\end{table}

\section{Application to fiscal policy}\label{sec:app}

We revisit \cite{james2015us}, focusing on the impact of increases in resource-based government revenues on various fiscal outcomes across U.S. states: non-resource tax revenues, income tax revenues, total expenditures, education expenditures, and public savings. The data comprises annual government revenues and expenditures, as well as private income for all U.S. states over the period from 1958 to 2008, so that $N=50$ and $T=51$.\footnote{The full dataset is available at \url{https://www.openicpsr.org/openicpsr/project/114577/version/V1/view}.}

As argued by \cite{james2015us}, following economic theory, resource-based tax revenue should have a negative effect on nonresource revenue and income tax revenue, but a positive impact on total expenditure, education expenditure, and savings. A potential confounding factor is the business cycle $\gamma_t$, which can influence both non-resource and resource-based revenues. During periods of high macroeconomic output, energy consumption and private income tend to rise, leading to higher revenues. This relationship can introduce omitted variable bias. Our estimation approach addresses this issue by allowing the effect of business cycles to vary nonlinearly across states and revenue types through $\alpha_i$, reflecting differences in tax schemes and economic structures. Arguably, the effect of most unobserved state-specific characteristics such as average population density, political preferences, wealth, unemployment, culture, and institutional quality deemed time-invariant in \cite{james2015us} might actually vary over 51 years.

\paragraph{Regression results.} \cite{james2015us} employs the within-estimator for a two-way fixed effects model, denoted $\widehat{\beta}^{\text{TWFE}}$, regressing the ratio of the various outcomes to private income in the state-year on the ratio of resource-based government revenues to private income. We consider seven estimators: $\widehat{\beta}$, $\widehat{\beta}^{\rm CF}$, $\widehat{\beta}^{\text{TWFE}}$, $\widehat{\beta}^{\rm Bai}$, $\widehat{\beta}^{\rm GFE}$, $\widehat{\beta}^{\text{FA}}$ and $\widehat{\beta}^{\text{CCE}}$. Note that for all estimators but $\widehat{\beta}^{\text{TWFE}}$, we first standardize the outcome variable and regressor before applying the methods. The estimated coefficients and standard errors are then rescaled to correspond to the original model.
All estimators and their standard errors are computed as in the simulations of Section \ref{sec:sim}, except that we use $10,000$ initializations for the kmeans algorithms of $\widehat{\beta}$ and $\widehat{\beta}^{\rm CF}$.  Table \ref{tab:app4} reports the results for all outcomes and estimators. Table \ref{tab:app41} presents the values of $\widehat{G}$ and $\widehat{C}$ for the different outcomes.

The proposed estimator $\widehat{\beta}$ always has the sign predicted by economic theory. The results for $\widehat{\beta}$ are also significant for 4 of the 5 outcomes. In contrast, each of the alternative estimators has a sign in disagreement with the theory for at least one outcome. For the outcome ``Savings", $\widehat{\beta}^{\rm GFE}$ and $\widehat{\beta}^{\rm CCE}$ yield estimates larger than $1$, which are difficult to justify from an economic perspective. Overall, the estimator's conclusions often differ from the alternatives, demonstrating its ability to provide unique insights. 

\begin{table}[ht]        
    \centering
 \adjustbox{max width=1.\textwidth}{\begin{threeparttable}
    \caption{Estimates of the effect of resource-based tax revenues\textsuperscript{a}}
    \label{tab:app4}
    \setlength{\tabcolsep}{12pt} 
    \begin{tabular}{c *{7}{S[table-format=-1.3]}}
  \hline
  \hline
  \rule{0pt}{15pt}
Outcome & {$\widehat{\beta}$} & {$\widehat{\beta}^{\rm CF}$} & {$\widehat{\beta}^{\rm TWFE}$} & {$\widehat{\beta}^{\rm Bai}$} & {$\widehat{\beta}^{\rm GFE}$} & {$\widehat{\beta}^{\rm FA}$}  &{$\widehat{\beta}^{\rm CCE}$} \\ 
\hline

Non-resource revenue & -0.376** & -0.478*** & 0.006 & 0.048 & 0.26 & 0.345*** & -0.038\\
 & (0.189) & (0.143) & (0.019) & (0.037) & (0.627) & (0.065) & (0.055)\\  [0.5em]
Income tax revenue & -0.083 & -0.035 & 0.018* & -0.008 & 0.224 & 0.151 & 0.002\\
 & (0.053) & (0.048) & (0.010) & (0.013) & (0.301) & (0.106) & (0.017) \\ [0.5em]
Total expenditures & 0.701*** & 0.527* & 0.397*** & -0.016 & 1.514** & 0.101 & -0.275**\\
 & (0.184) & (0.284) & (0.006) & (0.027) & (0.740) & (0.568) & (0.112)  \\ [0.5em]
Education expenditures & 0.208** & -0.041 & 0.063*** & 0.049*** & 0.231 & 0.113 & -0.004\\
 & (0.092) & (0.079) & (0.007) & (0.009) & (0.205) & (0.153) & (0.029) \\ [0.5em]
Savings & 0.480*** & 0.587*** & 0.609*** & 0.938*** & 14.965*** & -0.183 & 1.240*** \\
 & (0.108) & (0.228) & (0.021) & (0.041) & (4.645) & (0.369) & (0.187) \\ [1em]
 Observations & {2550} & {2550}  & {2550}  & {2550}  & {2550}  & {2550}  & {2550}  \\
\hline
\end{tabular}
            \begin{tablenotes}
            \footnotesize
                \item[a] 
                Statistical significance: ***$p<.01$, **$p<.05$, *$p<.10$. Standard errors clustered at the state level in parentheses.
            \end{tablenotes}
        \end{threeparttable}}

\end{table}

\begin{table}[ht]        
    \centering
  \adjustbox{max width=1.\textwidth}{  
  \begin{threeparttable}
      \renewcommand{\arraystretch}{1.5} 
    \caption{Number of estimated unit and time clusters, by regression outcome} 
    \label{tab:app41}
\begin{tabular}{c|ccccc}
 \hline
 \hline
 Outcome & Non-resource revenue & Income tax revenue & Total expenditures & Education expenditures & Savings\\
\hline
$\widehat{G}$ & 11 & 9 & 8 & 11 & 5\\
$\widehat{C}$ & 3 & 6 & 3 & 4 & 4\\
\hline
\end{tabular}

        \end{threeparttable}}

\end{table}

\begin{figure}[h!] 
    \centering
    \includegraphics[width=0.9\textwidth]{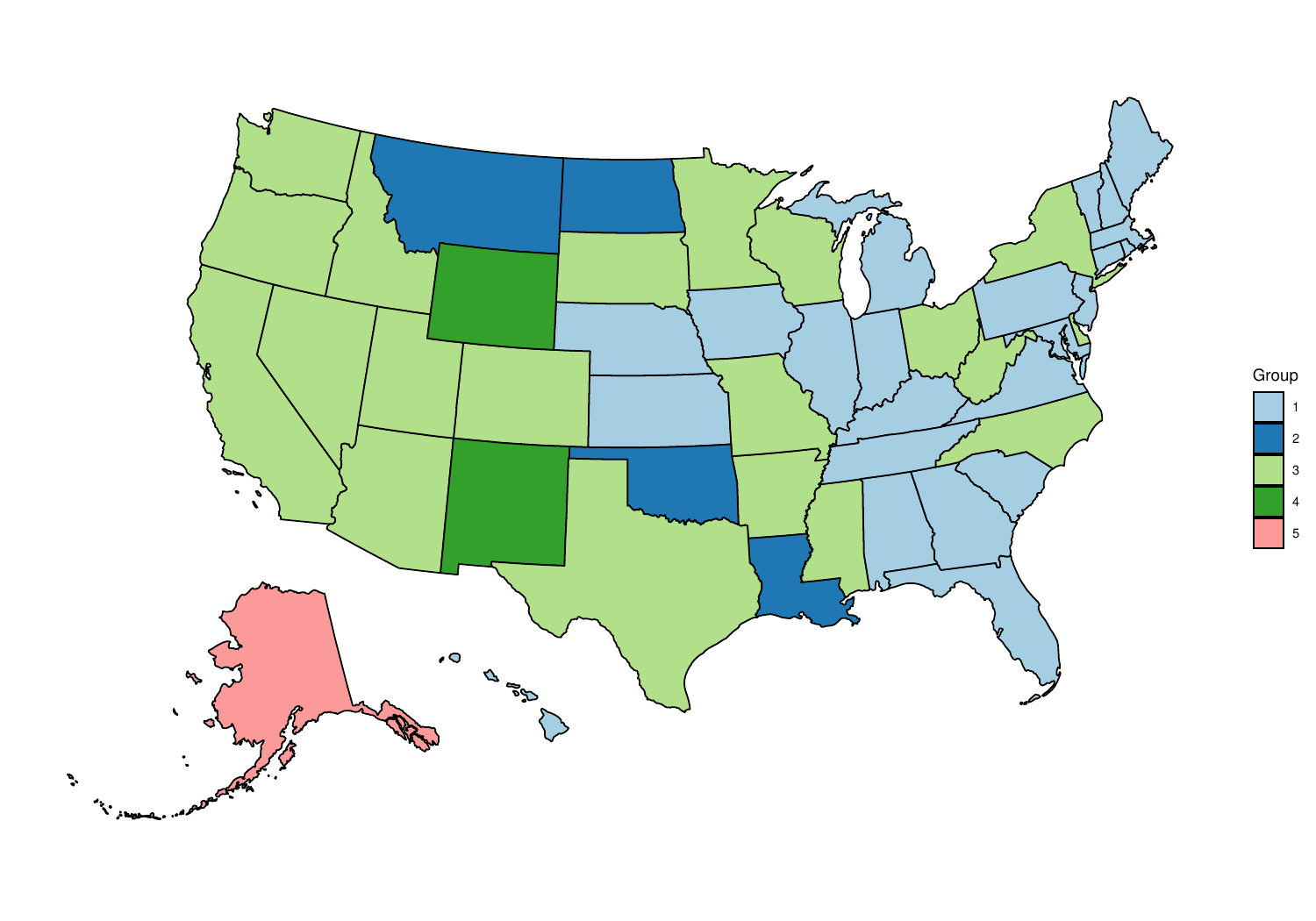} 
    \caption{Map of clusters of U.S.~states for the outcome ``Savings".}
    \label{fig:map}
\end{figure}

\paragraph{Clusters for the outcome ``Savings."} We now present the clusters obtained by our method for the outcome ``Savings."\footnote{We selected this outcome because it has the fewest clusters, making it easier to represent on a map.} Figure \ref{fig:map} displays the 5 unit clusters on a map, with their centers listed in Table \ref{tab:app41}. These clusters represent states with similar average savings and resource-based government revenues over the period. While there is no reason to expect that they should correspond to geographically close states (this assumption is not imposed in the data-driven estimation), it turns out that some geographical dependence is effectively captured as geographically close states often end up in the same estimated cluster. Interestingly, cluster 5 corresponds to Alaska, and cluster 4 consists of New Mexico and Wyoming. These three states are known to be particularly rich in natural resources. The information for the time clusters is given in Figure \ref{fig:time} and Table \ref{tab:app44}. We find similar patterns, with clusters seemingly capturing the business cycle.

\begin{center}
    \begin{minipage}{0.45\textwidth}
        \centering
        \includegraphics[width=0.9\textwidth]{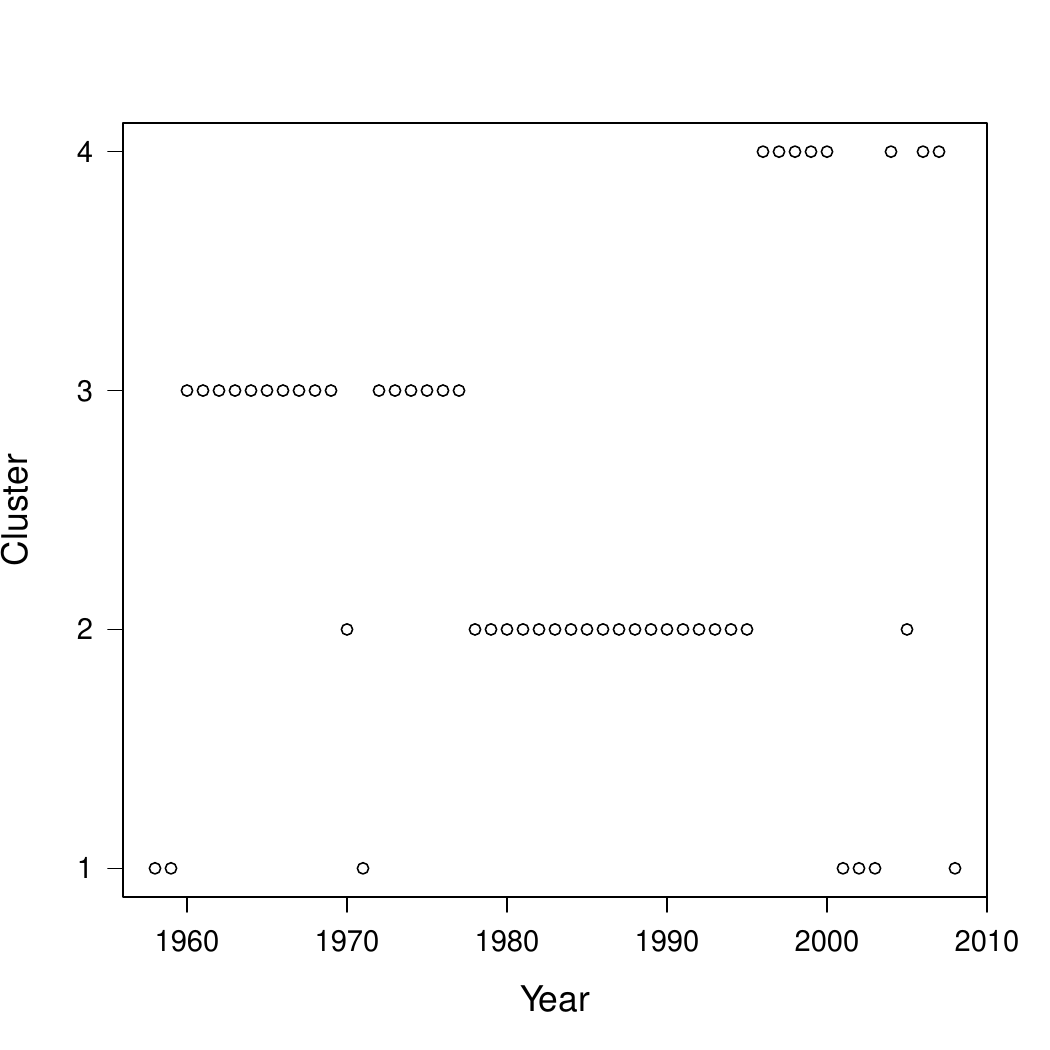}
        \captionof{figure}{Time clusters for the outcome ``Savings".}
        \label{fig:time}
    \end{minipage}
    \hfill
    \begin{minipage}{0.45\textwidth}
        \centering
        \begin{tabular}{c|c|c}
            \hline
            \hline
            Cluster & Savings & Resource-based revenues \\
            \hline
            1 & -0.674 & -0.065\\
            2 & 0.191 & 0.156\\
            3 & -0.258 & -0.179\\
            4 & 0.629 & 0.026\\
            \hline
        \end{tabular}
        \captionof{table}{Centers of time clusters for the outcome ``Savings".}
        \label{tab:app44}
    \end{minipage}
\end{center}

Overall, our results indicate that controlling for flexible patterns of time-varying unobserved heterogeneity does not refute the predictions made by economic theory. Estimated clusters confirm that unobserved heterogeneity is both spatially and temporally correlated.

\section{Conclusion} \label{sec:conc}
This paper shows how to use Neyman-orthogonal moments to build inference tools after discretizing time-varying unobserved heterogeneity in linear panel data models. The proposed procedure is intuitive and simple, but nevertheless exhibits excellent asymptotic properties and finite-sample performance. A natural extension is to consider heterogeneous slope parameters. While adapting the proposed estimation procedure to accommodate either unit- or time-specific slope coefficients is relatively straightforward, we leave the study of more flexible unit- and time-varying structures for further research.

\bibliographystyle{apalike}
\bibliography{paper}
\setcounter{section}{0}
\renewcommand{\thesection}{\Alph{section}}
\begin{center} \Huge{\bf{Appendix}}\end{center}
\section{Cross-fitting procedure}\label{sub.app_CF}
As for $\widehat{\beta}$, we estimate the group memberships in the first step via a clustering method and compute an OLS estimator in the second step. The main difference is that the data used in each of these two steps do not intersect but the final estimator still uses variation across the full dataset. The estimation procedure to obtain the resulting cross-fitted two-way grouped fixed effect estimator is as follows. For each fold, $d\in\{1,\dots,4\}$:
\begin{enumerate}
\item Apply k-means clustering to the data in $\{\mathcal{O}_{\tilde{d}}\}$, where 
$$\tilde{d}=\left\{\begin{array}{cc}2 & \text{ if $d=1$},\\
1 & \text{ if $d=2$},\\
4 & \text{ if $d=3$},\\
3 & \text{ if $d=4$},
\end{array}\right.$$
to obtain the unit cluster indicators $g_i^d\in\{1,\dots,G_d\}$.
\item Apply k-means clustering to the data in $\{\mathcal{O}_{\tilde{d}}\}$, where 
$$\tilde{d}=\left\{\begin{array}{cc}3 & \text{ if $d=1$},\\
4 & \text{ if $d=2$},\\
1 & \text{ if $d=3$},\\
2 & \text{ if $d=4$},
\end{array}\right.$$
to obtain the time cluster indicators $c_t^d\in\{1,\dots,C_d\}$.
\item Estimate $e_{it}$ and $u_{it}$ on fold $d$ by
 $\widehat{e}^d_{it}:=y_{it}-\bar{y}_{g_i^dt}- \bar{y}_{ic_t^d}+\bar{y}_{g_i^dc_t^d}$ and $\widehat{u}^d_{it}:=x_{it}-\bar{x}_{g_i^dt}- \bar{x}_{ic_t^d}+\bar{x}_{g_i^dc_t^d}$, and where, for any variable $w_{it}$, we define
\begin{align*}\bar{w}_{g_i^dt}&:=\frac{1}{N^d_{g_i^d}}\sum_{j\in\mathcal{N}_d} \mathbf{1}\{g^d_j=g^d_i\} w_{jt},\\
\bar{w}_{ic_t^d}&:=\frac{1}{T^d_{c_t^d}}\sum_{s\in\mathcal{T}_d} \mathbf{1}\{c^d_s=c^d_t\} w_{is},\\
\bar{w}_{g_i^dc_t^d}&:=\frac{1}{N^d_{g_i^d}T^d_{c_t^d}}\sum_{(j,s)\in\mathcal{O}_d} \mathbf{1}\{g^d_j=g^d_i\}\mathbf{1}\{c^d_s=c^d_t\} w_{js},
\end{align*}
with $N^d_{g_i^d}:= \sum_{j\in\mathcal{N}_d} \mathbf{1}\{g^d_j=g^d_i\}$ and $T^d_{c_t^d} :=\sum_{s\in\mathcal{T}_d} \mathbf{1}\{c^d_s=c^d_t\}$.
\end{enumerate}
The final estimator is the linear regression of the $\widehat{e}_{it}^d$ on the $\widehat{u}_{it}^d$,
$$\widehat{\beta}^{\rm CF}:=\left(\sum_{d=1}^4\sum_{(i,t)\in \mathcal{O}_d} \widehat{u}_{it}^d (\widehat{u}_{it}^d)^\top \right)^{-1}\sum_{d=1}^4\sum_{(i,t)\in \mathcal{O}_d}\widehat{u}_{it}^d \widehat{e}_{it}^d, $$
which is numerically equivalent to 
\begin{equation}\label{11224}
\argmin_{\beta\in\R^K}\min_{\delta}\min_{\nu}\sum_{i=1}^N\sum_{t=1}^T\left[y_{it}-x_{it}^\top\beta-\sum_{d=1}^4\mathbf1\{(i,t)\in\mathcal O_d\}\left(\delta_{i,c_t^d}^d+\nu_{t,g_i^d}^d\right)\right]^2.\end{equation}

In summary, to obtain the unit cluster indicators $g_i^d\in\{1,\dots,G_d\}$ (resp.~the time cluster indicators $c_t^d\in\{1,\dots,C_d\}$), we use the fold that contains the same units as $\mathcal{O}_d$ but different dates (resp.~the same dates as $\mathcal{O}_d$ but different units). A similar trick is used for clustering time periods. We then use these clusters to estimate $e_{it}$ and $u_{it}$, before running a linear regression on such estimates.\footnote{As for our baseline estimator, the second step~\eqref{11224} of our cross-fitted estimator corresponds to the second step of the cross-fitted estimator in \cite{freeman2023linear}. The clustering steps differ between the two papers.}

The clustering steps are carried out using straightforward adapted versions of the algorithm introduced in Section~\ref{sec:est} (indexing all relevant sample $d$-dependent variable by $d$)
, which we display below for completeness. 


\medskip
\noindent \textit{Clustering algorithm for units.} Let the empirical averages $  a_i^d:=\frac{1}{T_{\tilde{d}}} \sum_{t\in \mathcal T_{\tilde{d}}} z_{it}, \ i\in\mathcal{N}_d$ be computed on fold $\tilde{d}$. We use the algorithm 
$$\left(\widehat{a}^d(1),\dots,\widehat{a}^d(G_d),\{g_i^d,\ i\in\mathcal{N}_{d}\}\right)\in\argmin\limits_{\begin{array}{c}a(1),\dots,a(G_d)\in\R^{K+1}\\g_i\in\{1,\dots,G_d\},\ i\in\mathcal{N}_{d}\end{array}}\sum_{i\in\mathcal{N}_{d}} \left\| a_i^d-a(g_i)\right\|^2. $$

\medskip
\noindent \textit{Clustering algorithm for dates.} .Let the empirical averages $  b_t^d:=\frac{1}{N_{\tilde{d}}} \sum_{i\in \mathcal{N}_{\tilde{d}}} z_{it},\ t\in\mathcal{T}_d$ be computed on fold $\tilde{d}$. We use the algorithm 
$$\left(\widehat{b}^d(1),\dots,\widehat{b}^d(C_d),\{c_t^d,\  t\in\mathcal{T}_{d}\}\right)\in\argmin\limits_{\begin{array}{c}b(1),\dots,b(C_d)\in\R^{K+1}\\c_t\in\{1,\dots,C_d\},\ t\in\mathcal{T}_{d}\end{array}}\sum_{t\in\mathcal{T}_{d}} \left\| b_t^d-b(c_t)\right\|^2. $$
In practice, we use the data-driven rule outlined in Section~\ref{sec:est} to select the number of clusters $G_d$ and $C_d$ in the different folds $d\in\{1,\dots,4\}$.
\section{Pseudo-distance and hierachical clustering}\label{subsec.hiech_app}
In this section, we describe an alternative clustering algorithm for the first step based on the pseudo-distance of \cite{zhang2017estimating} and hierarchical clustering as in \cite{mugnier2024simple}. Let us explain how units are clustered with this approach. For $i,j\in\{1,\dots,N\}$, we define the pseudo-distance 
$$\widehat{d}_{\infty,1}(i,j)=\frac1T\max_{\ell \in\{1,\dots,N\}\backslash\{i,j\}}\left(\left|\sum_{t=1}^T(y_{it}-y_{jt})y_{\ell t}\right|+\sum_{k=1}^K\left|\sum_{t=1}^T(x_{itk}-x_{jtk})x_{\ell tk}\right|\right).$$
Then, to obtain the unit clusters, we apply a hierarchical clustering algorithm to the $N\times N$ matrix $\widehat{D}$ such that $\widehat{D}_{ij}=\widehat{d}_{\infty,1}(i,j).$ See \cite{mugnier2024simple} for a formal presentation. Similarly to \cite{mugnier2024simple}, we choose the threshold $c_{NT}$ for the maximum intragroup distance equal to
$$1.35\frac{\log(T)}{K\sqrt{\min(N,T)}}\check{\sigma},$$
where 
\begin{align*}
    \check{\sigma}= \max_{i\in\{1\dots,N\}}\min_{j\in\{1\dots,N\}, j\ne i}\frac{1}{2T}\sum_{t=1}^T(y_{it}-y_{jt})^2+ \sum_{k=1}^K \max_{i\in\{1\dots,N\}}\min_{j\in\{1\dots,N\}, j\ne i}\frac{1}{2T}\sum_{t=1}^T(x_{itk}-x_{jtk})^2.
\end{align*}
To avoid having 0 degrees of freedom, if this value of $c_{NT}$ gives more than $\lfloor 2N/5\rfloor $ clusters, we set the number of unit clusters to $\lfloor 2N/5\rfloor $. The time clusters are obtained symmetrically. The proposed approach circumvents averaging the data before clustering, which should allow avoiding the injectivity assumption; see also the discussion in \cite{athey2025identification}. 

We consider baseline $\widetilde{\beta}$ and cross-fitted $\widetilde{\beta}^{\rm CF}$ adaptations of the estimators that use  hierarchical clustering with an average linkage function on the pseudo-distance instead of k-means to obtain the clusters. In Table \ref{tab.reshiech}, we report the results of simulations with these alternative estimators, where the DGPs and standard errors are as in Section \ref{sec:sim}. Results suggest that $\widetilde{\beta}$ has poor performance, while $\widetilde{\beta}^{\rm CF}$ has almost nominal coverage but large confidence intervals compared to the version using k-means.

\begin{table}[htbp]
    \centering 
    \caption{Simulation results, $N=50$\textsuperscript{a}}
    \label{tab.reshiech}
    \adjustbox{max width=0.7\textwidth}{
        \begin{threeparttable}
            \renewcommand{\arraystretch}{1.3} 
            \setlength{\tabcolsep}{4pt} 
            \begin{tabular}{c|cccccc|cccc}
                \hline
                \hline
                & \multicolumn{6}{c|}{$\widetilde{\beta}$} & \multicolumn{4}{c|}{$\widetilde{\beta}^{\rm CF}$}\\
                \hline
                $T$ & Bias & Var & Cov & Wid & $\widehat{G}$ & $\widehat{C}$ & Bias & Var & Cov & Wid\\
                \hline
                \multicolumn{11}{c}{DGP 1,\ $\rho=0$,\ $\kappa=0$}\\
                \hline
               10 & -0.140 & 0.018 & 0.676 & 0.411 & 14.208 & 2.912 & 0.029 & 0.005 & 0.962 & 0.392\\
20 & -0.090 & 0.007 & 0.774 & 0.295 & 14.441 & 5.910 & 0.018 & 0.003 & 0.961 & 0.275\\
30 & -0.079 & 0.004 & 0.771 & 0.246 & 14.771 & 8.957 & 0.014 & 0.002 & 0.967 & 0.228\\
40 & -0.075 & 0.003 & 0.741 & 0.216 & 14.965 & 11.995 & 0.012 & 0.001 & 0.964 & 0.199\\
50 & -0.074 & 0.003 & 0.696 & 0.196 & 15.228 & 15.122 & 0.010 & 0.001 & 0.970 & 0.180\\
                \hline
                \hline
                \multicolumn{11}{c}{DGP 1,\ $\rho=0.7$,\ $\kappa=0$}\\
                \hline
              10 & -0.149 & 0.017 & 0.789 & 0.470 & 14.207 & 3.688 & 0.011 & 0.005 & 0.991 & 0.447\\
20 & -0.099 & 0.006 & 0.829 & 0.325 & 14.509 & 7.097 & 0.009 & 0.002 & 0.992 & 0.316\\
30 & -0.087 & 0.004 & 0.806 & 0.264 & 14.857 & 10.259 & 0.008 & 0.002 & 0.990 & 0.258\\
40 & -0.082 & 0.003 & 0.751 & 0.228 & 15.115 & 13.305 & 0.007 & 0.001 & 0.991 & 0.221\\
50 & -0.079 & 0.003 & 0.708 & 0.205 & 15.166 & 16.504 & 0.008 & 0.001 & 0.990 & 0.197\\
                \hline
                \hline
                \multicolumn{11}{c}{DGP 1,\ $\rho=0.7$,\ $\kappa=0.7$}\\
                \hline
10 & -0.105 & 0.016 & 0.891 & 0.509 & 13.849 & 3.670 & 0.016 & 0.007 & 0.982 & 0.497\\
20 & -0.083 & 0.008 & 0.882 & 0.379 & 14.369 & 7.049 & 0.012 & 0.004 & 0.989 & 0.381\\
30 & -0.071 & 0.006 & 0.862 & 0.314 & 14.685 & 10.099 & 0.011 & 0.003 & 0.989 & 0.326\\
40 & -0.064 & 0.004 & 0.860 & 0.277 & 15.174 & 13.204 & 0.011 & 0.002 & 0.990 & 0.290\\
50 & -0.059 & 0.003 & 0.867 & 0.250 & 15.190 & 16.320 & 0.010 & 0.002 & 0.990 & 0.263\\
                \hline
                \multicolumn{11}{c}{DGP 2,\ $\rho=0$,\ $\kappa=0$}\\
                \hline
               10 & -0.088 & 0.016 & 0.837 & 0.413 & 14.937 & 3.216 & 0.064 & 0.016 & 0.927 & 0.460\\
20 & -0.052 & 0.006 & 0.888 & 0.291 & 15.302 & 6.201 & 0.037 & 0.009 & 0.935 & 0.318\\
30 & -0.045 & 0.004 & 0.896 & 0.240 & 15.532 & 9.155 & 0.023 & 0.005 & 0.952 & 0.254\\
40 & -0.044 & 0.003 & 0.894 & 0.210 & 15.824 & 12.118 & 0.017 & 0.004 & 0.954 & 0.217\\
50 & -0.044 & 0.002 & 0.896 & 0.191 & 16.005 & 15.263 & 0.014 & 0.003 & 0.962 & 0.193\\
                \hline
                                \multicolumn{11}{c}{DGP 2,\ $\rho=0.7$,\ $\kappa=0$}\\
                \hline
10 & -0.113 & 0.014 & 0.871 & 0.459 & 14.956 & 3.840 & 0.026 & 0.008 & 0.982 & 0.482\\
20 & -0.069 & 0.005 & 0.925 & 0.317 & 15.431 & 7.225 & 0.014 & 0.004 & 0.989 & 0.341\\
30 & -0.059 & 0.003 & 0.912 & 0.255 & 15.614 & 10.327 & 0.009 & 0.002 & 0.992 & 0.273\\
40 & -0.054 & 0.002 & 0.901 & 0.221 & 15.858 & 13.402 & 0.007 & 0.002 & 0.990 & 0.233\\
50 & -0.052 & 0.002 & 0.884 & 0.197 & 16.014 & 16.354 & 0.006 & 0.001 & 0.991 & 0.206\\
                \hline
                                \multicolumn{11}{c}{DGP 2,\ $\rho=0.7$,\ $\kappa=0.7$}\\
                \hline
               10 & -0.101 & 0.017 & 0.923 & 0.536 & 14.669 & 3.845 & 0.054 & 0.016 & 0.954 & 0.552\\
20 & -0.078 & 0.008 & 0.930 & 0.400 & 15.094 & 7.190 & 0.022 & 0.007 & 0.981 & 0.427\\
30 & -0.069 & 0.005 & 0.921 & 0.334 & 15.494 & 10.281 & 0.012 & 0.004 & 0.987 & 0.365\\
40 & -0.063 & 0.004 & 0.910 & 0.291 & 15.709 & 13.229 & 0.010 & 0.003 & 0.989 & 0.320\\
50 & -0.059 & 0.003 & 0.905 & 0.264 & 15.996 & 16.291 & 0.008 & 0.003 & 0.991 & 0.289\\
                \hline
            \end{tabular}
            \begin{tablenotes}
                \small
                \item[a] Results are based on $10,000$ simulations. DGP 1 and 2 are described in Section~\ref{sec:sim}. Both estimators $\widetilde\beta$ and $\widetilde \beta^{\rm CF}$ are described in Section~\ref{subsec.hiech_app}.
            \end{tablenotes}
        \end{threeparttable}
    }
\end{table}

\section{Additional simulation results}\label{sec:sim_add}
\subsection{Results with large $N$ and small $T$}\label{subsec:sim_add1}
We present simulation results for the estimators when $N=500$ and $T\in\{10,20,30,40,50\}$ under the data-generating processes outlined in Section \ref{sec:sim}. The results are reported in Tables \ref{tab.res500} and \ref{tab.res.alt500}.

\begin{table}
    \caption{Simulation results for $\widehat{\beta}$, $\widehat{\beta}^{\rm CF}$, $\widehat{\beta}^{\rm Bai}$, $\widehat{\beta}^{\rm GFE}$ with $N=500$\textsuperscript{a}}
    \label{tab.res500}
    \adjustbox{max width=1.\textwidth}{
  \begin{threeparttable}
    \centering   
    \renewcommand{\arraystretch}{1.3} 
    \setlength{\tabcolsep}{4pt} 
    \begin{tabular}{c|cccccc|cccc|cccc|cccc}
    \hline 
    \hline
    \renewcommand{\arraystretch}{1.3}
     & \multicolumn{6}{c|}{$\widehat{\beta}$} & \multicolumn{4}{c|}{$\widehat{\beta}^{\rm CF}$}&\multicolumn{4}{c|}{$\widehat{\beta}^{\rm Bai}$} &\multicolumn{4}{c}{$\widehat{\beta}^{\rm GFE}$}\\
    \hline
    \renewcommand{\arraystretch}{1.}
         $T$ & Bias & Var & Cov & Wid  &$\widehat{G}$ & $\widehat{C}$& Bias & Var & Cov & Wid& Bias & Var & Cov & Wid& Bias & Var & Cov & Wid\\
        \hline 
        \multicolumn{19}{c}{DGP 1,\ $\rho=0$,\ $\kappa=0$}\\
        \hline
10 & 0.004 & 0.000 & 0.984 & 0.095 & 9.687 & 3.820 & 0.007 & 0.000 & 0.990 & 0.120 & 0.024 & 0.001 & 0.664 & 0.062 & 0.011 & 0.001 & 0.819 & 0.079\\
20 & 0.002 & 0.000 & 0.974 & 0.056 & 11.761 & 5.571 & 0.005 & 0.000 & 0.974 & 0.062 & 0.022 & 0.000 & 0.511 & 0.040 & 0.006 & 0.000 & 0.912 & 0.065\\
30 & 0.001 & 0.000 & 0.969 & 0.043 & 13.181 & 7.235 & 0.004 & 0.000 & 0.960 & 0.046 & 0.021 & 0.000 & 0.398 & 0.032 & 0.005 & 0.000 & 0.919 & 0.055\\
40 & 0.001 & 0.000 & 0.974 & 0.037 & 14.579 & 8.868 & 0.003 & 0.000 & 0.962 & 0.038 & 0.020 & 0.000 & 0.340 & 0.028 & 0.005 & 0.000 & 0.928 & 0.048\\
50 & 0.001 & 0.000 & 0.969 & 0.032 & 15.942 & 10.489 & 0.003 & 0.000 & 0.957 & 0.033 & 0.018 & 0.000 & 0.292 & 0.025 & 0.004 & 0.000 & 0.919 & 0.043\\
\hline
        \multicolumn{19}{c}{DGP 1,\ $\rho=0.7$,\ $\kappa=0$}\\
        \hline
10 & 0.003 & 0.000 & 0.980 & 0.099 & 15.494 & 3.454 & 0.004 & 0.000 & 0.984 & 0.114 & 0.019 & 0.001 & 0.734 & 0.063 & 0.010 & 0.001 & 0.823 & 0.080\\
20 & 0.002 & 0.000 & 0.978 & 0.057 & 18.980 & 5.230 & 0.004 & 0.000 & 0.979 & 0.062 & 0.017 & 0.000 & 0.634 & 0.041 & 0.006 & 0.000 & 0.909 & 0.065\\
30 & 0.001 & 0.000 & 0.971 & 0.044 & 20.820 & 6.801 & 0.004 & 0.000 & 0.965 & 0.046 & 0.017 & 0.000 & 0.529 & 0.033 & 0.005 & 0.000 & 0.919 & 0.055\\
40 & 0.001 & 0.000 & 0.973 & 0.037 & 22.393 & 8.334 & 0.003 & 0.000 & 0.964 & 0.038 & 0.016 & 0.000 & 0.465 & 0.028 & 0.004 & 0.000 & 0.932 & 0.048\\
50 & 0.000 & 0.000 & 0.972 & 0.033 & 22.719 & 10.378 & 0.003 & 0.000 & 0.973 & 0.033 & 0.016 & 0.000 & 0.405 & 0.025 & 0.004 & 0.000 & 0.935 & 0.043\\
\hline
        \multicolumn{19}{c}{DGP 1,\ $\rho=0.7$,\ $\kappa=0.7$}\\
        \hline
10 & 0.006 & 0.001 & 0.976 & 0.127 & 24.400 & 3.470 & 0.010 & 0.001 & 0.977 & 0.134 & 0.119 & 0.006 & 0.179 & 0.048 & 0.063 & 0.004 & 0.425 & 0.078\\
20 & 0.003 & 0.000 & 0.973 & 0.084 & 24.401 & 5.221 & 0.008 & 0.000 & 0.966 & 0.086 & 0.127 & 0.003 & 0.046 & 0.030 & 0.060 & 0.002 & 0.312 & 0.063\\
30 & 0.002 & 0.000 & 0.969 & 0.067 & 25.290 & 6.744 & 0.006 & 0.000 & 0.969 & 0.068 & 0.121 & 0.002 & 0.013 & 0.024 & 0.052 & 0.001 & 0.283 & 0.054\\
40 & 0.002 & 0.000 & 0.968 & 0.058 & 25.845 & 8.297 & 0.005 & 0.000 & 0.959 & 0.058 & 0.106 & 0.002 & 0.009 & 0.022 & 0.043 & 0.001 & 0.307 & 0.048\\
50 & 0.001 & 0.000 & 0.967 & 0.051 & 26.321 & 9.759 & 0.004 & 0.000 & 0.962 & 0.051 & 0.092 & 0.001 & 0.006 & 0.020 & 0.035 & 0.001 & 0.345 & 0.044\\

        \hline 
        \multicolumn{19}{c}{DGP 2,\ $\rho=0$,\ $\kappa=0$}\\
        \hline
10 & 0.002 & 0.001 & 0.996 & 0.139 & 5.963 & 5.730 & 0.005 & 0.001 & 0.999 & 0.196 & 0.033 & 0.001 & 0.450 & 0.054 & 0.064 & 0.004 & 0.412 & 0.089\\
20 & 0.001 & 0.000 & 0.989 & 0.070 & 7.056 & 8.437 & 0.003 & 0.000 & 0.994 & 0.083 & 0.026 & 0.000 & 0.302 & 0.034 & 0.036 & 0.001 & 0.553 & 0.071\\
30 & 0.001 & 0.000 & 0.985 & 0.051 & 8.087 & 10.684 & 0.002 & 0.000 & 0.988 & 0.057 & 0.023 & 0.000 & 0.235 & 0.027 & 0.024 & 0.001 & 0.646 & 0.059\\
40 & 0.001 & 0.000 & 0.981 & 0.042 & 9.021 & 12.722 & 0.002 & 0.000 & 0.981 & 0.045 & 0.021 & 0.000 & 0.190 & 0.023 & 0.017 & 0.000 & 0.705 & 0.051\\
50 & 0.001 & 0.000 & 0.977 & 0.036 & 9.869 & 14.533 & 0.001 & 0.000 & 0.971 & 0.038 & 0.020 & 0.000 & 0.145 & 0.020 & 0.013 & 0.000 & 0.760 & 0.045\\
        \hline 
        \multicolumn{19}{c}{DGP 2,\ $\rho=0.7$,\ $\kappa=0$}\\
        \hline
10 & 0.002 & 0.001 & 0.993 & 0.132 & 10.334 & 5.200 & 0.004 & 0.001 & 0.994 & 0.162 & 0.035 & 0.002 & 0.555 & 0.058 & 0.044 & 0.004 & 0.611 & 0.087\\
20 & 0.001 & 0.000 & 0.987 & 0.068 & 11.462 & 7.820 & 0.002 & 0.000 & 0.991 & 0.078 & 0.025 & 0.001 & 0.446 & 0.037 & 0.029 & 0.002 & 0.668 & 0.069\\
30 & 0.001 & 0.000 & 0.983 & 0.050 & 12.435 & 9.916 & 0.002 & 0.000 & 0.987 & 0.054 & 0.020 & 0.000 & 0.388 & 0.029 & 0.019 & 0.001 & 0.719 & 0.058\\
40 & 0.000 & 0.000 & 0.981 & 0.041 & 13.292 & 11.855 & 0.001 & 0.000 & 0.983 & 0.043 & 0.018 & 0.000 & 0.329 & 0.025 & 0.014 & 0.000 & 0.763 & 0.050\\
50 & 0.000 & 0.000 & 0.980 & 0.035 & 14.019 & 13.764 & 0.001 & 0.000 & 0.978 & 0.037 & 0.016 & 0.000 & 0.301 & 0.022 & 0.010 & 0.000 & 0.806 & 0.044\\
        \hline 
        \multicolumn{19}{c}{DGP 2,\ $\rho=0.7$,\ $\kappa=0.7$}\\
        \hline
10 & 0.004 & 0.001 & 0.991 & 0.160 & 15.738 & 5.203 & 0.011 & 0.002 & 0.984 & 0.182 & 0.125 & 0.005 & 0.071 & 0.050 & 0.106 & 0.010 & 0.277 & 0.091\\
20 & 0.001 & 0.000 & 0.986 & 0.096 & 14.834 & 7.768 & 0.004 & 0.000 & 0.989 & 0.102 & 0.109 & 0.003 & 0.015 & 0.031 & 0.107 & 0.006 & 0.174 & 0.076\\
30 & 0.001 & 0.000 & 0.984 & 0.074 & 14.942 & 9.854 & 0.003 & 0.000 & 0.983 & 0.077 & 0.094 & 0.002 & 0.007 & 0.024 & 0.102 & 0.004 & 0.127 & 0.066\\
40 & 0.001 & 0.000 & 0.980 & 0.062 & 15.337 & 11.783 & 0.002 & 0.000 & 0.983 & 0.064 & 0.081 & 0.002 & 0.009 & 0.020 & 0.095 & 0.003 & 0.095 & 0.058\\
50 & 0.000 & 0.000 & 0.980 & 0.055 & 15.741 & 13.607 & 0.001 & 0.000 & 0.982 & 0.056 & 0.072 & 0.001 & 0.007 & 0.017 & 0.088 & 0.002 & 0.076 & 0.052\\
\hline
    \end{tabular}
    \begin{tablenotes}
      \small
      \item[a]Results are based on $10,000$ simulations. DGP 1 and 2 and all considered estimators are described in Section~\ref{sec:sim}.
    \end{tablenotes}
  \end{threeparttable}
  }
\end{table}

\begin{table}[h!]
\centering
    \caption{Simulation results for $\widehat{\beta}^{\rm TWFE}$, $\widehat{\beta}^{\rm FA}$, $\widehat{\beta}^{\rm CCE}$, $\widehat{\beta}^{\rm 1}$, $\widehat{\beta}^{\rm 2}$ with $N=500$\textsuperscript{a}}
    \label{tab.res.alt500}
    \adjustbox{max width=1.\textwidth}{
        \begin{threeparttable}
    \renewcommand{\arraystretch}{1.3} 
    \setlength{\tabcolsep}{4pt} 
    \begin{tabular}{c|cccc|cccc|cccc|cccc|cccc}
    \hline 
    \hline
    \renewcommand{\arraystretch}{1.3}
     & \multicolumn{4}{c|}{$\widehat{\beta}^{\rm TWFE}$} & \multicolumn{4}{c|}{$\widehat{\beta}^{\rm FA}$}&\multicolumn{4}{c|}{$\widehat{\beta}^{\rm CCE}$} &\multicolumn{4}{c|}{$\widehat{\beta}^{\rm 1}$}&\multicolumn{4}{c}{$\widehat{\beta}^{\rm 2}$}\\
    \hline
    \renewcommand{\arraystretch}{1.}
         $T$ & Bias & Var & Cov & Wid  & Bias & Var & Cov & Wid& Bias & Var & Cov & Wid& Bias & Var & Cov & Wid& Bias & Var & Cov & Wid\\
        \hline 
        \multicolumn{21}{c}{DGP 1,\ $\rho=0$,\ $\kappa=0$}\\
       \hline
10 & 0.104 & 0.002 & 0.051 & 0.053 & 0.237 & 0.008 & 0.014 & 0.065 & 0.004 & 0.000 & 0.934 & 0.066 & 0.053 & 0.004 & 0.449 & 0.052 & 0.057 & 0.003 & 0.405 & 0.051\\
20 & 0.107 & 0.001 & 0.002 & 0.039 & 0.272 & 0.004 & 0.000 & 0.071 & 0.005 & 0.000 & 0.918 & 0.042 & 0.046 & 0.002 & 0.387 & 0.039 & 0.051 & 0.002 & 0.265 & 0.038\\
30 & 0.110 & 0.001 & 0.000 & 0.033 & 0.283 & 0.003 & 0.000 & 0.074 & 0.006 & 0.000 & 0.880 & 0.034 & 0.040 & 0.001 & 0.350 & 0.033 & 0.045 & 0.001 & 0.194 & 0.032\\
40 & 0.110 & 0.000 & 0.000 & 0.030 & 0.287 & 0.002 & 0.000 & 0.074 & 0.006 & 0.000 & 0.843 & 0.029 & 0.034 & 0.001 & 0.342 & 0.030 & 0.039 & 0.001 & 0.163 & 0.028\\
50 & 0.111 & 0.000 & 0.000 & 0.028 & 0.288 & 0.002 & 0.000 & 0.073 & 0.007 & 0.000 & 0.790 & 0.025 & 0.030 & 0.001 & 0.330 & 0.027 & 0.034 & 0.001 & 0.140 & 0.026\\
\hline
        \multicolumn{21}{c}{DGP 1,\ $\rho=0.7$,\ $\kappa=0$}\\
        \hline
10 & 0.063 & 0.003 & 0.329 & 0.055 & 0.142 & 0.014 & 0.224 & 0.061 & 0.004 & 0.000 & 0.938 & 0.066 & 0.026 & 0.003 & 0.635 & 0.054 & 0.032 & 0.003 & 0.619 & 0.053\\
20 & 0.080 & 0.002 & 0.124 & 0.039 & 0.198 & 0.012 & 0.067 & 0.058 & 0.005 & 0.000 & 0.913 & 0.042 & 0.035 & 0.003 & 0.584 & 0.040 & 0.041 & 0.002 & 0.451 & 0.039\\
30 & 0.087 & 0.002 & 0.045 & 0.033 & 0.226 & 0.010 & 0.020 & 0.059 & 0.006 & 0.000 & 0.869 & 0.034 & 0.034 & 0.002 & 0.530 & 0.033 & 0.040 & 0.002 & 0.340 & 0.032\\
40 & 0.091 & 0.001 & 0.019 & 0.030 & 0.242 & 0.008 & 0.008 & 0.062 & 0.006 & 0.000 & 0.836 & 0.029 & 0.032 & 0.002 & 0.508 & 0.030 & 0.038 & 0.002 & 0.275 & 0.029\\
50 & 0.096 & 0.001 & 0.007 & 0.028 & 0.252 & 0.006 & 0.006 & 0.062 & 0.007 & 0.000 & 0.800 & 0.025 & 0.033 & 0.002 & 0.482 & 0.027 & 0.038 & 0.002 & 0.230 & 0.026\\
\hline
        \multicolumn{21}{c}{DGP 1,\ $\rho=0.7$,\ $\kappa=0.7$}\\
     \hline
10 & 0.083 & 0.004 & 0.313 & 0.073 & 0.151 & 0.012 & 0.212 & 0.072 & 0.005 & 0.001 & 0.930 & 0.085 & -0.034 & 0.006 & 0.259 & 0.085 & -0.024 & 0.005 & 0.360 & 0.081\\
20 & 0.093 & 0.003 & 0.152 & 0.056 & 0.194 & 0.010 & 0.074 & 0.060 & 0.006 & 0.000 & 0.917 & 0.064 & 0.001 & 0.003 & 0.535 & 0.064 & 0.010 & 0.003 & 0.625 & 0.061\\
30 & 0.096 & 0.002 & 0.078 & 0.048 & 0.216 & 0.009 & 0.029 & 0.058 & 0.007 & 0.000 & 0.907 & 0.053 & 0.010 & 0.003 & 0.653 & 0.054 & 0.018 & 0.002 & 0.691 & 0.052\\
40 & 0.099 & 0.002 & 0.039 & 0.043 & 0.232 & 0.008 & 0.012 & 0.058 & 0.007 & 0.000 & 0.886 & 0.046 & 0.014 & 0.002 & 0.696 & 0.048 & 0.021 & 0.002 & 0.686 & 0.046\\
50 & 0.100 & 0.001 & 0.018 & 0.039 & 0.242 & 0.006 & 0.005 & 0.060 & 0.007 & 0.000 & 0.874 & 0.042 & 0.015 & 0.002 & 0.727 & 0.043 & 0.021 & 0.001 & 0.685 & 0.041\\
\hline
        \multicolumn{21}{c}{DGP 2,\ $\rho=0$,\ $\kappa=0$}\\
        \hline
10 & 0.388 & 0.032 & 0.001 & 0.120 & -0.152 & 0.230 & 0.014 & 0.217 & 0.035 & 0.001 & 0.530 & 0.068 & 0.383 & 0.149 & 0.150 & 0.183 & 0.386 & 0.144 & 0.138 & 0.181\\
20 & 0.417 & 0.021 & 0.000 & 0.120 & 0.204 & 0.352 & 0.001 & 0.217 & 0.045 & 0.001 & 0.157 & 0.045 & 0.288 & 0.071 & 0.085 & 0.135 & 0.292 & 0.069 & 0.053 & 0.132\\
30 & 0.427 & 0.016 & 0.000 & 0.122 & 0.431 & 0.314 & 0.000 & 0.224 & 0.050 & 0.000 & 0.041 & 0.037 & 0.222 & 0.040 & 0.061 & 0.108 & 0.227 & 0.038 & 0.023 & 0.106\\
40 & 0.434 & 0.013 & 0.000 & 0.123 & 0.549 & 0.247 & 0.000 & 0.225 & 0.052 & 0.000 & 0.011 & 0.033 & 0.184 & 0.027 & 0.048 & 0.093 & 0.188 & 0.025 & 0.012 & 0.090\\
50 & 0.442 & 0.011 & 0.000 & 0.124 & 0.620 & 0.185 & 0.000 & 0.224 & 0.055 & 0.000 & 0.002 & 0.031 & 0.159 & 0.020 & 0.040 & 0.082 & 0.163 & 0.019 & 0.004 & 0.080\\
        \hline 
        \multicolumn{21}{c}{DGP 2,\ $\rho=0.7$,\ $\kappa=0$}\\
        \hline
10 & 0.246 & 0.049 & 0.074 & 0.097 & -0.193 & 0.084 & 0.198 & 0.152 & 0.029 & 0.002 & 0.655 & 0.067 & 0.234 & 0.167 & 0.309 & 0.145 & 0.240 & 0.161 & 0.361 & 0.143\\
20 & 0.309 & 0.044 & 0.003 & 0.100 & -0.023 & 0.242 & 0.060 & 0.177 & 0.039 & 0.001 & 0.392 & 0.045 & 0.226 & 0.115 & 0.325 & 0.119 & 0.231 & 0.110 & 0.297 & 0.117\\
30 & 0.339 & 0.039 & 0.000 & 0.103 & 0.188 & 0.328 & 0.026 & 0.194 & 0.043 & 0.001 & 0.232 & 0.037 & 0.198 & 0.081 & 0.293 & 0.102 & 0.203 & 0.077 & 0.211 & 0.099\\
40 & 0.358 & 0.035 & 0.000 & 0.106 & 0.319 & 0.343 & 0.013 & 0.206 & 0.047 & 0.001 & 0.142 & 0.033 & 0.175 & 0.063 & 0.265 & 0.090 & 0.180 & 0.059 & 0.145 & 0.087\\
50 & 0.374 & 0.030 & 0.000 & 0.111 & 0.442 & 0.336 & 0.006 & 0.219 & 0.049 & 0.001 & 0.086 & 0.030 & 0.154 & 0.045 & 0.233 & 0.082 & 0.159 & 0.042 & 0.113 & 0.079\\
        \hline 
        \multicolumn{21}{c}{DGP 2,\ $\rho=0.7$,\ $\kappa=0.7$}\\
        \hline
10 & 0.307 & 0.059 & 0.063 & 0.116 & -0.234 & 0.073 & 0.220 & 0.176 & 0.050 & 0.005 & 0.598 & 0.087 & 0.093 & 0.202 & 0.100 & 0.182 & 0.106 & 0.193 & 0.119 & 0.177\\
20 & 0.347 & 0.048 & 0.006 & 0.112 & -0.196 & 0.147 & 0.077 & 0.187 & 0.053 & 0.002 & 0.419 & 0.066 & 0.138 & 0.128 & 0.185 & 0.144 & 0.148 & 0.121 & 0.243 & 0.139\\
30 & 0.368 & 0.041 & 0.000 & 0.113 & -0.081 & 0.226 & 0.031 & 0.191 & 0.054 & 0.002 & 0.302 & 0.055 & 0.137 & 0.089 & 0.258 & 0.119 & 0.145 & 0.084 & 0.345 & 0.116\\
40 & 0.380 & 0.036 & 0.000 & 0.114 & 0.077 & 0.304 & 0.016 & 0.199 & 0.055 & 0.001 & 0.203 & 0.049 & 0.124 & 0.066 & 0.336 & 0.104 & 0.131 & 0.062 & 0.413 & 0.100\\
50 & 0.388 & 0.030 & 0.000 & 0.117 & 0.230 & 0.343 & 0.008 & 0.211 & 0.055 & 0.001 & 0.151 & 0.045 & 0.109 & 0.048 & 0.368 & 0.091 & 0.116 & 0.045 & 0.434 & 0.088\\
\hline
    \end{tabular}
        \begin{tablenotes}
      \small
      \item[a] Results are based on $10,000$ simulations. DGP 1 and 2  and all considered estimators are described in Section~\ref{sec:sim}.
    \end{tablenotes}
    \end{threeparttable}}
\end{table}
\subsection{Sensitivity to the number of groups}\label{subsec:sim_add2}
\begin{figure}[htbp]
    \centering

    \begin{subfigure}{0.45\textwidth}
        \includegraphics[width=\linewidth]{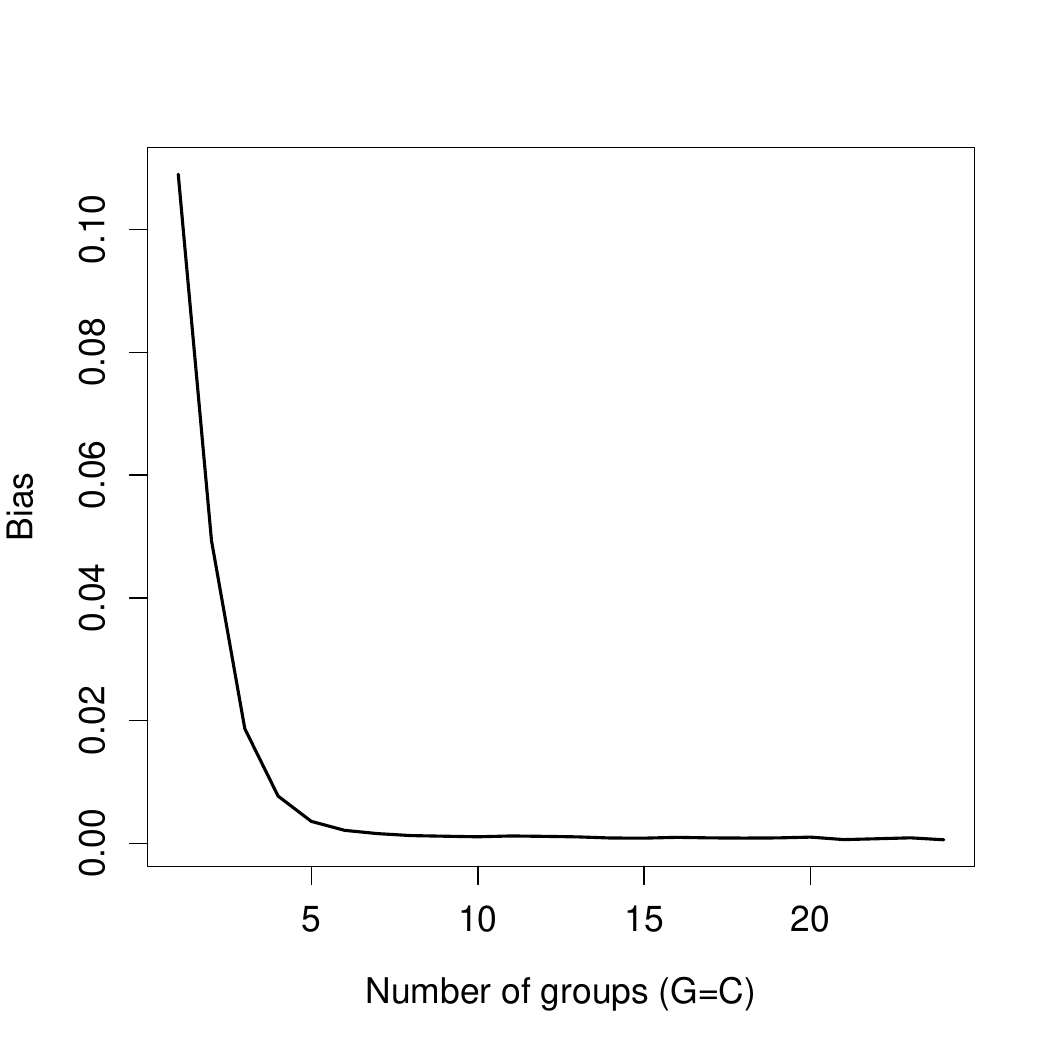}
        \caption{Bias}
        \label{fig:sub1}
    \end{subfigure}
    \hfill
    \begin{subfigure}{0.45\textwidth}
        \includegraphics[width=\linewidth]{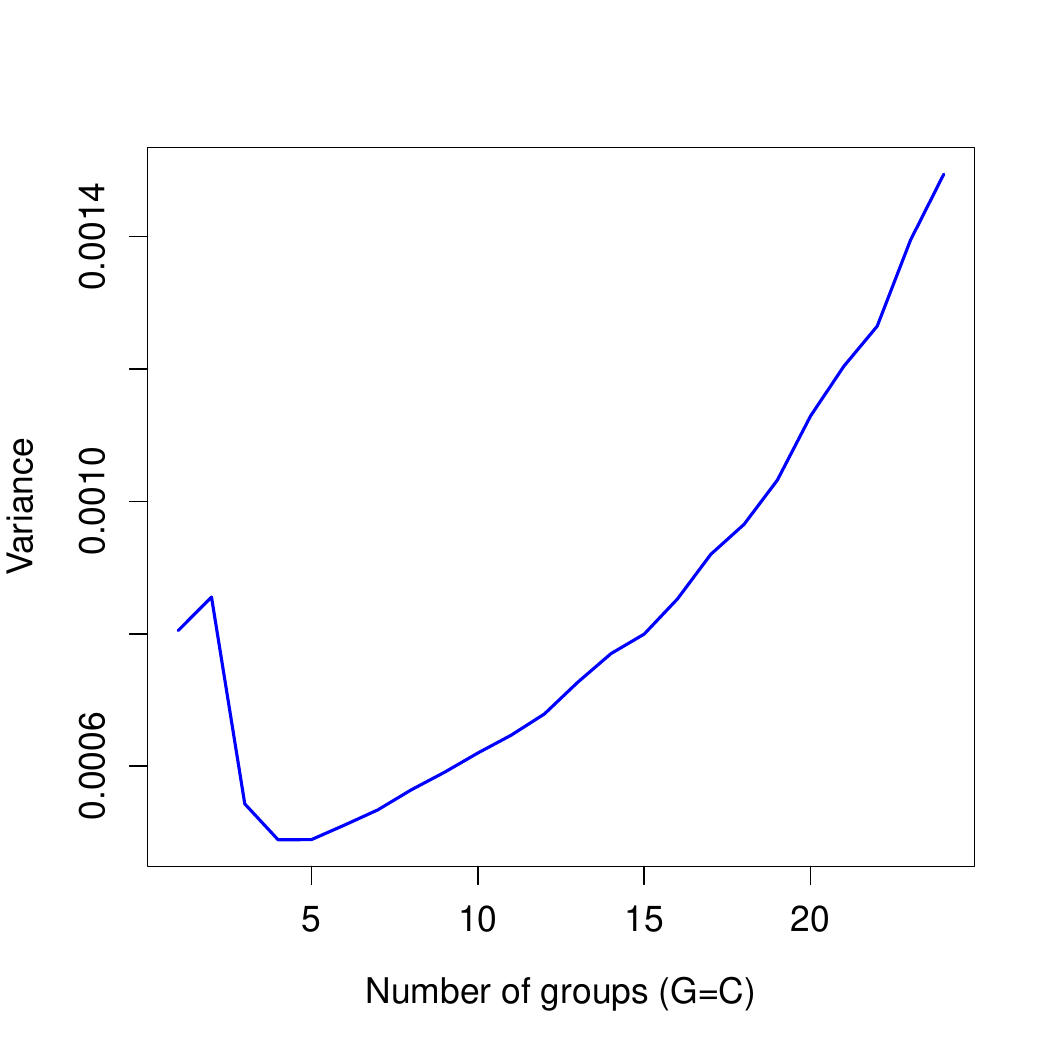}
        \caption{Variance}
        \label{fig:sub2}
    \end{subfigure}

    \vspace{0.5cm}

    \begin{subfigure}{0.45\textwidth}
        \includegraphics[width=\linewidth]{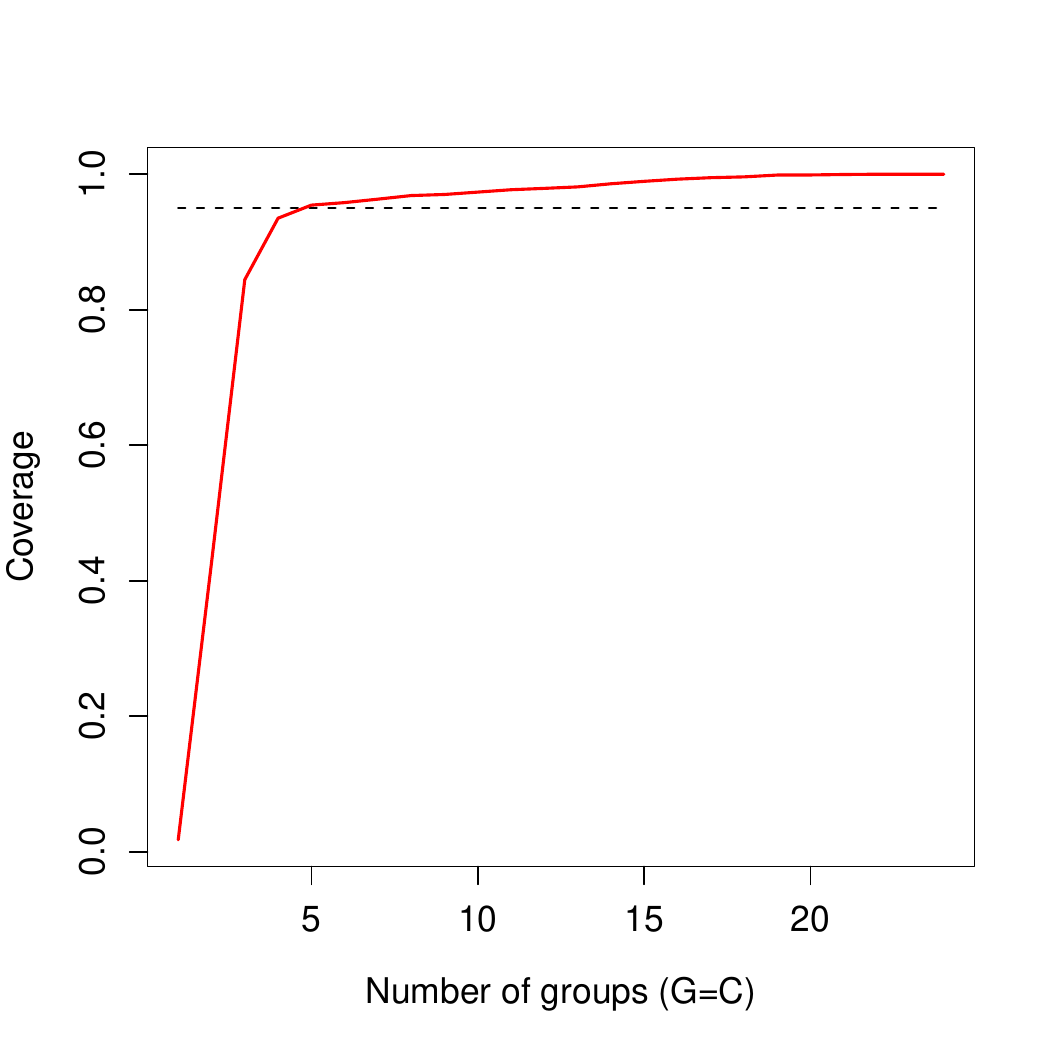}
        \caption{Coverage of 95\% confidence intervals}
        \label{fig:sub3}
    \end{subfigure}
    \hfill
    \begin{subfigure}{0.45\textwidth}
        \includegraphics[width=\linewidth]{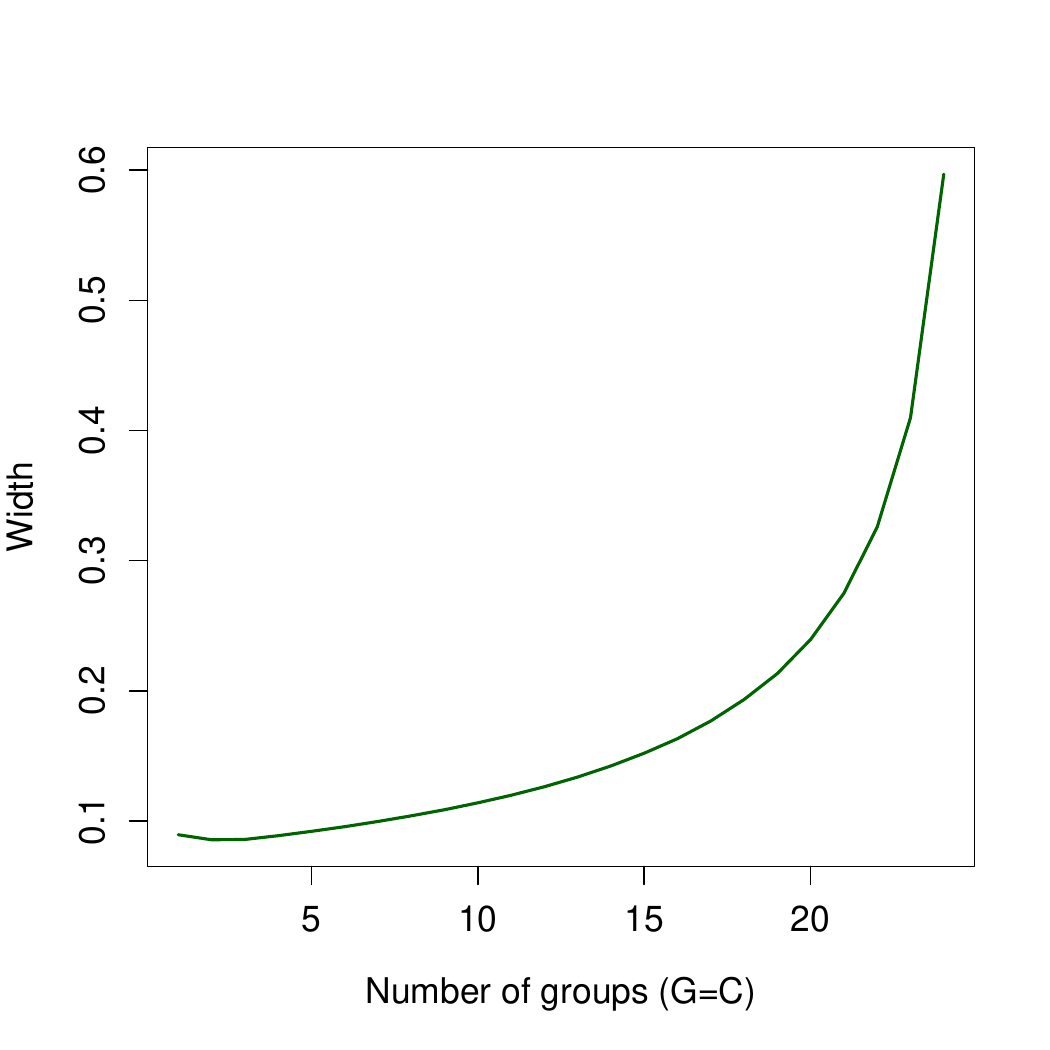}
        \caption{Width of 95\% confidence intervals}
        \label{fig:sub4}
    \end{subfigure}

    \caption{Bias, variance, coverage and width of 95\% confidence intervals for DGP 1}
 \label{fig.sens1}
\end{figure}

\begin{figure}[htbp]
    \centering

    \begin{subfigure}{0.45\textwidth}
        \includegraphics[width=\linewidth]{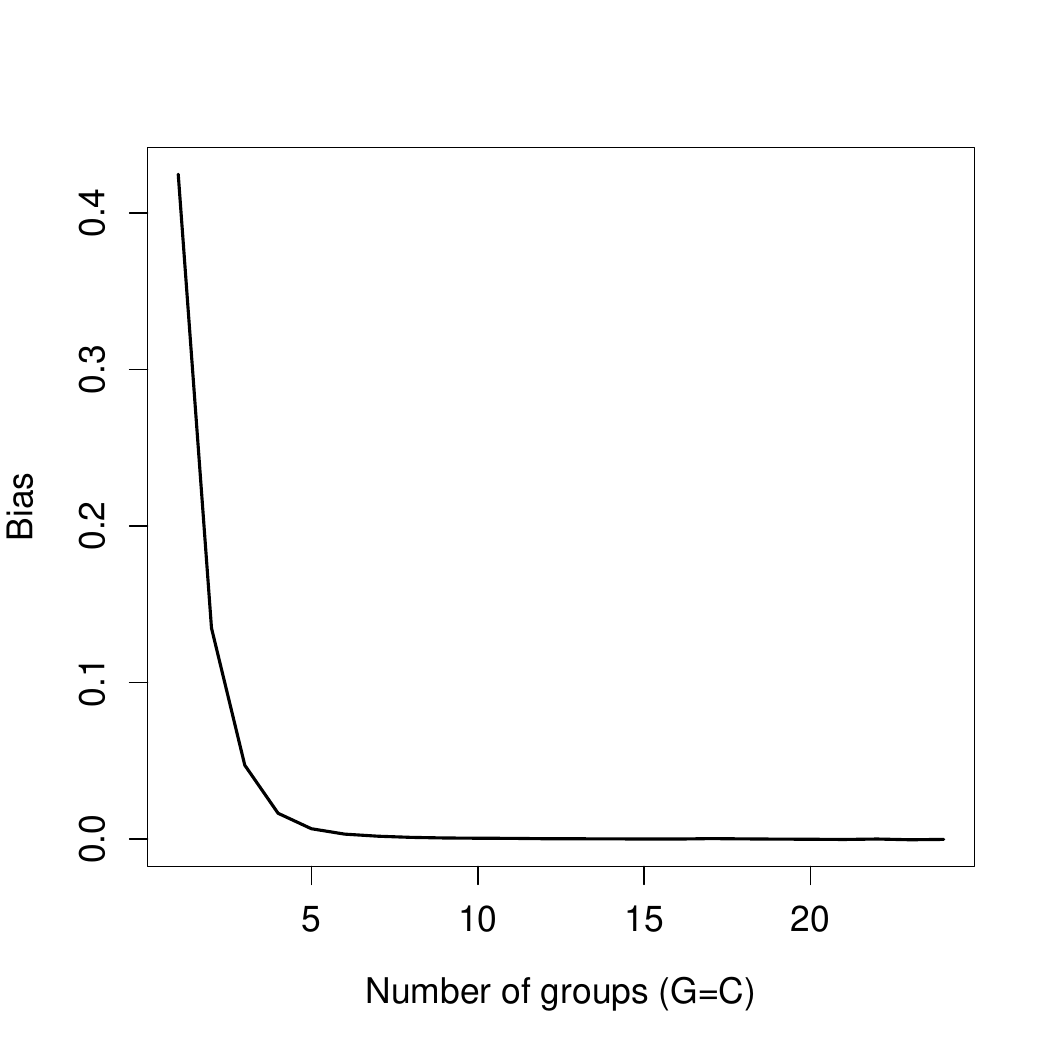}
        \caption{Bias}
        \label{fig:sub12}
    \end{subfigure}
    \hfill
    \begin{subfigure}{0.45\textwidth}
        \includegraphics[width=\linewidth]{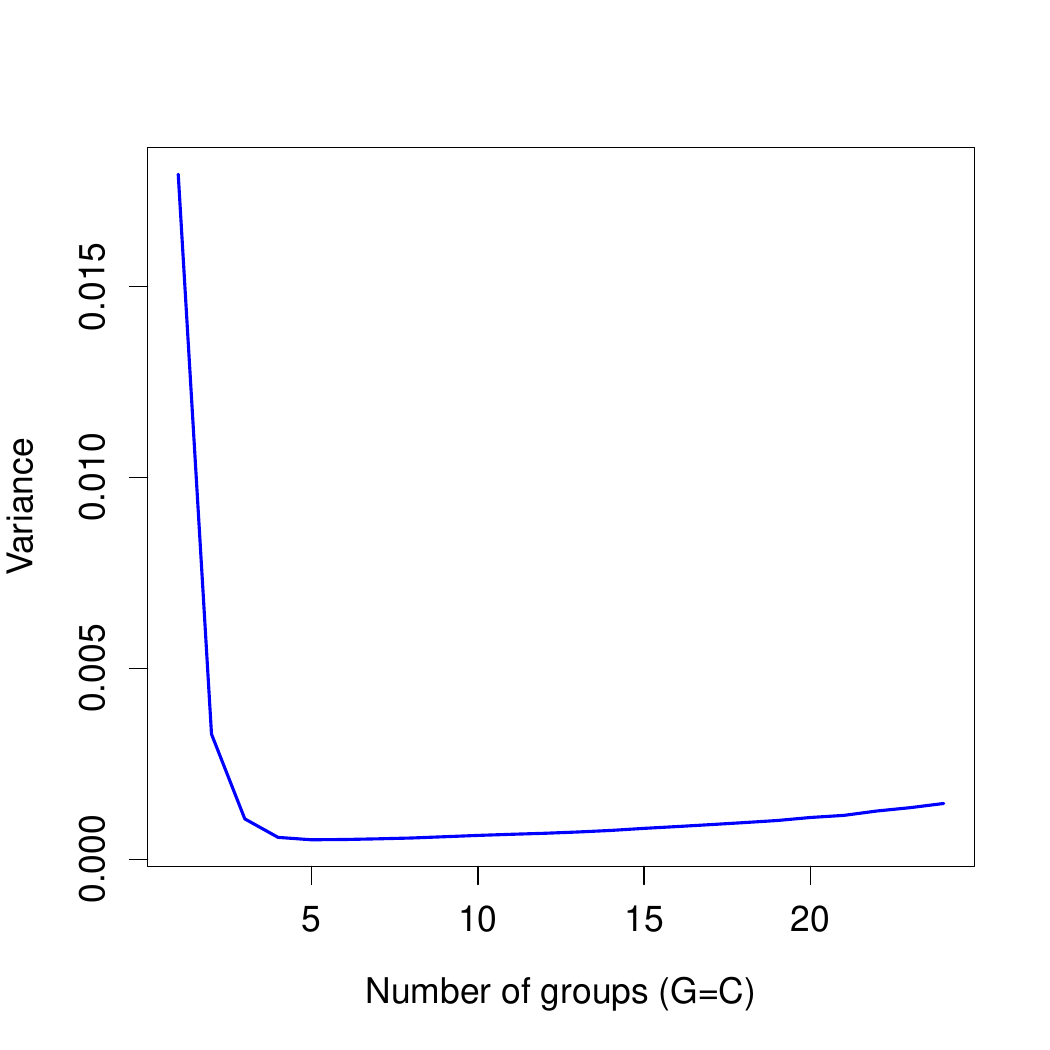}
        \caption{Variance}
        \label{fig:sub22}
    \end{subfigure}

    \vspace{0.5cm}

    \begin{subfigure}{0.45\textwidth}
        \includegraphics[width=\linewidth]{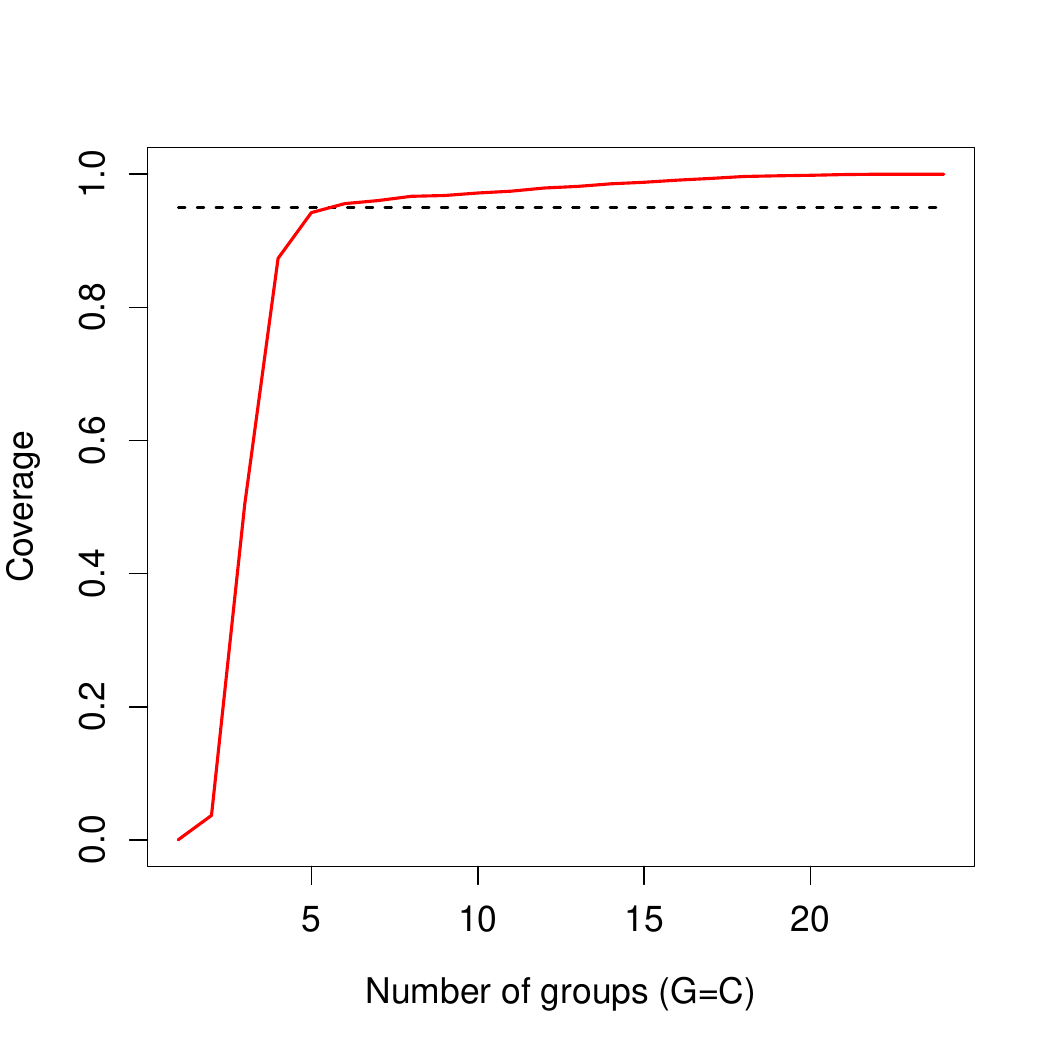}
        \caption{Coverage of 95\% confidence intervals}
        \label{fig:sub32}
    \end{subfigure}
    \hfill
    \begin{subfigure}{0.45\textwidth}
        \includegraphics[width=\linewidth]{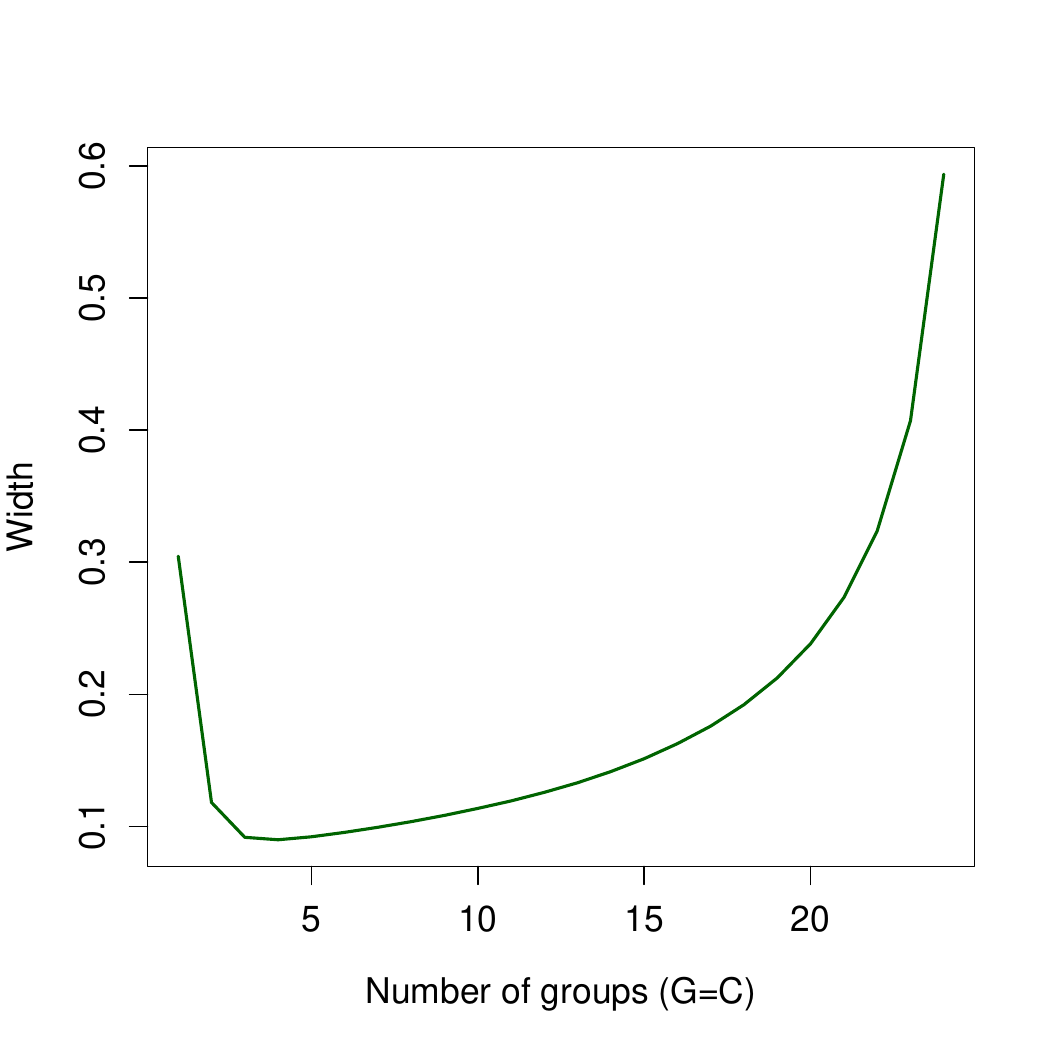}
        \caption{Width of 95\% confidence intervals}
        \label{fig:sub42}
    \end{subfigure}

    \caption{Bias, variance, coverage and width of 95\% confidence intervals for DGP 2}
  \label{fig.sens2}
\end{figure}

We also study the sensitivity of the estimator to the number of groups. To do so, we implement simulations under $N=T=50$ in DGP 1 and 2 (with $\kappa=\rho=0$) of Section \ref{sec:sim}. We simulate 10,000 datasets and compute the value of $\widehat{\beta}$ with a number of unit and time clusters $G=C$ varying between 1 and 24. The estimator and its standard error are computed as in Section~\ref{sec:sim}. Figures~\ref{fig.sens1} and~\ref{fig.sens2} present the average bias, variance, coverage and width of 95\% confidence intervals, in DGP 1 and 2, respectively. 
As long as the number of groups is larger than 5, the bias and coverage are insensitive to $G=C$. However, as $G=C$ increases, the variance and the width also become larger.

\section{Proof of Lemma \ref{lm.clus}}
We prove only the first statement, as the argument for the second is analogous. The proof proceeds in two steps.
\paragraph{Step 1.} In this step, we establish that 
\begin{equation}\label{2611240}
\frac{1}{N_d}\sum_{i\in\mathcal{N}_d} \left\|\varphi_d^\alpha(\alpha_i)-\frac{1}{N^d_{g_i^d}}\sum_{j\in\mathcal{N}_d} \mathbf{1}\{g_j^d= g_i^d\}\varphi_d^\alpha(\alpha_j)\right\|^2 = O_P\left(\frac{r_\alpha}{T}\right)+O_P\left(B_\alpha^d(G_d)\right).
\end{equation}
By the triangle inequality and the classical inequality $ab\le (a^2+b^2)/2$ for all $a,b\in\R$, 
\begin{equation}\label{2611241}
\begin{aligned}
    & \frac{1}{N_d}\sum_{i\in\mathcal{N}_d} \left\|\varphi_d^\alpha(\alpha_i)-\frac{1}{N^d_{g_i^d}}\sum_{j\in\mathcal{N}_d}\mathbf{1}\{g_j^d= g_i^d\}\varphi_d^\alpha(\alpha_j)\right\|^2 \\
    &\leq \frac{2}{N_d}\sum_{i\in\mathcal{N}_d} \left\|\varphi_d^\alpha(\alpha_i)-\widehat a^d(g_i^d)\right\|^2 +\frac{2}{N_d}\sum_{i\in\mathcal{N}_d} \left\|\widehat a^d(g_i^d)-\frac{1}{N^d_{g_i^d}}\sum_{j\in\mathcal{N}_d}\mathbf{1}\{g_j^d= g_i^d\}\varphi_d^\alpha(\alpha_j)\right\|^2.
\end{aligned}
\end{equation}
Under Assumption~\ref{as.clus}\eqref{clusi}, arguments analogous to those in the proof of Lemma 1 in \cite{bonhomme2022discretizing} yield
\begin{equation}\label{2611242}
    \frac{1}{N_d}\sum_{i\in\mathcal{N}_d} \left\|\varphi_d^\alpha(\alpha_i)-\widehat a^d(g_i^d)\right\|^2 = O_P\left(\frac{r_\alpha}{T}\right)+O_P\left(B_\alpha^d(G_d)\right).
\end{equation}
Next, using that $\widehat a^d(g_i^d)= \frac{1}{N^d_{g_i^d}}\sum_{j\in\mathcal N_d} \mathbf{1}\{g_j^d= g_i^d\}a_j^d $, we obtain
\begin{equation}\label{2611243}
\begin{aligned}
  &\frac{1}{N_d}\sum_{i\in\mathcal{N}_d} \left\|\widehat a^d(g_i^d)-\frac{1}{N^d_{g_i^d}}\sum_{j\in\mathcal N_d} \mathbf{1}\{g_j^d= g_i^d\}\varphi_d^\alpha(\alpha_j)\right\|^2\\
    & =\frac{1}{N_d}\sum_{i\in\mathcal{N}_d} \left\|\frac{1}{N^d_{g_i^d}}\sum_{j\in\mathcal N_d}\mathbf{1}\{g_j^d= g_i^d\}(a_j^d-\varphi_d^\alpha(\alpha_j))\right\|^2 \\
    &\le \frac{1}{N_d}\sum_{i\in\mathcal{N}_d} \max_{j\in\mathcal N_d}\left\|a_j^d-\varphi_d^\alpha(\alpha_j)\right\|^2  = \max_{j\in\mathcal N_d}\left\|a_j^d-\varphi_d^\alpha(\alpha_j)\right\|^2
     = O_P\left(\frac{r_\alpha}{T}\right),
\end{aligned}
\end{equation}
where the inequality follows from the triangle inequality, and the last equality is a consequence of Assumption \ref{as.clus}. Combining \eqref{2611241}--
\eqref{2611243}, we obtain \eqref{2611240}.

\paragraph{Step 2.} In this step, we establish the result stated in the lemma. We first note that
\begin{equation}\label{2611244}
\begin{aligned}
&  \frac{1}{N_d}\sum_{i\in\mathcal{N}_d} \left\|\alpha_i-\frac{1}{N^d_{g_i^d}}\sum_{j\in\mathcal N_d}\mathbf{1}\{g_j^d= g_i^d\}\alpha_j\right\|^2\\
  &=\frac{1}{N_d}\sum_{i\in\mathcal{N}_d} \left\|\frac{1}{N^d_{g_i^d}}\sum_{j\in\mathcal N_d}\mathbf{1}\{g_j^d= g_i^d\}(\alpha_i-\alpha_j)\right\|^2 \\
  &\le \frac{1}{N_d}\sum_{i\in\mathcal{N}_d} \frac{1}{\left(N^d_{g_i^d}\right)^2}\left(\sum_{j\in\mathcal N_d}\mathbf{1}\{g_j^d= g_i^d\}\left\|\alpha_i-\alpha_j\right\|\right)^2 \\
  &\le \frac{1}{N_d}\sum_{i\in\mathcal{N}_d} \frac{1}{N^d_{g_i^d}}\left(\sum_{j\in\mathcal N_d}\mathbf{1}\{g_j^d= g_i^d\}\left\|\alpha_i-\alpha_j\right\|^2\right), \\
\end{aligned}
\end{equation}
where the first inequality follows from the triangle inequality and the second inequality is a consequence of  the Cauchy--Schwarz inequality. Next, by Assumption \ref{as.clus}\eqref{clusi}, there exists a constant $L>0$ such that 
\begin{equation}\label{2611245}
\begin{aligned}
    &\frac{1}{N_d}\sum_{i\in\mathcal{N}_d} \frac{1}{N^d_{g_i^d}}\left(\sum_{j\in\mathcal N_d}\mathbf{1}\{g_j^d= g_i^d\}\left\|\alpha_i-\alpha_j\right\|^2\right)\\
    &=\frac{1}{N_d}\sum_{i\in\mathcal{N}_d}\frac{1}{N^d_{g_i^d}}\left(\sum_{j\in\mathcal N_d}\mathbf{1}\{g_j^d= g_i^d\}\left\|\psi^\alpha_d(\varphi^\alpha_d(\alpha_i))-\psi^\alpha_d(\varphi^\alpha_d(\alpha_j))\right\|^2\right)\\
    &\le  \frac{L}{N_d} \sum_{i\in\mathcal{N}_d} \frac{1}{N^d_{g_i^d}}\left(\sum_{j\in\mathcal N_d}\mathbf{1}\{g_j^d= g_i^d\}\left\|\varphi^\alpha_d(\alpha_i)-\varphi^\alpha_d(\alpha_j)\right\|^2\right).
\end{aligned}
\end{equation}
Moreover, we have
\begin{align*}
& 
\frac{1}{N_d}\sum_{i\in\mathcal{N}_d} \frac{1}{N^d_{g_i^d}}\left(\sum_{j\in\mathcal N_d}\mathbf{1}\{g_j^d= g_i^d\}\left\|\varphi_d^\alpha(\alpha_i)-\varphi_d^\alpha(\alpha_j)\right\|^2\right)\\
&=\frac{1}{N_d}\sum_{i\in \mathcal{N}_d}\frac{1}{N^d_{g_i^d}}\sum_{j\in\mathcal{N}_d}\mathbf{1}\{g^d_j=g^d_i\} \left(\varphi_d^\alpha(\alpha_i)^\top(\varphi_d^\alpha(\alpha_i)-\varphi_d^\alpha(\alpha_j))- \varphi_d^\alpha(\alpha_j)^\top(\varphi_d^\alpha(\alpha_i)-\varphi_d^\alpha(\alpha_j))\right)\\
&=\frac{2}{N_d}\sum_{i\in \mathcal{N}_d}\varphi_d^\alpha(\alpha_i)^\top\varphi_d^\alpha(\alpha_i)-\frac{2}{N_d}\sum_{i\in \mathcal{N}_d}\frac{1}{N^d_{g_i^d}}\sum_{j\in\mathcal N_d}\mathbf{1}\{g^d_j=g^d_i\}\varphi_d^\alpha(\alpha_i)^\top\varphi_d^\alpha(\alpha_j)\\
&= \frac{2}{N_d}\sum_{i\in \mathcal{N}_d}\varphi_d^\alpha(\alpha_i)^\top\left(\varphi_d^\alpha(\alpha_i)-
\frac{1}{N^d_{g_i^d}}\sum_{j\in\mathcal N_d}\mathbf{1}\{g^d_j=g^d_i\}\varphi_d^\alpha(\alpha_j)\right)\\
&=\frac{2}{N_d}\sum_{i\in \mathcal{N}_d}\left\|\varphi_d^\alpha(\alpha_i)-\frac{1}{N^d_{g_i^d}}\sum_{j\in\mathcal{N}_d}\mathbf{1}\{g^d_j=g^d_i\} \varphi_d^\alpha(\alpha_j)\right\|^2= O_P\left(\frac{r_\alpha}{T}+B_\alpha^d(G_d)\right),
\end{align*}
where in the last equality we used \eqref{2611240}. Combining the last result with \eqref{2611244}--\eqref{2611245}, we obtain the result of the lemma.

\section{On Theorem \ref{th.ols}}
This section concerns the proof of Theorem \ref{th.ols}. It is organized as follows. Section~\ref{subsec.not} introduces the notation used in the proof. Section~\ref{subsec.pr_th} contains the main body of the proof of Theorem~\ref{th.ols}, which relies on auxiliary lemmas stated and proved in Section~\ref{subsec.aux}. The proofs of these auxiliary lemmas, in turn, depend on technical lemmas stated and proved in Section~\ref{subsec.tech}.

\subsection{Notation}\label{subsec.not} For all $(i,t,k,d)\in\{1,\dots,N\}\times\{1,\dots,T\}\times\{1,\dots,K\}\times\{1,\ldots,4\}$, we let $h_{itk}:=h_k(\alpha_i,\gamma_t)$, $f_{it}:=f(\alpha_i,\gamma_t)$, and we use the notation
\begin{align*}\widetilde{h}_{itk}^d &:= h_{itk} -\left(\bar{h}_{g_i^dt}\right)_k- \left(\bar{h}_{ic_t^d}\right)_k+\left(\bar{h}_{g_i^dc_t^d}\right)_k,\\
\widetilde{f}_{it}^d&:= f_{it} -\bar{f}_{g_i^dt}- \bar{f}_{ic_t^d}+\bar{f}_{g_i^dc_t^d},\\
\widetilde{u}_{itk}^d &:= u_{itk} -\left(\bar{u}_{g_i^dt}\right)_k- \left(\bar{u}_{ic_t^d}\right)_k+\left(\bar{u}_{g_i^dc_t^d}\right)_k.
\end{align*}
\subsection{Proof of Theorem \ref{th.ols}}\label{subsec.pr_th}

 We have
\begin{align*}\notag \widehat{\beta}^{\rm CF}&=\left(\sum_{d=1}^4\sum_{(i,t)\in\mathcal{O}_d} \widehat{u}_{it}^d(\widehat{u}_{it}^d)^\top \right)^{-1}\sum_{d=1}^4\sum_{(i,t)\in\mathcal{O}_d}\widehat{u}_{it}^d \widehat{e}_{it}^d\\
&= \left(\sum_{d=1}^4\sum_{(i,t)\in\mathcal{O}_d} \widehat{u}_{it}^d (\widehat{u}_{it}^d)^\top \right)^{-1}\sum_{d=1}^4\sum_{(i,t)\in\mathcal{O}_d}\widehat{u}_{it}^d y_{it}.
\end{align*}
 Since $y_{it}=x_{it}^\top\beta+f_{it}+v_{it}$, this yields
\begin{align*}
\widehat{\beta}^{\rm CF}&= \beta + \left(\sum_{d=1}^4\sum_{(i,t)\in\mathcal{O}_d} \widehat{u}_{it}^d (\widehat{u}_{it}^d)^\top \right)^{-1}\sum_{d=1}^4\sum_{(i,t)\in\mathcal{O}_d}\widehat{u}_{it}^d f_{it} \\
&\quad +\left(\sum_{d=1}^4\sum_{(i,t)\in\mathcal{O}_d}\widehat{u}_{it}^d (\widehat{u}_{it}^d)^\top \right)^{-1}\sum_{d=1}^4\sum_{(i,t)\in\mathcal{O}_d}\widehat{u}_{it}^d v_{it}.
\end{align*}
We obtain
\begin{align*}
    \sqrt{NT}(\widehat{\beta}^{\rm CF}-\beta) 
    &= \left(\frac{1}{NT}\sum_{d=1}^4\sum_{(i,t)\in\mathcal{O}_d} \widehat{u}_{it}^d (\widehat{u}_{it}^d)^\top \right)^{-1}\frac{1}{\sqrt{NT}}\sum_{d=1}^4\sum_{(i,t)\in\mathcal{O}_d}\widehat{u}_{it}^d f_{it}  \\
   & \quad+\left(\frac{1}{NT}\sum_{d=1}^4\sum_{(i,t)\in\mathcal{O}_d}\widehat{u}_{it}^d (\widehat{u}_{it}^d)^\top \right)^{-1}\frac{1}{\sqrt{NT}}\sum_{d=1}^4\sum_{(i,t)\in\mathcal{O}_d}\widehat{u}_{it}^d v_{it}.
\end{align*}
By Lemmas \ref{lm.term1}--
\ref{lm.term3} and the continuous mapping theorem, we obtain
\begin{equation*}
\sqrt{NT}(\widehat{\beta}^{\rm CF}-\beta) 
=(\Sigma_U^{-1}+o_P(1))\frac{1}{\sqrt{NT}} \sum_{i=1}^N \sum_{t=1}^Tu_{it}v_{it}+o_P(1).
\end{equation*}
Under Assumption~\ref{as.errors}, by combining H\"older's and Markov's inequalities, it is not difficult to show that the following conditional Lindeberg condition holds: for all $\varepsilon>0$, as $N,T$ tend to infinity,
\[
    \frac{1}{NT}\sum_{i=1}^N\sum_{t=1}^T\E\left[\Vert v_{it}u_{it}\Vert^2\mathbf 1\left\{\Vert v_{it}u_{it}\Vert\geq \varepsilon\sqrt{NT}\right\}|\mathcal F_{NT}\right]\to 0.
\]
An application of the multivariate Lindeberg--Feller central limit theorem yields, for all $c\in\R^K$ and $z\in\R$, almost-surely, as $N$ and $T$ tend to infinity, 
\begin{equation*}
\left.\Pr\left(c^\top\Omega^{-1/2}\left(\frac{1}{\sqrt{NT}} \sum_{i=1}^N \sum_{t=1}^Tu_{it}v_{it}\right)\leq z \right|\mathcal F_{NT}\right)\to(2\pi \Vert c \Vert^2 )^{-1/2}\int_{-\infty}^z\exp\left(-\frac{t^2}{2\Vert c\Vert^2}\right)dt.
\end{equation*}
By the dominated convergence theorem, the sequence of unconditional cumulative distribution functions of $c^\top\Omega^{-1/2}\left(\frac{1}{\sqrt{NT}} \sum_{i=1}^N \sum_{t=1}^Tu_{it}v_{it}\right)$ evaluated at $z$ converges to the same limit. By the Cramer--Wold device, this yields, as $N$ and $T$ tend to infinity,
\[
    \frac{1}{\sqrt{NT}} \sum_{i=1}^N \sum_{t=1}^Tu_{it}v_{it}\overset{d}{\to}\mathcal N\left(0,\Omega\right).
\]
In particular, $\frac{1}{\sqrt{NT}} \sum_{i=1}^N \sum_{t=1}^Tu_{it}v_{it}=O_P(1)$ so that 
\begin{align*}
\sqrt{NT}(\widehat{\beta}^{\rm CF}-\beta) 
&=(\Sigma_U^{-1}+o_P(1))\frac{1}{\sqrt{NT}} \sum_{i=1}^N \sum_{t=1}^Tu_{it}v_{it}+o_P(1)\\
&=\Sigma_U^{-1}\frac{1}{\sqrt{NT}} \sum_{i=1}^N \sum_{t=1}^Tu_{it}v_{it}+o_P(1)
\end{align*}
and the result follows from Slutsky's lemma.

\subsection{Auxiliary lemmas}\label{subsec.aux}
\begin{Lemma}\label{lm.term1} Let Assumptions~\ref{as.fac}--
\ref{as.approx} hold. Then, for every fold $d\in\{1,\dots,4\}$, as $N,T,G_d,C_d$ tend to infinity, we have 
$$\frac{1}{N_dT_d} \sum_{(i,t)\in\mathcal{O}_d}\widehat{u}_{it}^d (\widehat{u}_{it}^d)^\top = \Sigma_U+o_P(1).$$
\end{Lemma}
\begin{Proof}
Fix $d\in\{1,\dots,4\}$ and $k,\ell\in\{1,\dots,K\}$. 
We have 
\begin{align*}
\frac{1}{N_dT_d} \sum_{(i,t)\in\mathcal{O}_d} \widehat{u}_{itk}^d \widehat{u}_{it\ell}^d
&=\frac{1}{N_dT_d} \sum_{(i,t)\in\mathcal{O}_d} \widehat{u}_{itk}^d x_{it\ell}\\
&=\frac{1}{N_dT_d}\sum_{(i,t)\in\mathcal{O}_d} \left(\widetilde{u}_{itk}^d+\widetilde{h}_{itk}^d\right) \left(u_{it\ell}+h_{it\ell}\right)\\
&=\frac{1}{N_dT_d}\sum_{(i,t)\in\mathcal{O}_d}  u_{itk}u_{it\ell}+ \frac{1}{N_dT_d}\sum_{(i,t)\in\mathcal{O}_d} \widetilde{h}_{itk}^d\widetilde{h}_{it\ell}^d+ \frac{1}{N_dT_d}\sum_{(i,t)\in\mathcal{O}_d} \widetilde{h}_{itk}^du_{it\ell} \\
&\quad + \frac{1}{N_dT_d} \sum_{(i,t)\in\mathcal{O}_d}  u_{itk}\widetilde{h}_{it\ell}^d+ \frac{1}{N_dT_d}\sum_{(i,t)\in\mathcal{O}_d}  (\widetilde{u}_{itk}^d-u_{itk})u_{it\ell}\\
&=\frac{1}{N_dT_d}\sum_{(i,t)\in\mathcal{O}_d}  u_{itk}u_{it\ell}+o_P(1),
\end{align*}
where we used Lemmas \ref{lm.aux1}--\ref{lm.aux2} and \ref{lm.aux4} and Assumptions \ref{as.rates}--\ref{as.approx} in the last equality. The result follows from the law of large numbers and the continuous mapping theorem.
\end{Proof}

\begin{Lemma}
    \label{lm.term2} Let Assumptions \ref{as.fac}--
    \ref{as.approx} hold. Then, for every fold $d\in\{1,\dots,4\}$, as $N,T,G_d,C_d$ tend to infinity, we have, we have
  $$\frac{1}{\sqrt{N_dT_d}}\sum_{(i,t)\in\mathcal{O}_d}\widehat{u}_{it}^d f_{it}=o_P\left(1\right).$$
\end{Lemma}
\begin{Proof}
    For any $k\in\{1,\dots,K\}$, it holds that
    \begin{align*}
        &\frac{1}{\sqrt{N_dT_d}}\sum_{(i,t)\in\mathcal{O}_d}\widehat{u}_{itk}^df_{it}\\
        &=  \frac{1}{\sqrt{N_dT_d}}\sum_{(i,t)\in\mathcal{O}_d}\left(\widetilde{u}_{itk}^d +\widetilde{h}_{itk}^d\right) f_{it}\\
        &=\frac{1}{\sqrt{N_dT_d}}\sum_{(i,t)\in\mathcal{O}_d}\widetilde{h}_{itk}^d\widetilde{f}_{it}^d+ \frac{1}{\sqrt{N_dT_d}}\sum_{(i,t)\in\mathcal{O}_d}u_{itk}\widetilde{f}_{it}^d\\
        &= \sqrt{N_dT_d}O_P\left( \left(\frac{r_\alpha}{T}\right)^2+\left(\frac{{r_\gamma}}{N}\right)^2+ B_\alpha^d(G_d)^2+ B_\gamma^d(C_d)^2\right)+o_P\left(1\right)
        =o_P\left(1\right),
    \end{align*}
 by Lemmas \ref{lm.aux1} and \ref{lm.aux4} and Assumptions \ref{as.rates}--\ref{as.approx}.   
\end{Proof}

\begin{Lemma}
    \label{lm.term3} Let Assumptions \ref{as.fac}--
    \ref{as.approx} hold. Then, for every fold $d\in\{1,\dots,4\}$, as $N,T,G_d,C_d$ tend to infinity, we have  
  $$\frac{1}{\sqrt{N_dT_d}}\sum_{(i,t)\in\mathcal{O}_d}\widehat{u}_{it}^d v_{it}=\frac{1}{\sqrt{N_dT_d}}\sum_{(i,t)\in\mathcal{O}_d}u_{it} v_{it}+o_P(1).$$
\end{Lemma}
\begin{Proof}
    For every $k\in\{1,\dots,K\}$, it holds that
    \begin{align*}
         &\frac{1}{\sqrt{N_dT_d}}\sum_{(i,t)\in\mathcal{O}_d}\widehat{u}_{itk}^d v_{it}\\
         &=  \frac{1}{\sqrt{N_dT_d}}\sum_{(i,t)\in\mathcal{O}_d}\left(\widetilde{u}_{itk}^d+\widetilde{h}_{itk}^d\right) v_{it}\\
        &= \frac{1}{\sqrt{N_dT_d}}\sum_{(i,t)\in\mathcal{O}_d}u_{itk} v_{it}+\frac{1}{\sqrt{N_dT_d}}\sum_{(i,t)\in\mathcal{O}_d}\left(\widetilde{u}_{itk}^d-u_{itk}\right) v_{it}+\frac{1}{\sqrt{N_dT_d}}\sum_{(i,t)\in\mathcal{O}_d}\widetilde{h}_{itk}^d v_{it}\\
        &=\frac{1}{\sqrt{N_dT_d}}\sum_{(i,t)\in\mathcal{O}_d}u_{itk} v_{it}+o_P(1),
    \end{align*}
   by Lemmas \ref{lm.aux3}--\ref{lm.aux4} and Assumptions \ref{as.rates}--\ref{as.approx}. 
\end{Proof}
\subsection{Technical lemmas}\label{subsec.tech}
\begin{Lemma}\label{lm.aux1}
    Let Assumptions \ref{as.fac}--\ref{as.clus} 
    hold. Then, for all $k\in\{1,\ldots,K\}$ and $d\in\{1,\dots,4\}$, as $N$ and $T$ tend to infinity, we have 
    $$\frac{1}{N_dT_d}\sum_{(i,t)\in\mathcal{O}_d}\left(\widetilde{h}_{itk}^d\right)^2=O_P\left(\left(\frac{r_\alpha}{T}\right)^2+\left(\frac{{r_\gamma}}{N}\right)^2+ B_\alpha^d(G_d)^2+ B_\gamma^d(C_d)^2\right)$$
    and    $$\frac{1}{N_dT_d}\sum_{(i,t)\in\mathcal{O}_d}\left(\widetilde{f}_{it}^d\right)^2=O_P\left(\left(\frac{r_\alpha}{T}\right)^2+\left(\frac{{r_\gamma}}{N}\right)^2+ B_\alpha^d(G_d)^2+ B_\gamma^d(C_d)^2\right).$$
\end{Lemma}
\begin{Proof} Fix $(k,d)\in\{1,\ldots,K\}\times\{1,\ldots,4\}$. By Assumption \ref{as.fac} and relying on analogous Taylor expansions as in the proof of Lemma 2 in \cite{freeman2023linear}, we have 
$$ \widetilde{h}_{itk}^d=O\left(\frac{1}{N^d_{g_i^d}}\sum_{j\in\mathcal{N}_d} \mathbf{1}\{g^d_j=g^d_i\}\left\|\alpha_i-\alpha_j\right\|^2 + \frac{1}{T^d_{c_t^d}}\sum_{s\in\mathcal{T}_d} \mathbf{1}\{c^d_s=c^d_t\}\left\|\gamma_t-\gamma_s\right\|^2\right),$$
uniformly in $i,t$. This implies
\begin{align*}
    &\frac{1}{N_dT_d}\sum_{(i,t)\in\mathcal{O}_d}\left|\widetilde{h}_{itk}^d\right|\\
    &= O\left(\frac{1}{N_d}\sum_{i\in \mathcal{N}_d}\frac{1}{N^d_{g_i^d}}\sum_{j\in \mathcal{N}_d}\mathbf{1}\{g^d_j=g^d_i\}\left\|\alpha_i-\alpha_j\right\|^2 + \frac{1}{T_d}\sum_{t\in \mathcal{T}_d}\frac{1}{T^d_{c_t^d}}\sum_{s\in\mathcal{T}_d} \mathbf{1}\{c^d_s=c^d_t\}\left\|\gamma_t-\gamma_s\right\|^2\right).
\end{align*}
Next, notice that 
\begin{align*}
&\frac{1}{N_d}\sum_{i\in \mathcal{N}_d}\frac{1}{N^d_{g_i^d}}\sum_{j\in\mathcal N_d}\mathbf{1}\{g^d_j=g^d_i\}\left\|\alpha_i-\alpha_j\right\|^2\\
&=\frac{1}{N_d}\sum_{i\in \mathcal{N}_d}\frac{1}{N^d_{g_i^d}}\sum_{j\in\mathcal N_d}\mathbf{1}\{g^d_j=g^d_i\}\left(\alpha_i^\top(\alpha_i-\alpha_j)- \alpha_j^\top(\alpha_i-\alpha_j)\right)\\
&=\frac{2}{N_d}\sum_{i\in \mathcal{N}_d}\alpha_i^\top\alpha_i-\frac{2}{N_d}\sum_{i\in \mathcal{N}_d}\frac{1}{N^d_{g_i^d}}\sum_{j\in\mathcal N_d}\mathbf{1}\{g^d_j=g^d_i\}\alpha_i^\top\alpha_j\\
&= \frac{2}{N_d}\sum_{i\in \mathcal{N}_d}\alpha_i^\top\left(\alpha_i-
\frac{1}{N^d_{g_i^d}}\sum_{j\in\mathcal N_d}\mathbf{1}\{g^d_j=g^d_i\}\alpha_j\right)\\
&=\frac{2}{N_d}\sum_{i\in \mathcal{N}_d}\left\|\alpha_i-\frac{1}{N^d_{g_i^d}}\sum_{j\in\mathcal{N}_d}\mathbf{1}\{g^d_j=g^d_i\} \alpha_j\right\|^2\\
&= O_P\left(\frac{r_\alpha}{T}+B_\alpha^d(G_d)\right),
\end{align*}
where we used Lemma \ref{lm.clus} to obtain the last equality.
 Similarly, we have 
 $$\frac{1}{T_d}\sum_{t\in \mathcal{T}_d}\frac{1}{T^d_{c_t^d}}\sum_{s\in\mathcal{T}_d} \mathbf{1}\{c^d_s=c^d_t\}\left\|\gamma_t-\gamma_s\right\|^2=O_P\left(\frac{r_\gamma}{N}+ B_\gamma^d(C_d)\right).$$
 This yields
 $$\frac{1}{N_dT_d}\sum_{(i,t)\in\mathcal{O}_d}\left|\widetilde{h}_{itk}^d\right|=O_P\left(\frac{r_\alpha}{T}+B_\alpha^d(G_d)+\frac{r_\gamma}{N}+B_\gamma^d(C_d)\right). $$
 We obtain the result using that $$\frac{1}{N_dT_d}\sum_{(i,t)\in\mathcal{O}_d}\left(\widetilde{h}_{itk}^d\right)^2\le \left(\frac{1}{N_dT_d}\sum_{(i,t)\in\mathcal{O}_d}\left|\widetilde{h}_{itk}^d\right|\right)^2.$$
 The proof of the second statement is similar and, therefore, omitted.
\end{Proof}

\begin{Lemma}\label{lm.aux2}
    Let Assumptions~\ref{as.fac}--
    \ref{as.errors} hold. Then, for all $k,\ell\in\{1,\dots,K\}$ and $d\in\{1,\dots,4\}$, as $N$ and $T$ tend to infinity, we have 
    $$\frac{1}{N_dT_d}\sum_{(i,t)\in\mathcal{O}_d}  (\widetilde{u}_{itk}^d-u_{itk})u_{it\ell}=O_P\left(\frac{G_d}{N}+\frac{C_d}{T}+ \frac{G_d C_d}{NT}\right).$$
\end{Lemma}

\begin{Proof}
Fix $(k,\ell,d)\in\{1,\ldots,K\}^2\times\{1,\ldots,4\}$. We have 

  \begin{align*}
  \frac{1}{N_dT_d}\sum_{(i,t)\in\mathcal{O}_d}  (\widetilde{u}_{itk}^d-u_{itk})u_{it\ell}&=
      \frac{1}{N_dT_d}\sum_{(i,t)\in\mathcal{O}_d}  \left(\bar{u}_{g_i^dt}^d+\bar{u}_{ic_t^d}^d-\bar{u}_{g_i^dc_t^d}^d\right)_ku_{it\ell} \\
      &= J_1+J_2+J_3,
  \end{align*}
  where 
  \begin{align*}
      J_1:=       \frac{1}{N_dT_d}\sum_{(i,t)\in\mathcal{O}_d}  \left(\bar{u}_{g_i^dt}^d\right)_ku_{it\ell},\
      J_2:= \frac{1}{N_dT_d}\sum_{(i,t)\in\mathcal{O}_d}  \left(\bar{u}_{ic_t^d}^d\right)_ku_{it\ell},\
      J_3 := \frac{1}{N_dT_d}\sum_{(i,t)\in\mathcal{O}_d}  \left(\bar{u}_{g_i^dc_t^d}^d\right)_ku_{it\ell}.
  \end{align*}
Let us bound $J_1$. It holds that
    \begin{align*}
  J_1&= \frac{1}{N_dT_d}\sum_{(i,t)\in\mathcal{O}_d}  \left(\frac{1}{N^d_{g_i^d}}\sum_{j\in\mathcal{N}_d} \mathbf{1}\{g^d_j=g^d_i\} u_{jtk}\right)u_{it\ell}\\
  &=\frac{1}{N_dT_d}\sum_{g=1}^{G_d} \sum_{(i,t)\in\mathcal{O}_d}  \left(\frac{1}{N^d_{g}}\sum_{j\in\mathcal{N}_d} \mathbf{1}\{g^d_j=g\} u_{jtk}\right)\mathbf{1}\{g^d_i=g\}u_{it\ell}\\
  &= \frac{1}{N_dT_d}\sum_{g=1}^{G_d} \sum_{t\in\mathcal{T}_d}  \left(\frac{1}{\sqrt{N^d_{g}}}\sum_{j\in\mathcal{N}_d} \mathbf{1}\{g^d_j=g\} u_{jtk}\right)\left(\frac{1}{\sqrt{N^d_{g}}}\sum_{j\in\mathcal{N}_d} \mathbf{1}\{g^d_j=g\} u_{jt\ell}\right).
  \end{align*}
  By the triangle inequality,
  \begin{align*}
      |J_1|&\le \frac{1}{N_dT_d}\sum_{g=1}^{G_d} \sum_{t\in\mathcal{T}_d} \left|\frac{1}{\sqrt{N^d_{g}}}\sum_{j\in\mathcal{N}_d} \mathbf{1}\{g^d_j=g\} u_{jtk}\right|\left|\frac{1}{\sqrt{N^d_{g}}}\sum_{j\in\mathcal{N}_d} \mathbf{1}\{g^d_j=g\} u_{jt\ell}\right| \\
      & \le \frac{1}{N_dT_d}\sum_{k=1}^K\sum_{g=1}^{G_d} \sum_{t\in\mathcal{T}_d}\left(\frac{1}{\sqrt{N^d_{g}}}\sum_{j\in\mathcal{N}_d} \mathbf{1}\{g^d_j=g\} u_{jtk}\right)^2.
  \end{align*}
 Next, by Assumption \ref{as.errors}, since conditional on $\mathcal F_{NT}$, $(u_{jtk})_{j\in\mathcal N_d}$ are mean-zero independent random variables, independent of $(g_j^d)_{j\in\mathcal N_d}$, we have
  \begin{align*}
    \E\left[\left(\frac{1}{\sqrt{N^d_{g}}}\sum_{j\in\mathcal{N}_d} \mathbf{1}\{g^d_j=g\} u_{jtk}\right)^2\right]&= \E\left[\frac{1}{N^d_g}\sum_{j\in\mathcal{N}_d} \mathbf{1}\{g^d_j=g\} u_{jtk}^2\right]\\
    &=\E\left[\frac{1}{N^d_g}\sum_{j\in\mathcal{N}_d}\E\left[\mathbf{1}\{g^d_j=g\} |\mathcal F_{NT}\right]\E\left[u_{jtk}^2|\mathcal F_{NT}\right]\right] \\
    &\leq M \E\left[\frac{1}{N^d_g}\sum_{j\in\mathcal{N}_d}\mathbf{1}\{g^d_j=g\}\right]
    =M.
  \end{align*}
  As a result, we get
  $\E[|J_1|]\le (KMG_d)/N_d.$
  This yields $J_1=O_P\left(G_d/N\right).$
Similarly, we have $J_2= O_P\left(C_d/T\right).$
Moreover, it holds that 
\begin{align*}
    J_3&= \frac{1}{N_dT_d}\sum_{(i,t)\in\mathcal{O}_d}  \left(\frac{1}{N^d_{g_i^d}T^d_{c_t^d}}\sum_{(j,s)\in\mathcal{O}_d} \mathbf{1}\{g^d_j=g^d_i\}\mathbf{1}\{c^d_s=c^d_t\} u_{jsk}\right)u_{it\ell}\\
  &=\frac{1}{N_dT_d}\sum_{g=1}^{G_d} \sum_{c=1}^{C_d}\Biggr[\left(\frac{1}{\sqrt{N^d_gT^d_c}}\sum_{(j,s)\in\mathcal{O}_d} \mathbf{1}\{g^d_j=g\}\mathbf{1}\{c^d_s=c\} u_{jsk}\right)\\
  &\quad \times \left(\frac{1}{\sqrt{N^d_gT^d_c}}\sum_{(j,s)\in\mathcal{O}_d} \mathbf{1}\{g^d_j=g\}\mathbf{1}\{c^d_s=c\} u_{js\ell}\right)\Biggr].
\end{align*}
Then, by arguments similar to the ones allowing to bound $J_1$, we obtain 
$J_3=O_P\left((G_d C_d)/(NT)\right).$
The result follows from combining the bounds on $J_1,J_2$, and $J_3$.
\end{Proof}
\begin{Lemma}\label{lm.aux3}
    Let Assumptions~\ref{as.fac}--
    \ref{as.errors} hold. Then, for all $k\in\{1,\dots,K\}$ and $d\in\{1,\dots,4\}$, as $N$ and $T$ tend to infinity, we have 
    $$\frac{1}{\sqrt{N_dT_d}}\sum_{(i,t)\in\mathcal{O}_d}  (\widetilde{u}_{itk}^d-u_{itk})v_{it}=O_P\left(\sqrt{\frac{G_d}{N}}+\sqrt{\frac{C_d}{T}}+ \sqrt{\frac{G_d C_d}{NT}}\right).$$
\end{Lemma}
\begin{Proof} Fix $(k,d)\in\{1,\ldots,K\}\times\{1,\ldots,4\}$. We have
    \begin{align*}
  \frac{1}{\sqrt{N_dT_d}}\sum_{(i,t)\in\mathcal{O}_d}  (\widetilde{u}_{itk}^d-u_{itk})v_{it}&=
     \frac{1}{\sqrt{N_dT_d}}\sum_{(i,t)\in\mathcal{O}_d}  \left(\bar{u}_{g_i^dt}^d+\bar{u}_{ic_t^d}^d-\bar{u}_{g_i^dc_t^d}^d\right)_kv_{it} \\
      &= J_1+J_2+J_3,
  \end{align*}
  where 
  \begin{align*}
      J_1&:=         \frac{1}{\sqrt{N_dT_d}}\sum_{(i,t)\in\mathcal{O}_d} \left(\bar{u}_{g_i^dt}^d\right)_kv_{it},\\
      J_2&:=   \frac{1}{\sqrt{N_dT_d}}\sum_{(i,t)\in\mathcal{O}_d}  \left(\bar{u}_{ic_t^d}^d\right)_k v_{it},\\
      J_3 &:=   \frac{1}{\sqrt{N_dT_d}}\sum_{(i,t)\in\mathcal{O}_d}  \left(\bar{u}_{g_i^dc_t^d}^d\right)_kv_{it}.
  \end{align*}
Let us bound $J_1$. First, notice that by Assumption \ref{as.errors}, conditional on $\mathcal F_{NT}$, $(v_{it})_{(i,t)\in\mathcal O_d}$ is a sequence of mean-zero independent random variables mutually independent of $(\bar{u}^d_{g_i^dt})_{(i,t)\in\mathcal O_d}$. Hence, we have $\E[J_1]=0$ and
\begin{align*}
    \E[J_1^2]&= \E\left[\left(  \frac{1}{\sqrt{N_dT_d}}\sum_{(i,t)\in\mathcal{O}_d}\left(\bar{u}_{g_i^dt}^d\right)_kv_{it}\right)^2\right]\\
    &= \E\left[\frac{1}{N_dT_d}\sum_{(i,t)\in\mathcal{O}_d}  \left(\bar{u}^d_{g_i^dt}\right)_k^2v_{it}^2\right]\leq M\E\left[\frac{1}{N_dT_d}\sum_{(i,t)\in\mathcal{O}_d}  \left(\bar{u}^d_{g_i^dt}\right)_k^2\right].
\end{align*}
Second, by the same arguments, it holds that 
\begin{align*}
   & \E\left[\frac{1}{N_dT_d}\sum_{(i,t)\in\mathcal{O}_d}  \left(\bar{u}^d_{g_i^dt}\right)_k^2\right]\\
   &= \E\left[\frac{1}{N_dT_d}\sum_{(i,t)\in\mathcal{O}_d}  \left(\frac{1}{N^d_{g_i^d}}\sum_{j\in\mathcal{N}_d} \mathbf{1}\{g^d_j=g^d_i\} u_{jtk}\right)^2\right]\\
    &=\E\left[\frac{1}{N_dT_d}\sum_{(i,t)\in\mathcal{O}_d}\left(\frac{1}{N^d_{g_i^d}}\right)^2\sum_{j\in\mathcal{N}_d} \mathbf{1}\{g^d_j=g_i^d\} u_{jtk}^2\right]\\
    &=\E\left[\frac{1}{N_dT_d}\sum_{g=1}^{G_d}\sum_{(i,t)\in\mathcal{O}_d}\left(\frac{1}{N^d_{g}}\right)^2\sum_{j\in\mathcal{N}_d} \mathbf{1}\{g^d_j=g\}\mathbf 1\{g_i^d=g\} u_{jtk}^2\right]\\
    &\leq M\E\left[\frac{1}{N_dT_d}\sum_{g=1}^{G_d}\sum_{(i,t)\in\mathcal{O}_d}\left(\frac{1}{N^d_{g}}\right)^2\sum_{j\in\mathcal{N}_d} \mathbf{1}\{g^d_j=g\}\mathbf 1\{g_i^d=g\}\right]=\frac{MG_d}{N_d}.
\end{align*}
This yields
$J_1=O_P\left(\sqrt{G_d/N}\right).$
Similarly, we have 
$J_2=O_P\left(\sqrt{C_d/T}\right).$
Finally, following the arguments used to bound $J_1$, we have $\E[J_3]=0$ and 
$$\E[J_3^2]\leq M\E\left[\frac{1}{N_dT_d}\sum_{(i,t)\in\mathcal{O}_d}  \left(\bar{u}^d_{g_i^dc_t^d}\right)_k^2\right].$$
Next, notice that 
\begin{align*}
   & \E\left[\frac{1}{N_dT_d}\sum_{(i,t)\in\mathcal{O}_d}  \left(\bar{u}^d_{g_i^dc_t^d}\right)_k^2\right]\\
   &= \E\left[\frac{1}{N_dT_d}\sum_{(i,t)\in\mathcal{O}_d}  \left(\frac{1}{N^d_{g_i^d}T^d_{c_t^d}}\sum_{(j,s)\in\mathcal{O}_d} \mathbf{1}\{g^d_j=g^d_i\} \mathbf{1}\{c^d_s=c^d_t\}u_{jsk}\right)^2\right]\\
         &= \E\left[\frac{1}{N_dT_d}\sum_{(i,t)\in\mathcal{O}_d}  \left(\frac{1}{N^d_{g_i^d}T^d_{c_t^d}}\right)^2\sum_{(j,s)\in\mathcal{O}_d} \mathbf{1}\{g^d_j=g^d_i\} \mathbf{1}\{c^d_s=c^d_t\}u_{jsk}^2\right]\\
         &= \E\left[\frac{1}{N_dT_d}\sum_{g=1}^{G_d}\sum_{c=1}^{C_d}\sum_{(i,t),(j,s)\in\mathcal{O}_d}  \left(\frac{1}{N^d_{g}T^d_{c}}\right)^2 \mathbf{1}\{g^d_j=g^d_i=g\} \mathbf{1}\{c^d_s=c^d_t=c\}u_{jsk}^2\right]\\
         &\leq M \E\left[\frac{1}{N_dT_d}\sum_{g=1}^{G_d}\sum_{c=1}^{C_d}1\right]\leq \frac{MG_dC_d}{N_dT_d}.
\end{align*}
This yields $J_3=O_P\left(\sqrt{(G_dC_d)/(NT)}\right).$
We obtain the result by combining the bounds on $J_1,J_2,$ and $J_3$.
\end{Proof}

\begin{Lemma}\label{lm.aux4}
    Let Assumptions~\ref{as.fac}--
    \ref{as.approx} hold. Then, for all $k,\ell\in\{1,\dots,K\}$ and $d\in\{1,\dots,4\}$, as $N,T,G_d,C_d$ tend to infinity, we have 
    \begin{align*}
        \frac{1}{\sqrt{N_dT_d}}\sum_{(i,t)\in\mathcal{O}_d}\widetilde{h}_{itk}^d v_{it}&=o_P\left(1\right),       \\ \frac{1}{\sqrt{N_dT_d}}\sum_{(i,t)\in\mathcal{O}_d}u_{itk}\widetilde{f}_{it}^d&=o_P\left(1\right),\\
        \frac{1}{\sqrt{N_dT_d}}\sum_{(i,t)\in\mathcal{O}_d} \widetilde{h}_{itk}^du_{it\ell}&=o_P\left(1\right).
    \end{align*}
\end{Lemma}
\begin{Proof}
    We only prove the first statement, as the proofs of the other two are similar. By Assumption~\ref{as.errors}, conditional on $\mathcal F_{NT}$,  $(v_{it})_{(i,t)\in\mathcal O_d}$ are mean-zero random variables independent of $(\widetilde{h}_{itk}^d)_{(i,t)\in\mathcal O_d}$. Hence, we have 
    $$\E\left[\frac{1}{\sqrt{N_dT_d}}\sum_{(i,t)\in\mathcal{O}_d}\widetilde{h}_{itk}^d v_{it}\right]=0.$$
    Moreover, it holds that 
    \begin{align*}
        \E\left[\left(\frac{1}{\sqrt{N_dT_d}}\sum_{(i,t)\in\mathcal{O}_d}\widetilde{h}_{itk}^d v_{it}\right)^2\right]
        &=\E\left[\frac{1}{N_dT_d}\sum_{(i,t)\in\mathcal{O}_d}\left(\widetilde{h}_{itk}^d \right)^2v_{it}^2\right] \\
        &\leq M\E\left[\frac{1}{N_dT_d}\sum_{(i,t)\in\mathcal{O}_d}\left(\widetilde{h}_{itk}^d \right)^2\right]. 
    \end{align*}
    By Lemma \ref{lm.aux1} and Assumptions~\ref{as.rates}--\ref{as.approx}, we have 
   $$\frac{1}{N_dT_d}\sum_{(i,t)\in\mathcal{O}_d}\left(\widetilde{h}_{itk}^d\right)^2=O_P\left(\left(\frac{r_\alpha}{T}\right)^2+\left(\frac{r_\gamma}{N}\right)^2+ B^d_\alpha(G_d)^2+ B_\gamma^d(C_d)^2\right)=o_P(1).$$
   Since $ \left(\widetilde{h}_{itk}^d \right)^2$ is bounded (because $h_k$ is bounded itself by Assumption \ref{as.fac}), this yields
   $$\E\left[\frac{1}{N_dT_d}\sum_{(i,t)\in\mathcal{O}_d}\left(\widetilde{h}_{itk}^d \right)^2\right]=o(1).$$ We obtain the result since this implies $$\E\left[\left(\frac{1}{\sqrt{N_dT_d}}\sum_{(i,t)\in\mathcal{O}_d}\widetilde{h}_{itk}^d v_{it}\right)^2\right]=o(1).$$
\end{Proof}
\section{Proof of Lemma \ref{lmm.data.driven}}

Fix $d\in\{1,\dots,4\}$. We only show the result for $\widehat{G}_d$; the proof for $\widehat{C}_d$ is similar and, therefore, omitted.
We have 
\begin{align*}
     B_\alpha^d(G_d)&= \min\limits_{\begin{array}{c}\alpha(1),\dots,\alpha(G_d)\in\R^{K_\alpha}\\\tilde g_i\in\{1,\dots,G_d\},\ i\in\mathcal{N}_{d}\end{array}}\frac{1}{N_d}\sum_{i\in\mathcal{N}_{d}} \left\|\alpha_i-\alpha(\tilde g_i)\right\|^2\\
     &=  \min\limits_{\begin{array}{c}\tilde g_i\in\{1,\dots,G_d\},\ i\in\mathcal{N}_{d}\end{array}}\frac{1}{N_d}\sum_{i\in\mathcal{N}_{d}} \left\|\alpha_i-\frac{1}{\sum_{j\in\mathcal{N}_d}\mathbf{1}\{\tilde{g}_j=\tilde{g}_i\}}\sum_{j\in\mathcal{N}_d}\mathbf{1}\{\tilde{g}_j=\tilde{g}_i\}\alpha_j\right\|^2.
\end{align*}
Following the arguments of Step 2 of the proof of Lemma \ref{lm.clus}, we obtain that there exists $L>0$ such that 
\begin{align*}
     B_\alpha^d(G_d)&\le 2L\left[\min\limits_{\begin{array}{c}\tilde g_i\in\{1,\dots,G_d\},\ i\in\mathcal{N}_{d}\end{array}}\frac{1}{N_d}\sum_{i\in\mathcal{N}_{d}}\left\|\varphi_d^\alpha(\alpha_i)-\frac{1}{N^d_{\tilde g_i}}\sum_{j\in\mathcal{N}_d}\mathbf{1}\{\tilde g_j=\tilde g_i\} \varphi_d^\alpha(\alpha_j)\right\|^2\right].
\end{align*}
Moreover, for all $\tilde g_i\in\{1,\dots,G_d\},\ i\in\mathcal{N}_{d} $, we have 
\begin{align*}
&\frac{1}{N_d}\sum_{i\in\mathcal{N}_{d}} \left\|\varphi_d^\alpha(\alpha_i)-\frac{1}{N^d_{\tilde g_i}}\sum_{j\in\mathcal{N}_d}\mathbf{1}\{\tilde g_j=\tilde g_i\} \varphi_d^\alpha(\alpha_j)\right\|^2\\
&= \frac{1}{N_d}\sum_{i\in\mathcal{N}_{d}} \left\|\varphi_d^\alpha(\alpha_i)-a_i^d+a_{i}^d-\frac{1}{N^d_{\tilde g_i}}\sum_{j\in\mathcal{N}_d}\mathbf{1}\{\tilde g_j=\tilde g_i\} \{\varphi_d^\alpha(\alpha_j)-a_j^d+a_j^d\}\right\|^2\\
&=  \frac{3}{N_d}\sum_{i\in\mathcal{N}_{d}}\left\|\varphi_d^\alpha(\alpha_i)-a_i^d\right\|^2+\frac{3}{N_d}\sum_{i\in\mathcal{N}_{d}}\left\|\frac{1}{N^d_{\tilde g_i}}\sum_{j\in\mathcal{N}_d}\mathbf{1}\{\tilde g_j=\tilde g_i\} \{\varphi_d^\alpha(\alpha_j)-a_j^d\} \right\|^2\\
&\quad +\frac{3}{N_d}\sum_{i\in\mathcal{N}_{d}}\left\|a_i^d-\frac{1}{N^d_{\tilde g_i}}\sum_{j\in\mathcal{N}_d}\mathbf{1}\{\tilde g_j=\tilde g_i\} a_j^d\right\|^2,
\end{align*}
where we used the triangle inequality and the classical inequality $ab\le (a^2+b^2)/2$. 
By Assumption \ref{as.clus}\eqref{clusi}, this yields 
$
    B_\alpha^d(G_d) \le 6LQ_g^d(G_d)+ O_P\left(\frac{r_\alpha}{T}\right).
$
Since $ Q_g^d(\widehat{G}_d)\le \widehat{V}_g^d=O_P(1/T)$, we obtain, by Assumption \ref{as.rates}\eqref{ratesi}, 
$$B_\alpha(\widehat{G}_d)= O_P\left(\frac{r_\alpha}{T}\right)=o_P\left(\frac{1}{(NT)^{1/4}}\right).$$

\section{Sufficient conditions for Assumption \ref{as.clus}}\label{sec.suff_app}
\begin{Lemma}\label{lmm.suff_c}
    The following holds:
    \begin{enumerate}[\textup{(}i\textup{)}]  
\item\label{suffi}
    If, conditional on $\alpha_i$, $(z_{it})_{t\in\mathcal{T}_d}$ are independent sub-Gaussian random variables with common mean $\E[z_{it}|\alpha_i]$ and sub-Gaussian norm bounded uniformly in $t$ and the value of $\alpha_i$, then, as $N$ and $T$ tend to infinity,
    $$\max_{i\in\mathcal N_d}\left\|a_i^d-\varphi_d^\alpha(\alpha_i)\right\|^2=O_P\left(\frac{\log(N)}{T}\right),$$
    with $\varphi_d^\alpha(\alpha_i)=\E[z_{it}|\alpha_i]$.
\item\label{suffii}
       If, conditional on $\gamma_t$, $(z_{it})_{t\in\mathcal{N}_d}$ are independent sub-Gaussian random variables with common mean $\E[z_{it}|\gamma_t]$ and with sub-Gaussian norm bounded uniformly in $i$ and the value of $\gamma_t$, then, as $N$ and $T$ tend to infinity,
    $$\max_{t\in\mathcal T_d}\left\|b_t^d-\varphi_d^\gamma(\gamma_t)\right\|^2=O_P\left(\frac{\log(T)}{N}\right),$$
     with $\varphi_d^\gamma(\gamma_t)=\E[z_{it}|\gamma_t]$.
\end{enumerate}
\end{Lemma}
\begin{Proof}
    We only show \eqref{suffi}, the proof of \eqref{suffii} being similar. Let $m>0$ be the bound on the sub-Gaussian norm of $z_{it}$ conditional on $\alpha_i$. By Theorem 2.6.2 in \cite{vershynin2018high}, there exists a constant $c>0$ such that, for all $k\in\{1,\dots, K+1\}$, and $\epsilon>0$, we have
    $$\P\left(\left.\left|\frac{1}{T_d}\sum_{t\in\mathcal{T}_d} z_{itk}- \E[z_{itk}|\alpha_i]\right|^2\ge \epsilon \right|\alpha_i\right)\le 2\exp\left(-\frac{c\epsilon T_d}{ m^2}\right).$$
    By the law of iterated expectations, we obtain 
        $$\P\left(\left|\frac{1}{T_d}\sum_{t\in\mathcal{T}_d} z_{itk}- \E[z_{itk}|\alpha_i]\right|^2\ge \epsilon\right)\le 2\exp\left(-\frac{c\epsilon T_d}{ m^2}\right).$$
    By the pigeonhole principle, this implies
       \begin{align*}\P\left(\left\|a_i^d- \E[z_{it}|\alpha_i]\right\|^2\ge \epsilon \right)&\le \sum_{k=1}^{K+1} \P\left(\left|\frac{1}{T_d}\sum_{t\in\mathcal{T}_d} z_{itk}- \E[z_{itk}|\alpha_i]\right|^2\ge \frac{\epsilon}{K+1} \right) \\
       &\le 2(K+1)\exp\left(-\frac{c\epsilon T_d}{ m^2(K+1)}\right).\end{align*}
       By the union bound, this yields
       $$\P\left(\max_{i\in\mathcal{N}_d}\left\|a_i^d- \E[z_{it}|\alpha_i]\right\|^2\ge \epsilon \right)\le 2(K+1)N_d\exp\left(-\frac{c\epsilon T_d}{m^2(K+1)}\right).$$
       We obtain the result by taking $\epsilon  \propto \log(N)/T.$
\end{Proof}

\end{document}